\documentclass[]{article}
\usepackage{amsmath,amssymb,amsbsy,graphicx,subfigure,multirow,hhline,epstopdf}
\usepackage[a4paper, left=2cm, right=2cm, top=2cm, bottom=2cm]{geometry}
\usepackage{natbib}
\usepackage[affil-it]{authblk}

\pdfoutput=1

\newcommand{\pdiff}[3][]{\dfrac{\partial^{#1} #2}{\partial {#3}^{#1}}}

\title{Air Entrainment in Dynamic Wetting: Knudsen Effects and the Influence of Ambient Air Pressure}

\author{James E. Sprittles}

\affil{Mathematics Institute, University of Warwick, Coventry, CV4 7AL, UK}

\begin{document}

\label{firstpage} \maketitle

\begin{abstract}
Recent experiments on coating flows and liquid drop impact both demonstrate that wetting failures caused by air entrainment can be suppressed by reducing the ambient gas pressure. Here, it is shown that non-equilibrium effects in the gas can account for this behaviour, with ambient pressure reductions increasing the gas' mean free path and hence the Knudsen number $Kn$.  These effects first manifest themselves through Maxwell slip at the gas' boundaries so that for sufficiently small $Kn$ they can be incorporated into a continuum model for dynamic wetting flows.  The resulting mathematical model contains flow structures on the nano-, micro- and milli-metre scales and is implemented into a computational platform developed specifically for such multiscale phenomena.  The coating flow geometry is used to show that for a fixed gas-liquid-solid system (a) the increased Maxwell slip at reduced pressures can substantially delay air entrainment, i.e.\ increase the `maximum speed of wetting', (b) unbounded maximum speeds are obtained as the pressure is reduced only when slip at the gas-liquid interface is allowed for and (c) the observed behaviour can be rationalised by studying the dynamics of the gas film in front of the moving contact line.  A direct comparison to experimental results obtained in the dip-coating process shows that the model recovers most trends but does not accurately predict some of the high viscosity data at reduced pressures. This discrepancy occurs because the gas flow enters the `transition regime', so that more complex descriptions of its non-equilibrium nature are required.  Finally, by collapsing onto a master curve experimental data obtained for drop impact in a reduced pressure gas, it is shown that the same physical mechanisms are also likely to govern splash suppression phenomena.
\end{abstract}

\section{Introduction}

The behaviour of liquid-solid-gas contact lines, at which an ambient gas is displaced by a liquid spreading over a solid, is the key element of many technological processes, ranging from coating flows used to apply thin liquid films to substrates \citep{weinstein04} through to 3D printers being developed to economically produce bespoke geometrically-complex structures \citep{derby10}.  Often, limitations on the operating devices are imposed by `wetting failure' which occurs when the liquid can no longer completely displace the gas at the contact line.  In coating technologies, this leads to the entrainment of gas bubbles into the liquid film which destroy its carefully tuned material properties whilst in drop-based flows entrainment of air under the moving contact line promotes the formation of splashes which, again, are usually undesirable.  Consequently, understanding the physical mechanisms which control wetting failure and, where possible, delaying this occurrence by manipulating the material parameters, flow geometry, etc, to ever increasing wetting speeds is one of the key aims of research into dynamic wetting.

Recent experimental studies have discovered new ways in which to delay the onset of wetting failures.  Specifically, in both coating flows  \citep{benkreira08,benkreira10}  and drop impact phenomena \citep{xu05,xu07}  it has been shown that a sufficient reduction in the pressure of the ambient gas, around a factor of ten in the former case and even less in the latter, can suppress wetting failures. This result was unexpected, as until \cite{xu05} it was assumed that the gas' dynamics can either be entirely neglected or, as found in coating research, only influences the system's behaviour through viscous forces which act in the thin gas film in front of the moving contact line \citep{kistler93}.  However, the significance of the viscous forces in the gas is usually characterised by the viscosity ratio between the gas and the liquid, and this remains unchanged by variations in gas pressure.  Why then, would reductions in ambient pressure have any influence on the contact line's dynamics?

An interpretation of the aforementioned phenomena, suggested in a number of different contexts, is that the reduction in ambient gas pressure, which leads to a corresponding decrease in gas density, increases the mean free path in this phase so that non-equilibrium gas dynamics \citep{chapman70}, also referred to as rarefied gas dynamics \citep{cercignani00}, become important \citep{marchand12,benkreira10,duchemin12}.  Specifically, boundary slip is generated in the gas whose magnitude is proportional to the mean free path and thus whose influence is increased by reductions in gas pressure. The increased slip allows gas trapped in thin films between the solid and the free-surface to escape more easily, so that the lubrication forces generated by the gas are reduced and its influence is negated.  

\begin{figure}
     \centering
\includegraphics[trim={0 15cm 1cm 7cm},clip,scale=0.75]{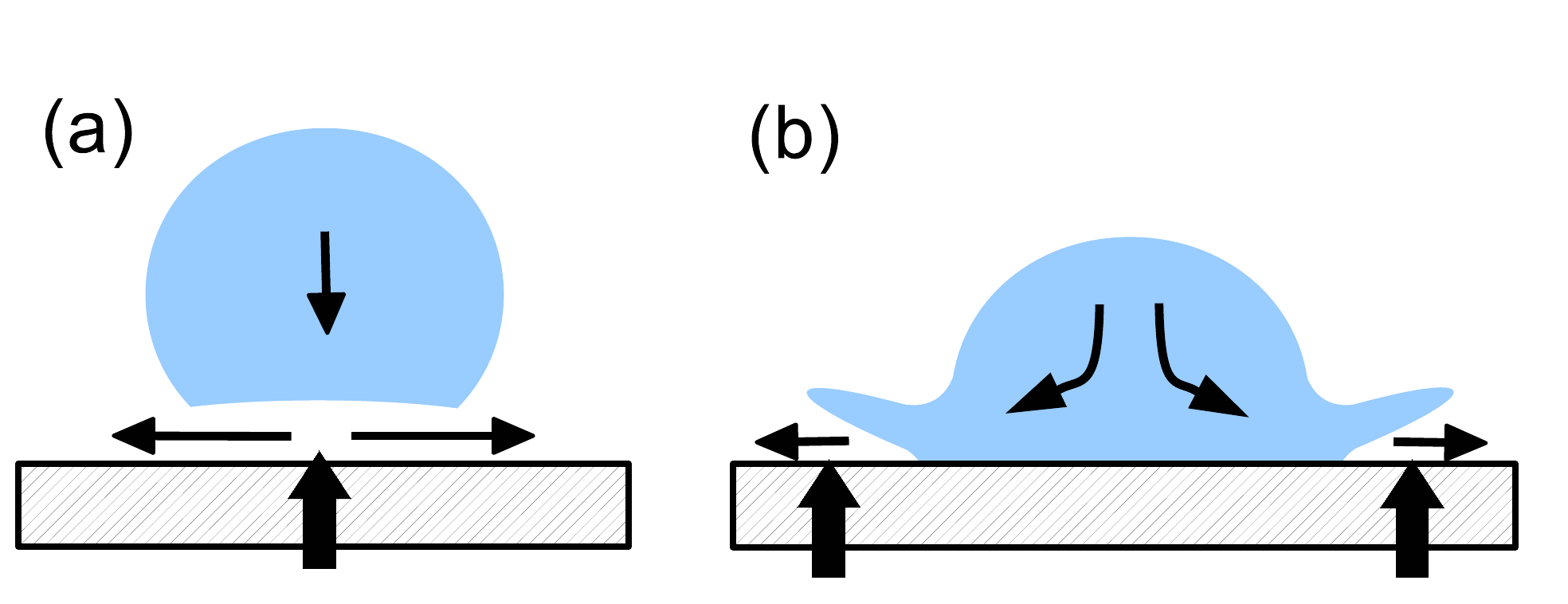}
 \caption{Sketch illustrating the influence of gas flow on the drop impact phenomenon.  The behaviour of the gas, in the regions indicated by block arrows, affects both (a) the pre-impact and (b) the post-impact spreading dynamics of the drop.}
 \label{F:drop}
\end{figure}

For drop impact phenomena, whose literature has been previously reviewed \citep{rein93,yarin06}, alterations in the lubrication forces could be important either (a) in the thin air cushion that exists under the drop \emph{before} it impacts the solid \citep{bouwhuis12,thoroddsen05a} and/or (b) in the gas film formed in front of the advancing contact line \emph{after} the drop has impacted the solid \citep{riboux14}, see Figure~\ref{F:drop}.  Theoretical work has also focussed on either (a) the pre-impact stage \citep{smith03,mani10,mandre12} or (b) the post-impact spreading phase where, with a few exceptions \citep{schroll10}, the gas is usually considered passive \citep{bussmann00,eggers10,sprittles_pof2}.   Distinguishing how each of these processes influences the spreading dynamics of the drop is a complex problem, both from a theoretical and experimental perspective, due to the inherently multiscale nature of the phenomenon.  As a result, establishing how reductions in ambient pressure suppress splashing remains an area of intensive debate \citep{driscoll11,kolinski12,ruiter12} with recent experimental results on the impact stage particularly interesting \citep{ruiter15,kolinski14}.  

In contrast to drop-based dynamics, for the dip-coating flow investigated in \cite{benkreira08} mechanism (a) does not exist, as up until the point of wetting failure, the liquid remains in contact with the solid and the motion is steady.  Therefore, studying this process allows us to isolate the influence of changes in ambient pressure on the behaviour of a moving contact line, i.e.\ effect (b), without any of the aforementioned complications that are inherent to the drop impact phenomenon. This is the path that will be pursued in this work with drop dynamics only re-considered again in \S\ref{S:drops} where, having characterised fully (b), our results enable us to infer some conclusions about drop splashing phenomena.

\subsection{Experimental Observations in Dip Coating}
\begin{figure}
     \centering
\includegraphics[trim={0 16cm 1cm 0},clip,scale=0.6]{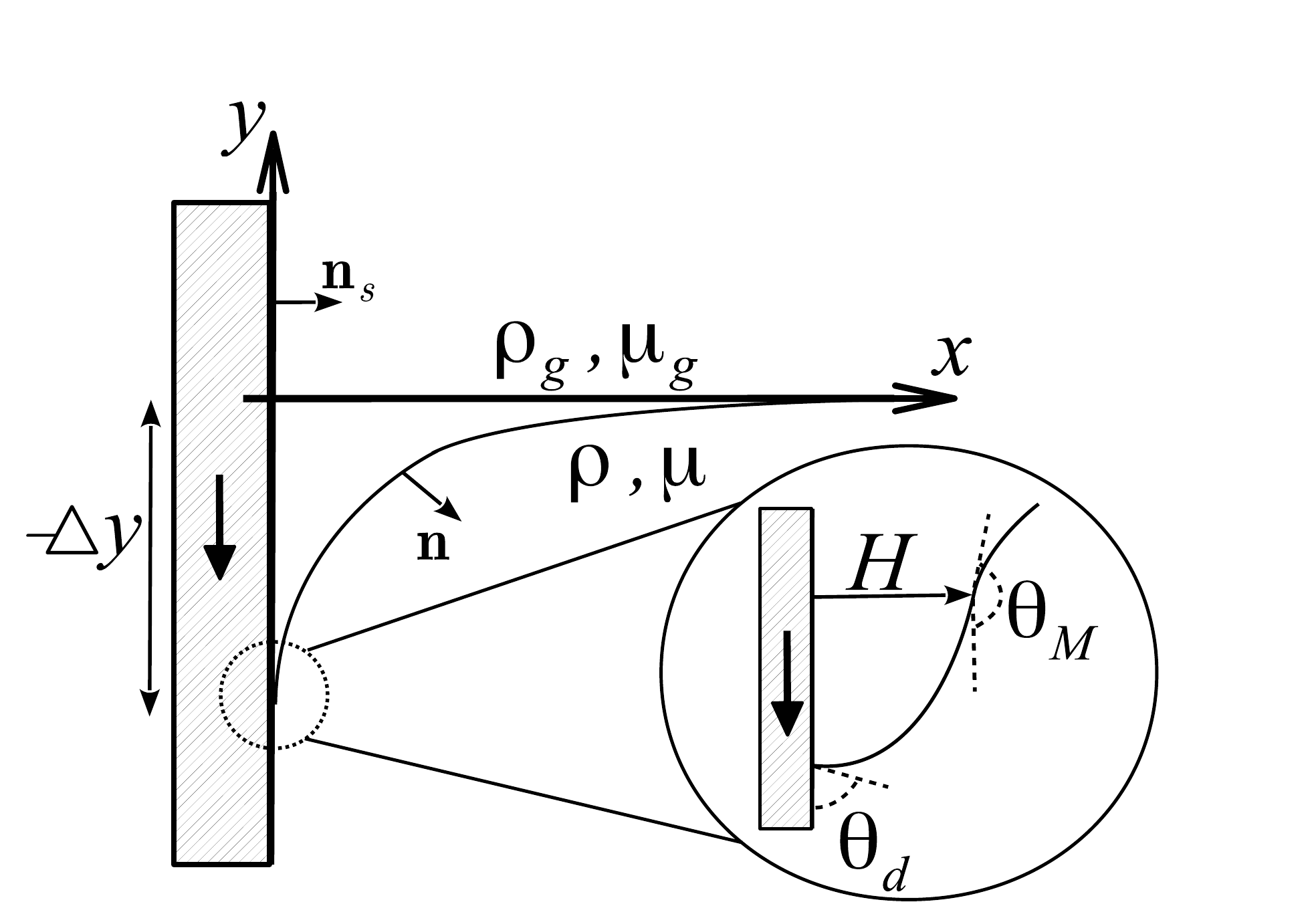}
 \caption{Sketch of the dip-coating flow configuration showing a free-surface formed between a liquid (below) and gas (above) which meets a solid surface at a contact line, which is a distance $\triangle y$ below the height of the surface ($y=0$).  The inset shows that the free-surface meets the solid at a contact angle $\theta_d$ but that a finite distance away has an apparent angle $\theta_M$ at an inflection point in the surface profile where the film has height $H$.}
 \label{F:sketch}
\end{figure}
The experimental setup constructed in \cite{benkreira08} is for dip-coating, where a plate/tape is continuously run through a liquid bath (Figure~\ref{F:sketch}) and gradually increased in speed until air is entrained into the liquid.  The speed at which this first occurs is called the `maximum speed of wetting' and will be denoted by $U_c^{\star}$, where stars will be used throughout to denote dimensional quantities.  In contrast to previous works \citep{burley76,blake02}, the ambient gas pressure $P_{g}^{\star}$ can be reduced from its atmospheric value by up to a factor of a hundred by mounting this apparatus inside a vacuum chamber. Silicone oils of differing viscosities are used as the coating liquid, because their relatively low volatility (compared to, say, water-glycerol mixtures) ensures no significant evaporation when the ambient pressure is reduced.  This work is extended in \cite{benkreira10} to consider also a range of different ambient gases (air, carbon dioxide and helium).  The key results from \cite{benkreira08} can be summarised as follows:\\
\begin{itemize}
  \item For a given liquid-solid-gas system, $U_c^{\star}$ can be increased by reducing $P_{g}^{\star}$.
  \item There is very little variation in $U_{c}^{\star}$ until $P_{g}^{\star}$ is approximately a tenth of its atmospheric value. Further reductions in $P_{g}^{\star}$ lead to sharp increases in $U_{c}^{\star}$.
  \item For sufficiently low $P_{g}^{\star}$, it is possible for high viscosity liquids to have a larger value of $U_{c}^{\star}$ than lower viscosity ones (the opposite of what occurs at atmospheric pressure).\\
\end{itemize}

In addition to these findings, in \cite{benkreira10} it is shown that:
\begin{itemize}
  \item For a given value of $P_{g}^{\star}$, using a gas with a larger mean free path increases $U_{c}^{\star}$.\\
  \end{itemize}   

In \cite{benkreira10}, the aforementioned experimental observations are assumed to be caused by non-equilibrium effects in the gas which become significant at sufficiently low pressures.  These effects, most notably slip at the gas film's boundaries, are interpreted in terms of an `effective gas viscosity' whose magnitude is reduced with decreases in ambient gas pressure. Using appropriate values for this parameter, obtained from measurements of the resistance of a gas trapped between parallel oscillating plates \citep{andrews95},  allows the data to be collapsed onto a single curve (Figure~9b in \cite{benkreira10}) and highlights the influence of non-equilibrium effects.

Although the effective viscosity may be a useful concept when trying to qualitatively interpret the experimental findings, as an \emph{integral} property of the system it will not enable us to formulate a \emph{local} fluid mechanical model for this problem.  In particular, the dependence of the effective gas viscosity on ambient gas pressure which is used in \cite{benkreira10} was obtained in \cite{andrews95} for a gas film of constant height $2\mu$m and will be different for other film profiles.  In reality, the gas film's height will vary all the way from the contact line to the far field, where a film can no longer be defined so that the notion of an effective viscosity no longer makes sense. Instead, we must look to capture the local physical mechanisms which alter the gas film's dynamics at reduced pressures.

\subsection{Mathematical Modelling of Dynamic Wetting Phenomena}

Consider the various levels of description that can be used to describe dynamic wetting phenomena depending on how the gas phase's dynamics are accounted for, with the simplest case first.

\subsubsection{Dynamic Wetting in a Passive Gas}

In many cases, the viscosity and density ratios indicate that the gas can be treated as dynamically passive.  Then, mathematical models for the liquid's dynamics must address two key aspects.  Firstly, the classical fluid mechanical model, with no-slip at fluid-solid boundaries, has no solution \citep{huh71,shik06}.  This is the so-called `moving contact line problem'.  And, secondly, the contact angle, which is seen to vary at experimental resolutions \citep{hoffman75,blake02}, must be prescribed as a boundary condition on the free-surface shape.  A review of the various classes of models proposed to address these issues can be found in Ch.~3 of \cite{shik07}.

To overcome the moving contact line problem, initially identified in \cite{huh71}, it is often assumed that some degree of slip occurs at the fluid-solid boundary \citep{dussan76}.  Methods for predicting the `slip length' of an arbitrary liquid-solid interface are not well developed but estimates suggest it is in the range $1$--$10$~nm for simple liquids \citep{lauga07}.  

The treatment of the dynamic contact angle remains a highly contentious issue.  The main question is whether or not the experimentally observed variation in the `apparent angle' \citep{wilson06} is caused entirely by the bending of the free surface below the resolution of the experiment, i.e.\ by the `viscous bending' mechanism quantified in \cite{cox86} and measured in \cite{dussan91,rame96}, or whether actual contact angle (sometimes referred to as the `microscopic angle') also varies.  In either case, the apparent angle depends on the capillary number, which is the dimensionless contact line speed, but in the former case the actual angle remains fixed whilst in the latter this angle varies, as in the molecular kinetic theory \citep{blake69}.  In more complex models, the angle can also depend on the flow field itself, as in the interface formation model \citep{shik07}.  

Models in which the contact angle is considered constant have been popular in recent years for investigating air entrainment phenomena and have produced a number of promising results \citep{vandre12,vandre13,vandre14,marchand12,chan13}.  What is clear from simulations is that viscous bending of the free surface becomes significant as the point of air entrainment is approached and cannot be neglected.  What remains unclear is whether this mechanism is sufficient to account for the experimentally observed dynamics of the apparent contact angle, particularly given that precise measurements of the magnitude of slip on the liquid-solid interface are lacking.  These difficulties have stimulated fundamental research into the contact line region's dynamics using molecular dynamics techniques \citep{deconinck08} but, as can be seen from a recent collection of papers \citep{velarde11} and review articles \citep{blake06,snoeijer13}, intense debate remains about the topic. 

The focus of this work is to characterise the effects which the gas dynamics can have on dynamic wetting phenomena and, consequently, the simplest possible model for the other aspects will be used.  Specifically, slip at the liquid-solid interface will be captured using the Navier condition \citep{navier23}, as commonly used in dynamic wetting models \citep{hocking76,huh77}, and the contact angle will be taken as a parameter that is fixed at its equilibrium value.  The result is a `conventional model' and a comprehensive review of all such models which can be `built' is given in Ch.~9 of \cite{shik97}. More complex models can easily be constructed on top of the model used here, but in the present context would distract from the main focus of this work.  

\subsubsection{In a Viscous Gas}

It is now well recognised that at sufficiently high coating speeds a thin film of gas forms in front of the moving contact line, as observed experimentally by laser-Doppler velocimetry in \cite{mues89}.  The thin nature of the film means that despite the large viscosity ratio, lubrication forces enhance the influence of the gas' dynamics so that they cannot be neglected. As coating speeds are increased, the resistance this film creates to contact line motion eventually becomes sufficiently large that it is entrained into the liquid \citep{vandre13,marchand12}. Therefore, to study the phenomenon of air entrainment and wetting failures, any theory developed must account for the dynamics of the viscous gas as well as the liquid.

In formulating a model for the gas' dynamics one again encounters the moving contact line problem, namely that if no-slip is used then the contact line can't move.  Consequently, slip has also often been applied on the solid-gas boundary with a magnitude, measured by the slip length, either explicitly stated to be the same as in the liquid \citep{vandre13} or implicitly assumed to be so \citep{cox86}.  In other words, slip is usually used as a mathematical fix to circumvent the moving contact line problem, rather than as a physical effect whose magnitude can significantly alter the flow.

\subsubsection{In a Non-Equilibrium Gas}

Experimental observations in \cite{marchand12}, where the film entrained by a plunging plate was measured using optical methods, suggest that for relatively viscous liquids the characteristic film thickness is rather small, in the range $H^{\star}\sim1$--$15~\mu$m. For flows where there is an external load on the film, such as in curtain coating, the film's thickness is likely to be even smaller.  Consequently, the film's dimensions become comparable to the mean free path in the gas $\ell^{\star}$ so that the Knudsen number $Kn_H=\ell^{\star}/H^{\star}$, characterising the importance of non-equilibrium gas effects, becomes non-negligible, particularly at reduced pressures.  For example, a factor of ten decrease in the ambient pressure of air, achieved in both \cite{benkreira08} and \cite{xu05}, gives a value of $\ell^{\star} = 0.7~\mu$m so that $Kn_H\sim0.1$. In this case, one may expect that non-equilibrium effects in the gas film will have a significant impact on its dynamics, and we must consider how this can be incorporated into our dynamic wetting model.

Non-equilibrium effects first manifest themselves at the gas' boundaries and can be accounted for by relaxing the no-slip condition to allow for boundary slip whilst continuing to use the the Navier-Stokes equations in the bulk. This approach remains accurate for $Kn_H<0.1$, during which gas flow is said to be in the `slip regime', after which Knudsen layers will begin to occupy the entire gas film, so that the scale separation between boundary and bulk effects no longer exists.

In \cite{maxwell79}, a slip condition was derived for the `micro-slip' at the actual boundary of the gas whose mathematical form remains the same, when $Kn_H \ll 1$, if `macro-slip' across a Knudsen layer of size $\sim\ell^{\star}$ is also accounted for, see \S3.5 of \cite{cercignani00}.  We shall henceforth refer to the discontinuity in tangential velocity as `Maxwell slip' which at an impermeable boundary is generated by the stress acting on that interface from the gas phase.  For two-dimensional isothermal flow, the component of velocity $u_g^{\star}$, where subscripts $g$ will denote properties of the gas, adjacent to a stationary planar surface located at $y=0$ is, in dimensional terms, given by:
\begin{equation}\label{slipa}
a~\ell^{\star}~\pdiff{u^{\star}_g}{y^{\star}}= u^{\star}_g
\end{equation}
where the coefficient of proportionality $a$ can depend on both the properties of the micro-slip, which will depend on the fraction of molecules which reflect diffusively and those which a purely specular, and the macro-slip generated across a Knudsen layer.  In practise, unless surfaces are molecularly smooth we will have $a\sim1$ \citep{agrawal08,millikan23,allen82} and, for simplicity, we will henceforth take $a=1$.

\begin{figure}
     \centering
\includegraphics[trim={0 9cm 0cm 7cm},clip,scale=0.5]{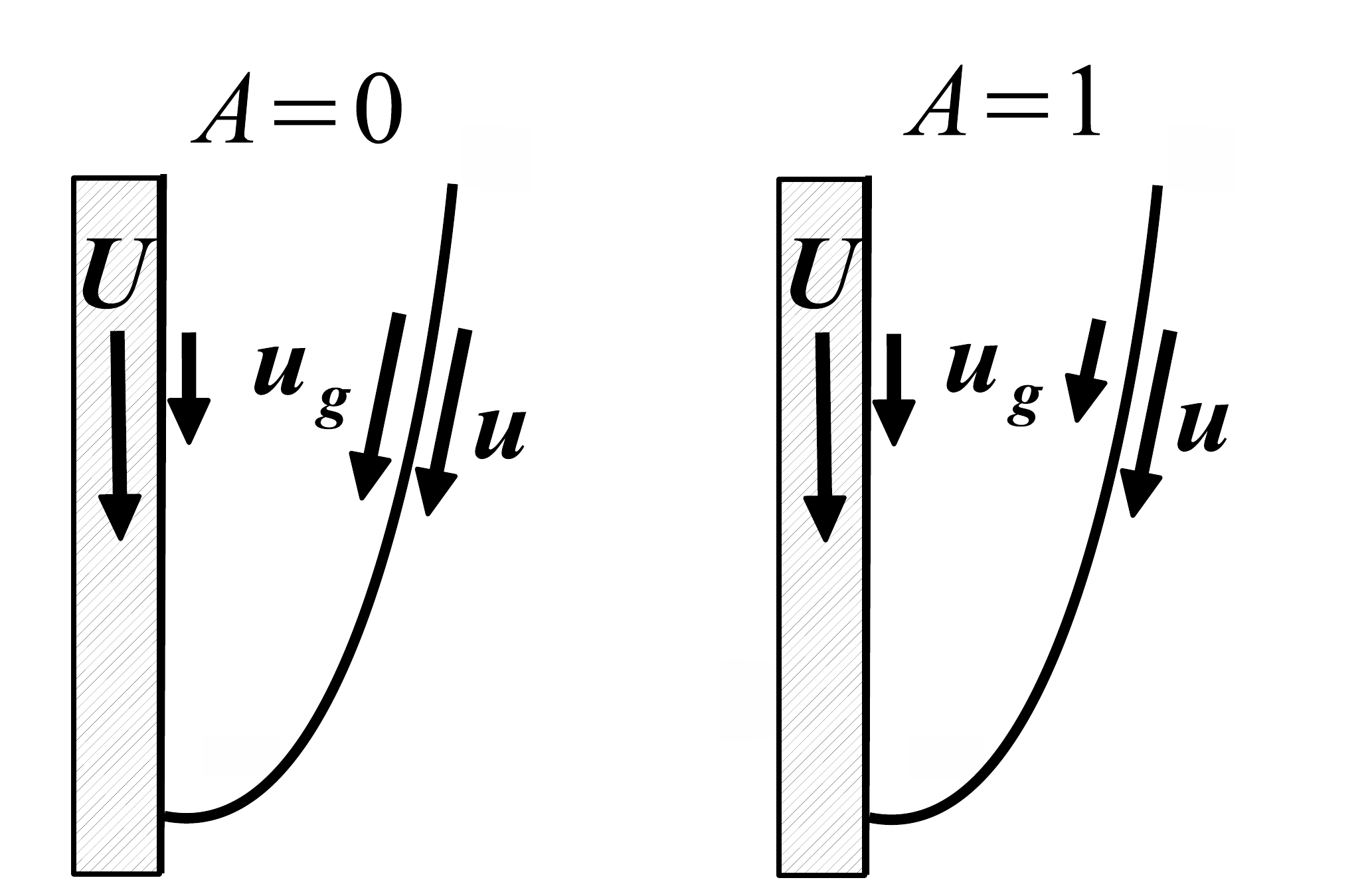}
 \caption{The effect of Maxwell-slip on the gas flow.  When $A=0$ slip is present at the gas-solid boundary but turned off at the gas-liquid boundary, so that $\mathbf{u}_g=\mathbf{u}_g$, whilst for $A=1$ it is present on both interfaces.}
 \label{F:slippy}
\end{figure}

The Maxwell slip condition (\ref{slipa}) has the same form as the slip conditions which have often been used in dynamic wetting flows to circumvent the moving contact line problem in the gas but occurs naturally here from the consideration of non-equilibrium gas effects.  Notably, this means that accurate values of the slip length can easily be obtained by knowing the mean free path in the gas, in contrast to the slip length in the liquid which is notoriously difficult to determine.  Furthermore, the approach taken here suggests that slip will also be present at the gas-liquid boundary (i.e.\ the free-surface), as demonstrated by experiments dating back to \cite{millikan23}.  This effect is shown in Figure~\ref{F:slippy}, where the parameter $A$ is used to change between the usual condition of continuity of velocity across the gas-liquid interface ($A=0$) and Maxwell-slip ($A=1$). 

Slip at the gas-liquid boundary has been accounted for in some models for the gas film in drop impact \citep{duchemin12,mani10}, and in great detail for drop collisions \citep{sundararajakumar96}, but is yet to be routinely incorporated into dynamic wetting models.  In most cases where slip on the gas-solid interface is accounted for, the same effect on the gas-liquid surface is not considered \citep{vandre13,riboux14}.  The only progress in this direction has been achieved in \cite{marchand12}, where non-equilibrium effects are accounted for by modifying the viscosity in the gas.  However, this is achieved by using an empirical expression for the effective viscosity which suffers from the same issues mentioned previously, namely a reliance on the lubrication approximation that does not enable one to fully formulate a uniformly-valid local fluid mechanical problem.

In order to apply Maxwell-slip at curved moving surfaces, the tensorial expression is required \citep{lockerby04}.  Specifically, for gas flow adjacent to a surface which moves with velocity $\mathbf{u}^{\star}$ and has normal $\mathbf{n}$ pointing into the gas, we have 
\begin{equation}\label{slip}
\ell^{\star}~\mathbf{n}\cdot\left(\nabla\mathbf{u}_g^{\star}+[\nabla\mathbf{u}_g^{\star}]^{T}\right)\cdot\left(\mathbf{I}-\mathbf{n}\mathbf{n}\right) = \mathbf{u}^{\star}_{g||}-\mathbf{u}^{\star}_{||},
\end{equation}
where the subscripts $\parallel$ denote the components of a vector tangential to the surface which can be obtained for an arbitrary vector $\mathbf{a}$ using $\mathbf{a}_{||}=\mathbf{a}\cdot\left(\mathbf{I}-\mathbf{n}\mathbf{n}\right)$, where $\mathbf{I}$ is the metric tensor of the coordinate system. 

If the gas is modelled as being composed of `hard spheres', then the mean free path is related to the temperature $T^{\star}$ and ambient pressure $P^{\star}_g$ by $\ell^{\star} = k^{\star}_{B}T^{\star}/(\sqrt{2}\pi P^{\star}_g (\l^{\star}_{mol})^2)$ where $l^{\star}_{mol}$ is the molecular diameter and $k^{\star}_B$ is Boltzmann's constant.  Notably, if local variations in gas pressure about $P^{\star}_{g}$ are small, which will be the case here, then for a given ambient pressure $\ell^{\star}$ is fixed throughout the entire gas phase. In practise, when considering the dependence of the mean free path of a gas on the ambient pressure at a fixed temperature it is convenient to use the expression
\begin{equation}\label{mfp}
\bar{\ell}\equiv\frac{\ell^{\star}}{\ell^{\star}_{atm}} = \frac{P^{\star}_{g,atm}}{P^{\star}_g} \equiv 1/\bar{P}_g 
\end{equation}
where $\bar{P}_g$ is the gas pressure normalised by its atmospheric value.  At atmospheric pressure, air has a mean free path $\ell^{\star}_{atm}=0.07~\mu$m and other commonly encountered gases take similar values $\ell^{\star}_{atm}\sim0.1~\mu$m.    

Although the Knudsen number in the gas film $Kn_H$ is of most interest, this can only be obtained once our computations have allowed us to find $H^{\star}$. Before then, the Knudsen number based on our characteristic scales $Kn=\ell^{\star}/L^{\star}$,  where $L^{\star}\sim1$~mm is the capillary length, must be used. Noting that the dimensionless mean free path at atmospheric pressure is $\ell_{atm}=\ell^{\star}_{atm}/L^{\star}$, the resulting expression for $Kn$ takes the form
\begin{equation}\label{kn}
Kn = \frac{\ell_{atm}}{\bar{P}_g}
\end{equation}
so that the explicit dependence on pressure reduction becomes clear.  

%

\subsection{Overview}

This work will answer the following questions:\\
\begin{itemize}

\item Can increases in Maxwell slip, caused by experimentally-attainable reductions in ambient gas pressure, significantly raise the maximum speed of wetting?

\item Does slip on the liquid-gas interface, usually neglected in this class of flows, have a substantial effect on the dynamic wetting process?

\item Will the magnitude of the Knudsen number based on the gas film's height $Kn_H$ be small enough for the gas flow to remain in the `slip regime'?

\item Is it possible to collapse the data from  \cite{xu05} onto a master curve by assuming that non-equilibrium gas effects are responsible for splash suppression in drop impact?\\

\end{itemize}

The paper is organised as follows.  In \S\ref{S:form} the dip-coating problem will be formulated and dimensionless parameters will be identified.  The mathematical model for this process will be seen to be inherently multiscale, with flow dynamics on the nano-, micro- and mm-scale caused by, respectively, slip in the liquid, slip in the gas and capillarity.  The computation of such multiscale flows is challenging and neglecting to resolve all scales in the problem yields inaccurate results \citep{sprittles_ijnmf}.  In \S\ref{S:comp}, the application of a computational framework developed in \cite{sprittles_ijnmf,sprittles_jcp} to this problem is described, which resolves all scales,  and our method for defining a maximum speed of wetting is shown.  Benchmark computations performed at atmospheric pressure are provided in the Appendix.  In \S\ref{S:anal} the effects of the non-equilibrium gas dynamics on a base state are analysed and the gas flow is described in a lubrication setting. After, in \S\ref{S:para}, a parametric study of the new system is conducted which allows us to identify the importance of the various competing physical mechanisms in the experimentally-verifiable regimes before, in \S\ref{S:exp}, directly comparing our computations to experimental data.  The implications of these findings for the drop impact phenomenon are considered in \S\ref{S:drops}. Finally, in \S\ref{S:discussion}, potential avenues of enquiry for future theoretical and experimental development in this area are highlighted. 

\section{Problem Formulation}\label{S:form}

Consider the flow generated by the steady motion of a smooth chemically-homogeneous solid surface which is driven through a liquid-gas free surface in a direction aligned with gravity at a constant speed $U^{\star}$ (see Figure~\ref{F:sketch}). A characteristic length scale for this problem is given by the capillary length $L^{\star}=\sqrt{\sigma^{\star}/((\rho^{\star}-\rho^{\star}_g)g^{\star})}\simeq \sqrt{\sigma^{\star}/(\rho^{\star} g^{\star})}$, where $\sigma^{\star}$ is the surface tension of the liquid-gas surface, which is assumed constant; $g^{\star}$ is the acceleration due to gravity; and $\rho^{\star},\rho_g^{\star}$ are the densities in the liquid and gas, respectively, with $\rho_g^{\star}/\rho^{\star} \ll 1$ in the liquid-gas systems of interest so that the capillary length can be determined solely from the liquid's density.  A scale for pressure is $\mu^{\star} U^{\star}/L^{\star}$, where $\mu^{\star}$ is the viscosity of the liquid and $\mu^{\star}_g$ will be its value in the gas. 
%

\subsection{Bulk equations}

For compressibility effects in the liquid and gas to be negligible we require that the Mach number $Ma=U^{\star}/c^{\star}$ in each phase remains small, where $c^{\star}$ is the speed of sound.  Given that $U^{\star}<1$~ms$^{-1}$ and $c^{\star}>100$~ms$^{-1}$ in all cases considered, we have $Ma<0.01$. However, this is not a sufficient condition for the flow to be considered incompressible, especially when there are thin films of gas present along which pressure can vary significantly, see \S4.5 of \cite{gadelhak06}. Here, we will assume that the flow in both phases can be described by the steady incompressible Navier-Stokes equations and in \S\ref{S:incom} the validity of this assumption will be confirmed from a-posteriori calculations.  Therefore, we have
\begin{equation}\label{ns}
\nabla\cdot\mathbf{u} = 0,\qquad Ca~\mathbf{u}\cdot\nabla\mathbf{u} = Oh^2\left(\nabla\cdot \mathbf{P} + \hat{\mathbf{g}}/Ca\right),
\end{equation}
\begin{equation}\label{ns1}
\nabla\cdot\mathbf{u}_g = 0,\qquad \bar{\rho}~Ca~\mathbf{u}_g\cdot\nabla\mathbf{u}_g  = Oh^2\left(\nabla\cdot\mathbf{P}_g + \bar{\rho}~\hat{\mathbf{g}}/Ca\right),\end{equation}
where the stress tensors in the liquid and gas are, respectively,
\begin{equation}\label{stresst}
\mathbf{P} = -p\mathbf{I} + \left[\nabla\mathbf{u}+\left(\nabla\mathbf{u}\right)^T\right] \qquad\hbox{and}\qquad \mathbf{P}_g = -p_g\mathbf{I} + \bar{\mu}\left[\nabla\mathbf{u}_g+\left(\nabla\mathbf{u}_g\right)^T\right].
\end{equation}
Here, $\mathbf{u}$ and $\mathbf{u}_g$ are the velocities in the liquid and gas; $p$ and $p_g$ are the local pressures in the liquid and gas, in contrast to uppercase $P$'s which are used for the ambient pressure in the gas; $Ca=\mu^{\star} U^{\star}/\sigma^{\star}$ is the capillary number based on the viscosity of the liquid; the viscosity ratio is $\bar{\mu}=\mu^{\star}_g/\mu^{\star}$; the ratio of gas to liquid densities is $\bar{\rho}=\rho^{\star}_g/\rho^{\star}$; and $\hat{\mathbf{g}}$ is a unit vector aligned with the gravitational field. The Ohnesorge number $Oh=\mu^{\star}/\sqrt{\rho^{\star}\sigma^{\star} L^{\star}}$ has been chosen as a dimensionless parameter, instead of the Reynolds number $Re=Ca/Oh^2=\rho^{\star} U^{\star}L^{\star}/\mu^{\star}$, as for a given liquid-solid-gas combination $Oh$ will remain constant, i.e.\ it will depend solely on material parameters and be independent of contact-line speed.  

\subsection{Boundary Conditions at the Gas-Solid Surface}

Conditions of impermeability and Maxwell slip give
\begin{equation}\label{sg_slip}
\mathbf{u}_g\cdot\mathbf{n}_s=0,\qquad  Kn~\mathbf{n}_s\cdot\mathbf{P}_g\cdot\left(\mathbf{I}-\mathbf{n}_s\mathbf{n}_s\right) = \mathbf{u}_{g\parallel}- \mathbf{U}_{\parallel},
\end{equation}
where $\mathbf{U}$ is the velocity of the solid and the normal to the solid surface is denoted as $\mathbf{n}_s$.  

\subsection{Boundary Conditions at the Liquid-Gas Free-Surface}

On the free-surface, whose location must be obtained as part of the solution, for a steady flow the kinematic equation and continuity of the normal component of velocity give that
\begin{equation}\label{fs_kin}
\mathbf{u}\cdot\mathbf{n}=\mathbf{u}_g\cdot\mathbf{n} = 0,
\end{equation}
where $\mathbf{n}$ is the normal to the surface pointing into the liquid phase.  These equations are combined with the standard balance of stresses with capillarity that give
\begin{equation}\label{fs_stress}
 \mathbf{n}\cdot\left(\mathbf{P}-\mathbf{P}_g\right)\cdot\left(\mathbf{I}-\mathbf{n}\mathbf{n}\right) =\mathbf{0},\qquad
\mathbf{n}\cdot\left(\mathbf{P}-\mathbf{P}_g\right)\cdot\mathbf{n} = \nabla\cdot\mathbf{n}/Ca.
\end{equation}

Equations (\ref{fs_kin}) and (\ref{fs_stress}) are usually combined with a condition stating that the velocity tangential to the interface is continuous across it $\mathbf{u}_{g\parallel}=\mathbf{u}_{\parallel}$.  However, when Maxwell slip is accounted for this is replaced with
\begin{equation}\label{fs_slip}
-A~Kn~\mathbf{n}\cdot\mathbf{P}_g\cdot\left(\mathbf{I}-\mathbf{n}\mathbf{n}\right) = \mathbf{u}_{g\parallel}- \mathbf{u}_{\parallel},
\end{equation}
where the minus sign occurs because the normal points into the liquid.  By setting $A=0$, Maxwell slip can be `turned off' at the gas-liquid interface, see Figure~\ref{F:slippy}, as has been the case in previous dynamic wetting works which only consider slip at the solid boundary.
  
\subsection{Boundary Conditions at the Liquid-Solid Surface}

The standard conditions of impermeability and Navier slip give
\begin{equation}\label{ls_slip}
\mathbf{u}\cdot\mathbf{n}_s=0,\qquad l_s~\mathbf{n}\cdot\mathbf{P}\cdot\left(\mathbf{I}-\mathbf{n}_s\mathbf{n}_s\right) = \mathbf{u}_{\parallel}- \mathbf{U}_{\parallel},
\end{equation}
where $l_s=l^{\star}_s/L^{\star}$ is the dimensionless slip length based on the (dimensional) slip length at the liquid-solid interface $l^{\star}_s$.  This parameter has no relation to the slip coefficient in the gas which was based on the mean free path $\ell^{\star}$.  In other words, although Maxwell slip and Navier slip have the same mathematical form, their physical origins differ and thus there is no reason to expect their coefficients to have similar magnitudes.

\subsection{Liquid-Solid-Gas Contact Line}

Equation (\ref{fs_stress}) requires a boundary condition at the contact line which is given by prescribing the contact angle $\theta_d$ satisfying
\begin{equation}\label{angle}
\mathbf{n}\cdot\mathbf{n}_s = -\cos\theta_d.
\end{equation}
Here, our focus is on understanding the effects of the gas dynamics on the dynamic wetting flow so that we will take the simplest option of assuming $\theta_d$ to be a constant.  More complex implementations, such as those in \cite{sprittles_jcp}, can be considered in future research.

\subsection{`Far Field' Conditions}

Dip coating is usually conducted in a tank of finite size whose dimensions are large enough not to alter the dynamics close to the contact line, as confirmed in \cite{benkreira08}.  The influence of system geometry, or `confinement', on the dynamic wetting process are well known \citep{ngan82} and, in particular, have been used to delay air entrainment \citep{vandre12}, but these effects are not the focus of this work.  Therefore, the `far field' boundaries of our domain, which are assumed to be no-slip solids $\mathbf{u}=\mathbf{u}_g=\mathbf{0}$, are set sufficiently far from the contact line that they have no influence on the dynamic wetting process.  Where the free-surface meets the far-field it is assumed to be perpendicular to the gravitational field, i.e.\ `flat', so that $\mathbf{n}\cdot\hat{\mathbf{g}}=1$.  In practise, setting the far field (dimensionally) to be twenty capillary lengths from the contact line in both the $x$ and $y$ directions was found to be sufficiently remote. 

\section{Computational Framework}\label{S:comp}

Dynamic wetting and dewetting flows have often been investigated using the simplifications afforded by the lubrication approximation \citep{voinov76,eggers04a}. However, for the two phase flow considered here, this approximation cannot simultaneously be valid in each phase, i.e.\ the liquid and gas phases can't both be thin films.  As shown in recent papers \citep{marchand12,vandre13}, plausible extensions in the spirit of the lubrication approach can be attempted, but their accuracy can only be ascertained from comparisons to full computations and is unlikely to be acceptable at larger capillary numbers where the contact line region cannot be considered as `localised'.

At present, to obtain accurate solutions to the mathematical model formulated in \S\ref{S:form}, one is left with no option other than to use computational methods. 

Computationally, resolving all scales in the problem is particularly important to ensure that our contact line equation (\ref{angle}) is applied to the `actual' or `microscopic' angle $\theta_d$, rather than the `apparent one' $\theta_{app}=\theta_{app}(s)$, i.e.\ the angle of the free surface at a finite distance $s$ from the contact line.  Although the apparent angle can, in certain circumstances, be related to the actual angle \citep{cox86}, this will not be true at the high capillary numbers observed in coating flows, particularly when non-standard gas dynamics are included. 

Here, gas dynamics will be built into a finite element framework developed in \cite{sprittles_ijnmf,sprittles_jcp} to capture dynamic wetting problems and then extended to consider two-phase flow problems in our work on the coalescence of liquid drops \citep{sprittles14_jfm2}. Notably, this code captures all physically-relevant length scales in the problem, from the slip length $l^{\star}_s\sim10$~nm up to the capillary length $L^{\star}\sim1$~mm, meaning that at least six orders of magnitude in spatial scale are resolved.  As a step-by-step user-friendly guide to the implementation has been provided in \cite{sprittles_ijnmf}, only the main details will be recapitulated here alongside some aspects which are specific to the current work.    

The code uses the arbitrary Lagrangian Eulerian scheme, based on the method of spines \citep{ruschak80,kistler83},  to capture the evolution of the free surface in two-dimensional or three-dimensional axisymmetric flows.  Both coating flows, which are often time-independent, as well as unsteady flows such as drop impact \citep{sprittles_pof} and drop coalescence \citep{sprittles_pof2,sprittles14_jfm2,sprittles14_pre} have been considered.  It has been confirmed that the code accurately captures viscous, inertial and capillarity effects, even when the mesh undergoes large deformations which inevitably occur when $Ca\sim 1$.  

The mesh is graded so that small elements can be used near the moving contact line to resolve the slip length whilst larger elements are used where only scales associated with the bulk flow are present.  Consequently, the computational cost is relatively low so that even for the highest resolution meshes used in this work, for a given liquid-solid-gas combination the entire map of, say, $Ca$ vs contact line elevation ($\triangle y$), can be mapped in just a few hours on a standard laptop.

Computations at high $Ca$, roughly those for which $Ca>0.1$, are notoriously difficult as (a) the free-surface becomes increasingly sensitive to the global flow and (b) regions of high curvature at the contact-line require spatial resolution on scales at or below the slip length $l_s$. Both factors hinder the ease at which converged solutions can be obtained and these issues are compounded when $Oh$ is small, indicating the importance of non-linear inertial effects.  For all parameter sets computed, converged solutions can always be obtained for $Ca\le2$, which compares very favourably to previous works, so that this is chosen as an upper cut-off for the results of our parametric study.  

\subsection{Benchmark Calculations for the Maximum Speed of Wetting}

In coating flows, one would like to know for a given liquid-solid-gas combination the wetting speed $U_c^{\star}$ at which the contact line motion becomes unstable so that gas is entrained into the liquid either through bubbles forming at the cusps of a `sawtooth' wetting-line \citep{blake79} or by an entire film of gas being dragged into the liquid \citep{marchand12}. This is referred to as the `maximum speed of wetting' and, in dimensionless terms, is represented by a critical capillary number $Ca_c=\mu^{\star} U_c^{\star}/\sigma^{\star}$.  Our method for calculating $Ca_c$ is explained in the Appendix alongside benchmark calculations for its value which are compared to the results of \cite{vandre13} across a range of viscosity ratios.  Excellent agreement is obtained between the results in \cite{vandre13} and those found here.  As these calculations distract from the main emphasis of this work, but could be useful as a benchmark for future investigations in the field, they are provided in the Appendix.  

\section{Effect of Gas Pressure on the Maximum Speed of Wetting: Analysis of a Base State}\label{S:anal}

Having demonstrated the accuracy of our code and shown how the critical capillary number $Ca_c$ is calculated, we now investigate the relation between the gas pressure $\bar{P}_g$ and $Ca_c$.  Here, we will analyse a base state, varying only gas pressure and $A$, in an attempt to deepen our understanding of the influence of non-equilibrium gas effects on the dynamic wetting process before, in \S\ref{S:para}, performing a parametric study of the system of interest.

To ensure we are studying an experimentally attainable regime, we use the system considered in \cite{benkreira08} to provide a base state about which our parameters can be varied.  In this work, a range of silicone oils of different viscosities $\mu^{\star}=20$--$200$~mPa~s are used whilst the density $\rho^{\star}=950$~kg~m$^{-3}$, surface tension $\sigma^{\star}=20$~mN~m$^{-1}$ and equilibrium contact angle $\theta_e=10^{\circ}$ remain approximately constant.  Then the characteristic (capillary) length $L^{\star}=1.5$~mm and $Oh=6\times10^{-3}\mu^{\star}$~(mPa~s)$^{-1}$ depends only on the viscosity $\mu^{\star}$. 

A base case is chosen by taking $\mu^{\star}=50$~mPa~s so that $Oh_{0}=0.3$, where base state values will be denoted with a subscript $0$.  Taking air as the displaced fluid with $\mu_g^{\star}=18~\mu$Pa~s, which remains independent of the ambient pressure \citep{maxwell67}, gives a viscosity ratio of $\bar{\mu}_{0}=3.6\times10^{-4}$.  The density of the gas $\rho_g^{\star}$ depends on the ambient gas pressure, with its maximum value of $\rho_g^{\star}=1.2$~kg~m$^{-3}$ at atmospheric pressure making $\bar{\rho}_{max}=1.3\times10^{-3}$.  However, for all calculations performed, the value of  $\bar{\rho}$ has a negligible effect on $Ca_c$ when $\bar{\rho}\le\bar{\rho}_{max}$ and so henceforth, without loss of generality, we take $\bar{\rho}=0$.

The slip length of the liquid-solid interface is fixed at $l^{\star}_s=10$~nm, which is well within the range of experimentally observed values \citep{lauga07}, so that the dimensionless parameter $l_s=6.8\times10^{-6}$.  Assuming the gas is air, we have $Kn = 4.8\times 10^{-5}/\bar{P_g}$ so that the only parameter which remains to be specified is $A$ characterising whether or not there is slip at the gas-liquid boundary.

\subsection{Effect of Gas Pressure}\label{S:gasp}

\begin{figure}
     \centering
\includegraphics[scale=0.3]{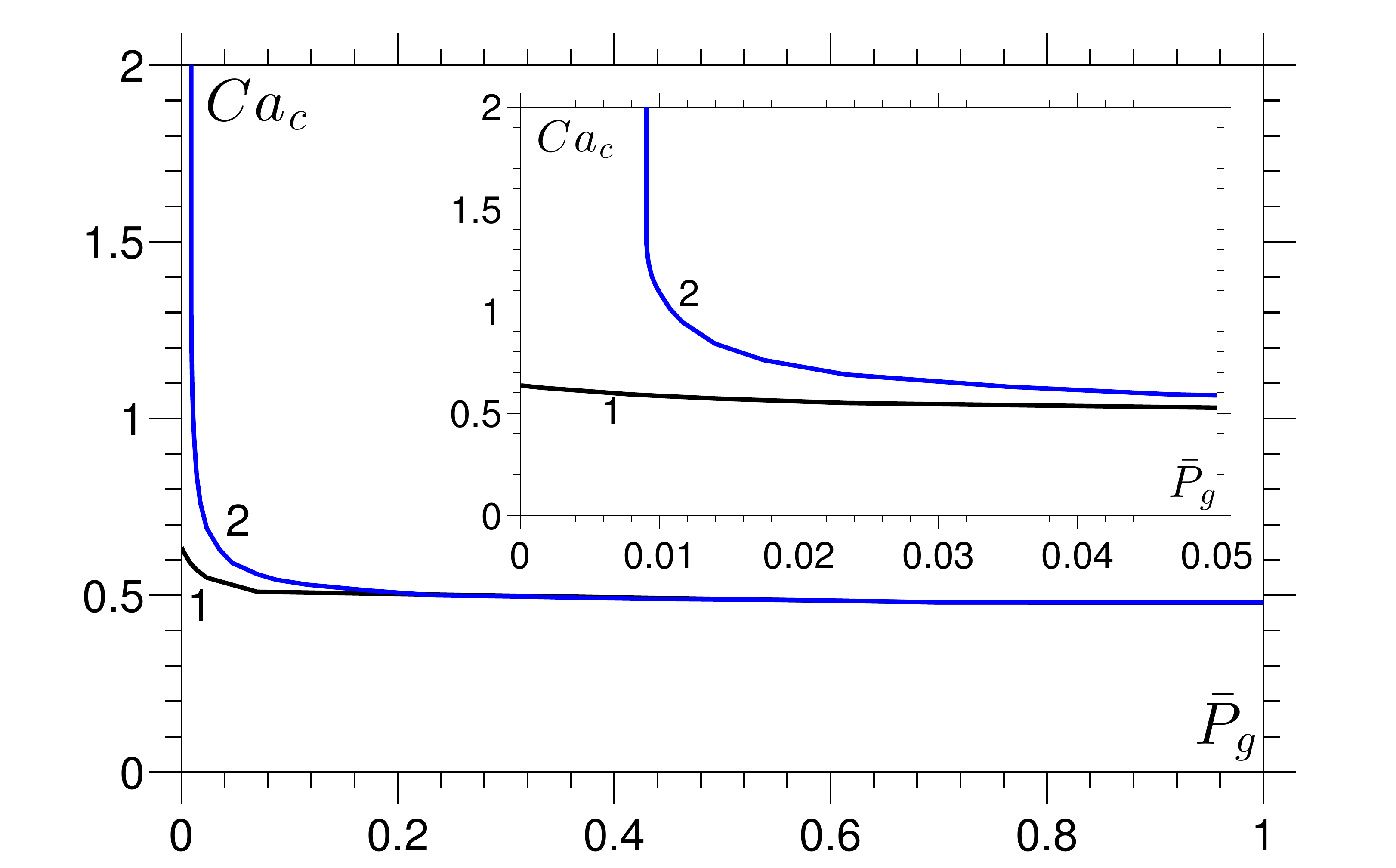}
 \caption{The dependence of the critical capillary number $Ca_c$ on the gas pressure $\bar{P}_g$, for our base parameters, with curve 1: $A=0$ (in black) and 2: $A=1$ (in blue). The inset shows how curve 1 ($A=0$) tends to a finite value of $Ca_c$ as $\bar{P}_g\rightarrow0$ whilst curve 2 appear to increase without bound. }
 \label{F:Pg_vs_cac}
\end{figure}

In Figure~\ref{F:Pg_vs_cac}, the effect on $Ca_c$ of reducing $\bar{P}_g$  is computed for $A=0$ and $A=1$.  In each case, for $\bar{P}_g > 0.1$ there is only a slight increase in $Ca_c$ from its base value of $0.47$, which is independent of $A$.  For $\bar{P}_g < 0.1$, dramatic changes in $Ca_c$ are observed, whose form depends on $A$, so that $\bar{P}_g \approx 0.1$ appears to be the point at which non-equilibrium effects become important.

Notably, as can be seen most clearly in the inset of Figure~\ref{F:Pg_vs_cac}, for $A=0$ as $\bar{P}_g\rightarrow 0$  we have $Ca_c \rightarrow 0.64$ whereas for $A=1$ computations suggest that $Ca_c\rightarrow\infty$.   In fact, in the latter case, once a critical pressure $\bar{P}_{g,c}$ is approached, $Ca_c$ appears to increase without bound.  For $A=1$ this occurs at $\bar{P}_{g,c}=9\times 10^{-3}$; just above this value, at $\bar{P}_{g,c}= 10^{-2}$, we have $Ca_c=1.09$ whilst for all $\bar{P}_{g}\le\bar{P}_{g,c}$ we find $Ca_c>2$.  

It is impossible to show rigorously that $Ca_c\rightarrow\infty$ as $\bar{P}_{g}\rightarrow \bar{P}_{g,c}$ without either the computation of higher values of $Ca_c$, which would allow some scaling to be inferred, or, ideally, the development of an analytic framework for this problem.  Unfortunately, in the former case (see \S3) computational constraints permit us from reaching $Ca_c>2$ whilst in the latter we are confined by the fact that all analytic development, in particular those of \cite{cox86}, are only valid for small $Ca_c$.  Despite this, the rapid increase of $Ca_c$ as $\bar{P}_{g,c}$ is approached is striking and $Ca_c=2$ is still a large value in the context of dip-coating phenomena.

To summarise, it has been shown that when Maxwell slip is accounted for on both the gas-solid and gas-liquid interfaces the maximum speed of wetting appears to be unbounded as the ambient pressure is reduced, whereas if slip on the latter surface is neglected ($A=0$), the maximum speed remains finite. In other words, the error associated with neglecting Maxwell slip on the free-surface is extremely large at reduced pressures and appears to be infinite in the limit $\bar{P}_g\rightarrow 0$. 


\subsection{Flow Kinematics}\label{S:kinematics}

\begin{figure}
     \centering
\subfigure[]{\includegraphics[scale=0.4]{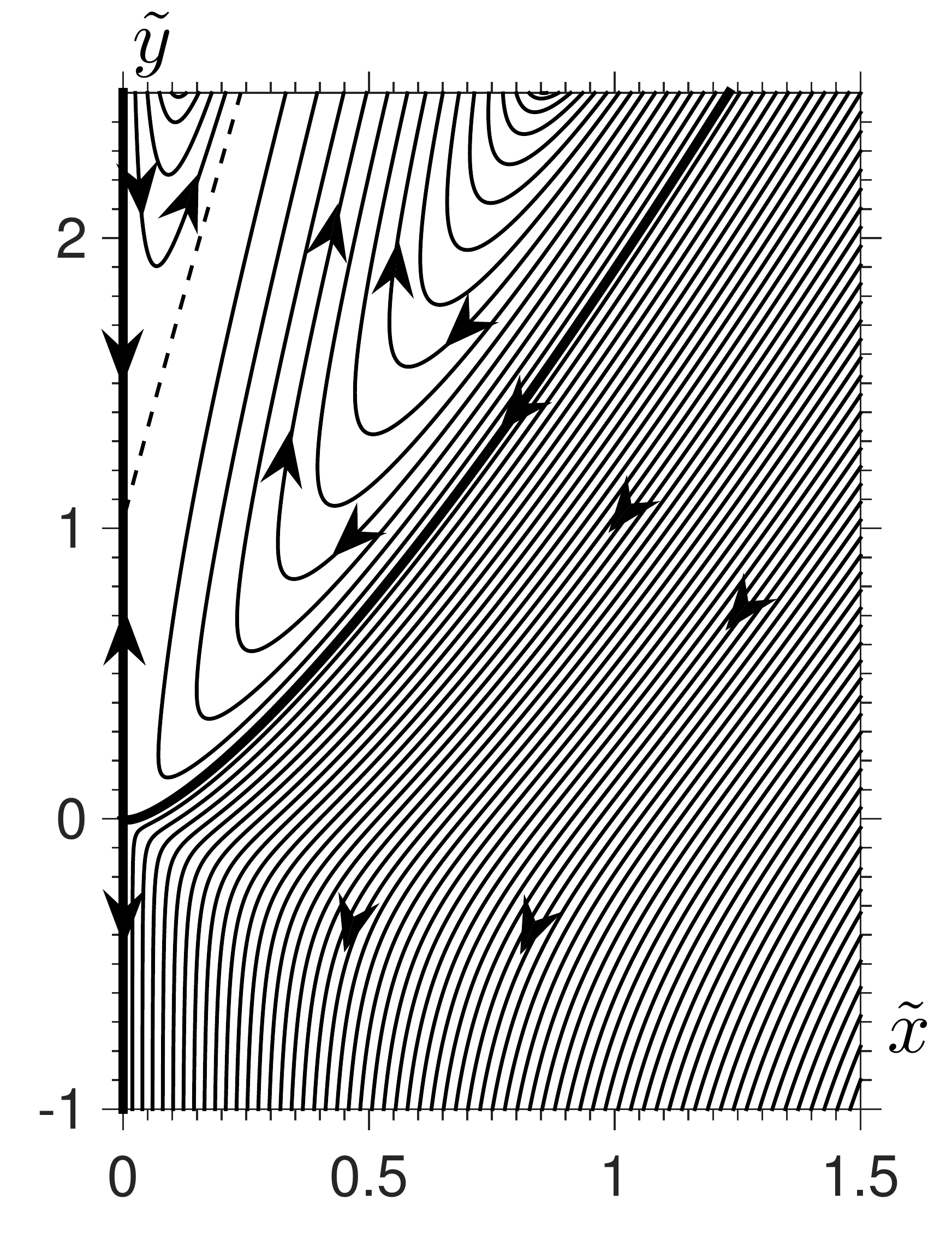}}
\subfigure[]{\includegraphics[scale=0.4]{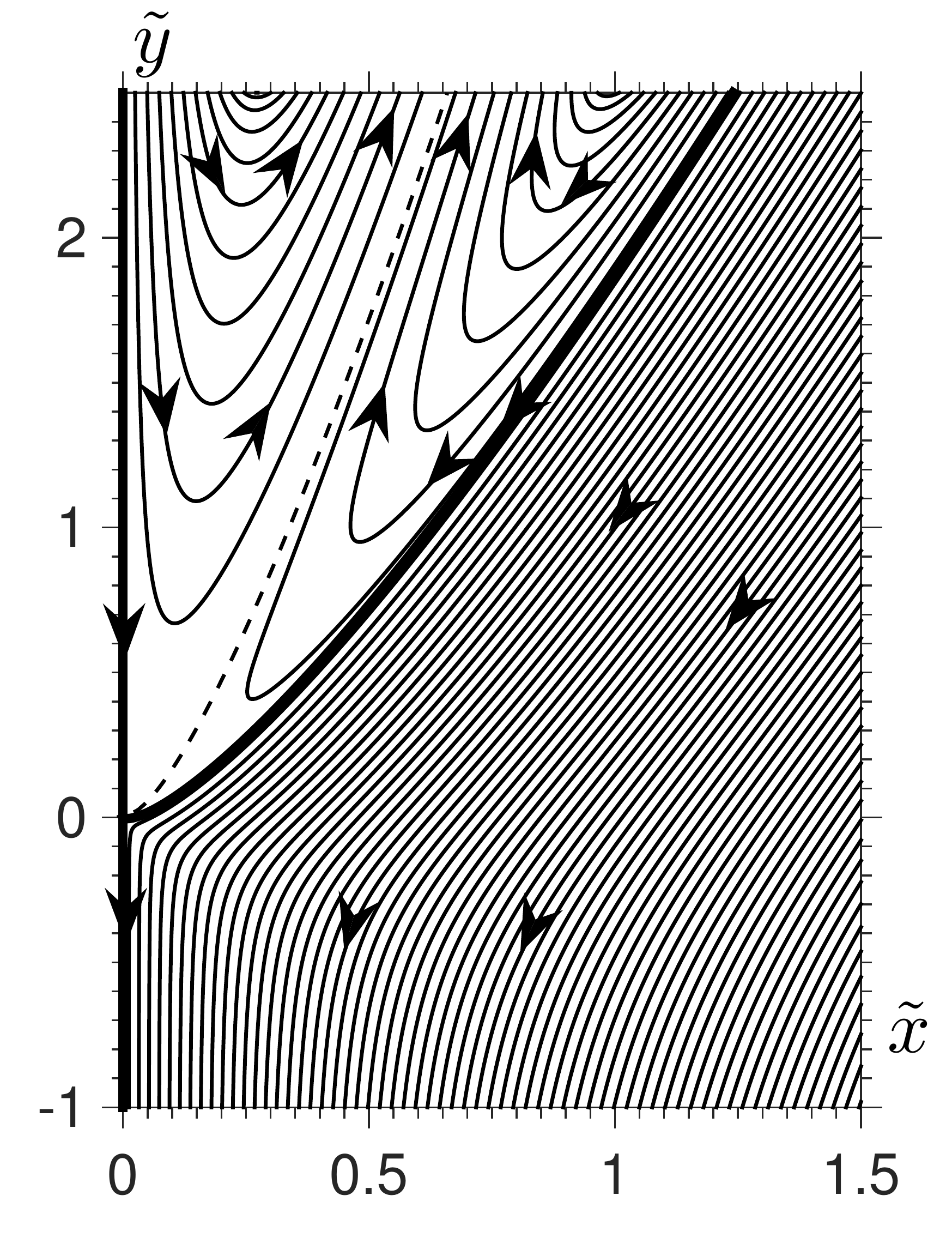}}
\subfigure[]{\includegraphics[scale=0.4]{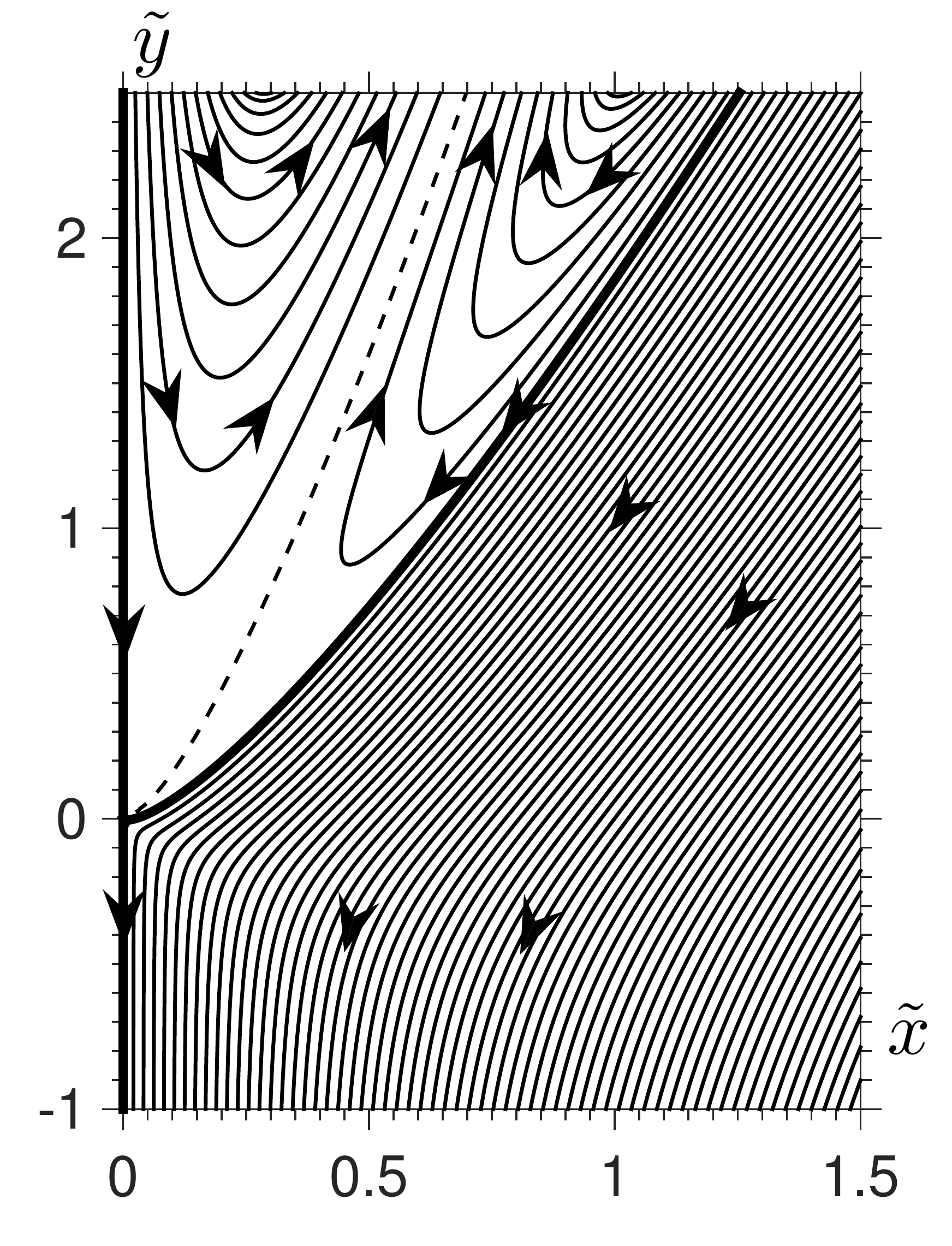}}
 \caption{Streamlines computed at $Ca=0.4$ for different ambient gas pressures $\bar{P}_g$ both with ($A=1$) and without ($A=0$) Maxwell slip at the free-surface.  In (a):$~(\bar{P}_g,A) = (0.14,0)$, (b):$~(0.14,1)$ and (c):$~(0.014,1)$.  The local `zoomed in' coordinates used are $\tilde{x}=x\times10^3$ and $\tilde{y}=(y-y_c)\times10^{3}$, where $y_c$ is the contact line position, and streamlines emanate from equally spaced points across $\tilde{y}=2.5$ with the dividing streamline dashed.}
 \label{F:stream}
\end{figure}

In Figure~\ref{F:stream}, streamlines of the flow near the contact line are shown at $Ca=0.4$ in three different regimes: (a) at a pressure where non-equilibrium effects are weak $\bar{P}_g=0.14$ with $A=0$, (b) $\bar{P}_g=0.14$ with $A=1$ and (c) at a pressure where non-equilibrium effects are becoming influential $\bar{P}_g=0.014$ with $A=1$.  Given that $L^{\star}\sim~1$mm for typical liquids, dimensionally the scale in Figure~\ref{F:stream} is a few microns. In all cases, the flow of the liquid, which is below the free-surface (represented by a thick black line), remains virtually unchanged, with the motion of the solid, located at $\tilde{x}=0$, dragging liquid downwards which, to conserve mass, is continually replenished from above.  If one notes that the free-surface meets the solid at an equilibrium contact angle of $10^\circ$, and yet any apparent angle defined on the scale seen in Figure~\ref{F:stream} would be obtuse, it is clear that at this relatively high capillary number there is significant deformation of the free-surface on a scale below what is visible here.

\begin{figure}
     \centering
\subfigure[]{\includegraphics[scale=0.3]{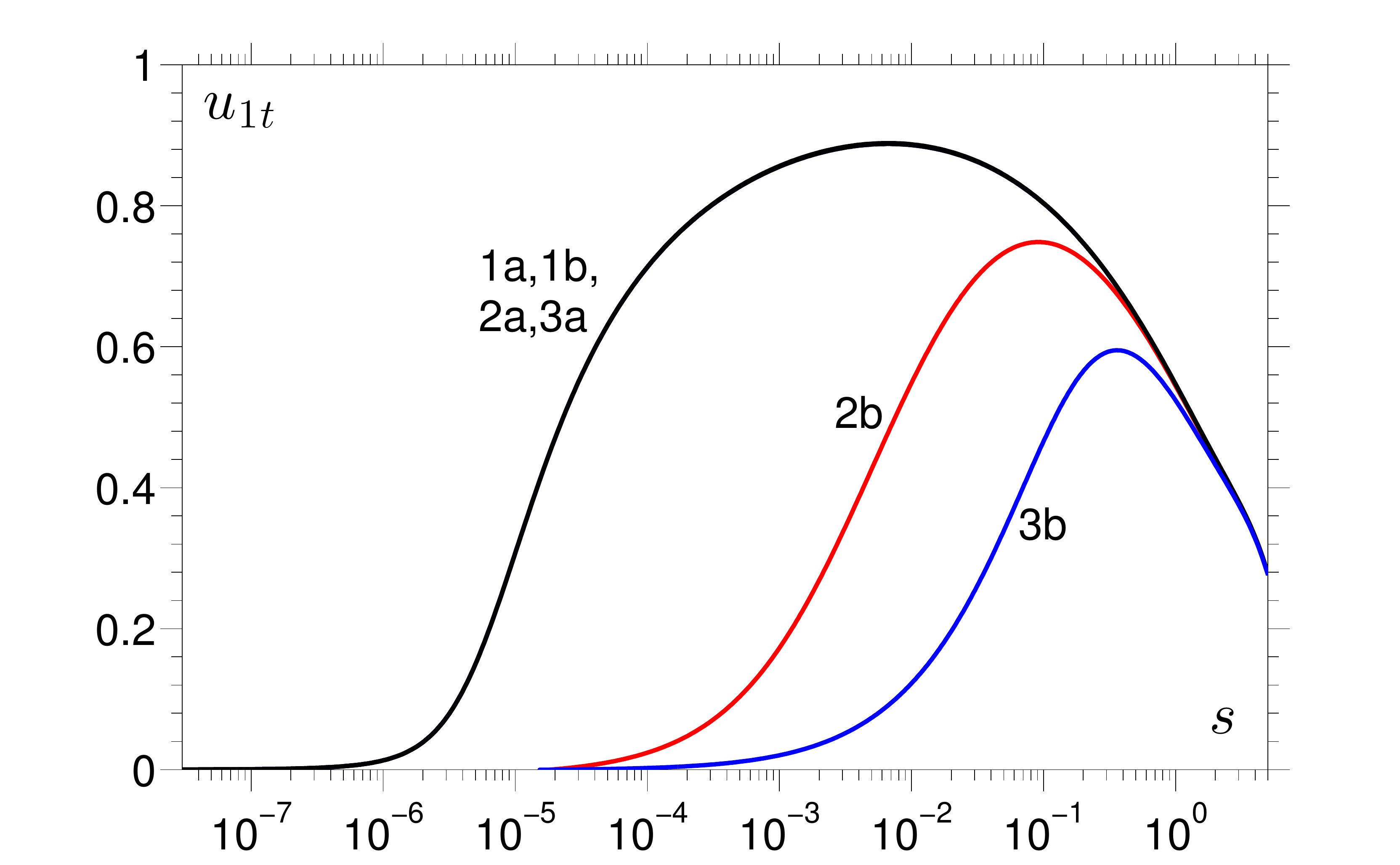}}
\subfigure[]{\includegraphics[scale=0.3]{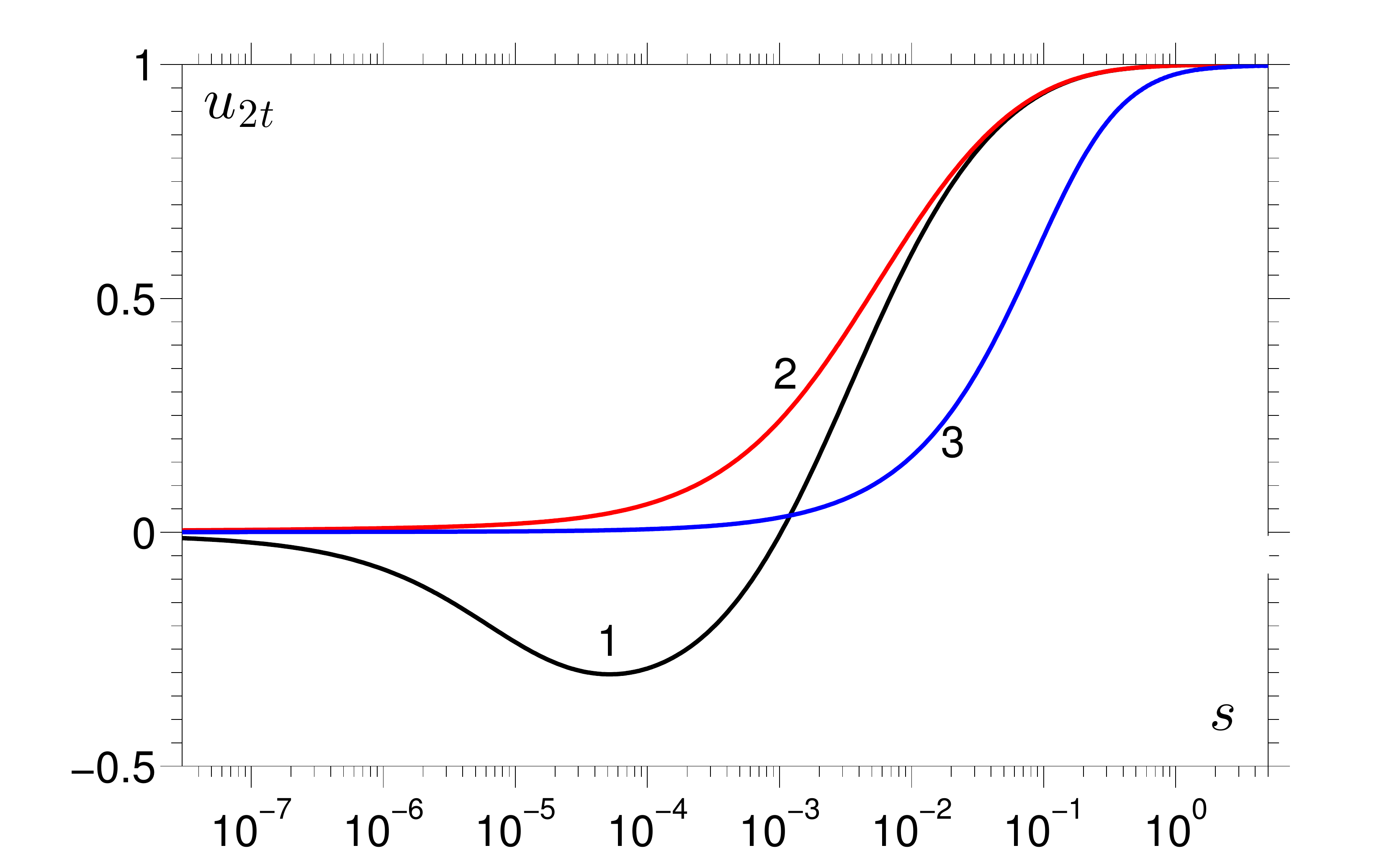}}
 \caption{Tangential velocities as a function of distance from the contact line $s$ along the (a) liquid-gas and (b) gas-solid boundaries computed at $Ca=0.4$ with curve 1:~$(\bar{P}_g,A) = (0.14,0)$, 2:$~(0.14,1)$ and 3:$~(0.014,1)$.  The velocity tangential to the free-surface and pointing towards the contact line is $u_{1t}$, with subscripts `a' for the liquid-facing side of the interface and subscripts `b' for the gas side.  The tangential velocity on the gas-solid interface pointing towards the contact line is $u_{2t}$ and the speed of the solid is $1$.}
 \label{F:u1t}
\end{figure}

In all three cases, the flow of liquid, which seems immune to the gas' dynamics, results in an almost identical velocity tangential to the free-surface on the liquid-facing side of this interface, as shown from curves $1a,2a,3a$ in Figure~\ref{F:u1t}a, where the velocities tangential to the gas-solid and liquid-gas interfaces have been plotted.

At the top $\tilde{y}=2.5$ of the figures, the gas flow field is qualitatively similar in all cases.  The motion of both the liquid and the solid drives gas towards the contact line region which, to ensure continuity of mass, results in a `split ejection' type flow, as observed in immiscible liquid-liquid systems \citep{dussan74,dussan77}, with an upwards flux of gas through the middle of this domain.   For $A=1$ the split ejection flow is maintained right up to the contact line (Figure~\ref{F:stream}b,c); however, for $A=0$ a flow reversal occurs on the solid-gas interface at $\tilde{y}\approx 1$ so that the direction of the gas flow actually opposes the solid's for $\tilde{y}<1$ (Figure~\ref{F:stream}a). The flow reversal can clearly be seen from curve~1 in Figure~\ref{F:u1t}b for $s<10^{-3}$, with a substantial minimum of $u_{2t}=-0.3$, and can also be seen in previous works, such as Figure~10b of \cite{vandre13}.  

The asymmetric gas flow observed for $A=0$ appears because in this case the gas' velocity is continuous across the liquid-gas free-surface, see curves $1a$ and $1b$ in Figure~\ref{F:u1t}a, but is allowed to slip across the solid boundary according to Maxwell's condition, see curve $1$ in Figure~\ref{F:u1t}b. This results in the liquid driving a gas flow towards the contact line which, when combined with the requirement of mass conservation, leads to such a large build of tangential stress at the solid, that the slip across the interface, determined by (\ref{sg_slip}), is sufficient to reverse the flow direction.  This effect will be further considered, in a lubrication setting, in \S\ref{S:lube}.

When $A=1$, slip is also allowed at the gas-liquid interface, see curves $2a$ and $2b$ in Figure~\ref{F:u1t}a, so that, as can be seen from curve $2$ in Figure~\ref{F:u1t}b, there is no flow reversal on either of the gas' boundaries.  This more symmetric flow field remains when the gas pressure is decreased to $\bar{P}_g=0.014$, see Figure~\ref{F:stream}c and curves labelled $3$ in Figure~\ref{F:u1t}.  Although the flow remains qualitatively the same, one can clearly see from Figure~\ref{F:u1t} that lowering the pressure significantly increases slip at the gas' boundaries.  For example, at $s=10^{-2}$, for $\bar{P}_g=0.14$ the velocity on the solid $u_{2t}=0.65$ whereas for $\bar{P}_g=0.014$ the velocity is just $u_{2t}=0.16$.   Therefore, less gas is dragged into the contact line region when (a) slip is accounted for on the gas-liquid boundary ($A=1$) and/or (b) slip is increased through reductions in pressure.  Both of these mechanisms reduce the gas' resistance to contact line motion, as shown analytically in \S\ref{S:lube}, and contribute to postponing the point of wetting failure.

\subsection{Characteristics of the Gas Film}\label{S:gasfilm}

\begin{figure}
     \centering
\includegraphics[scale=0.3]{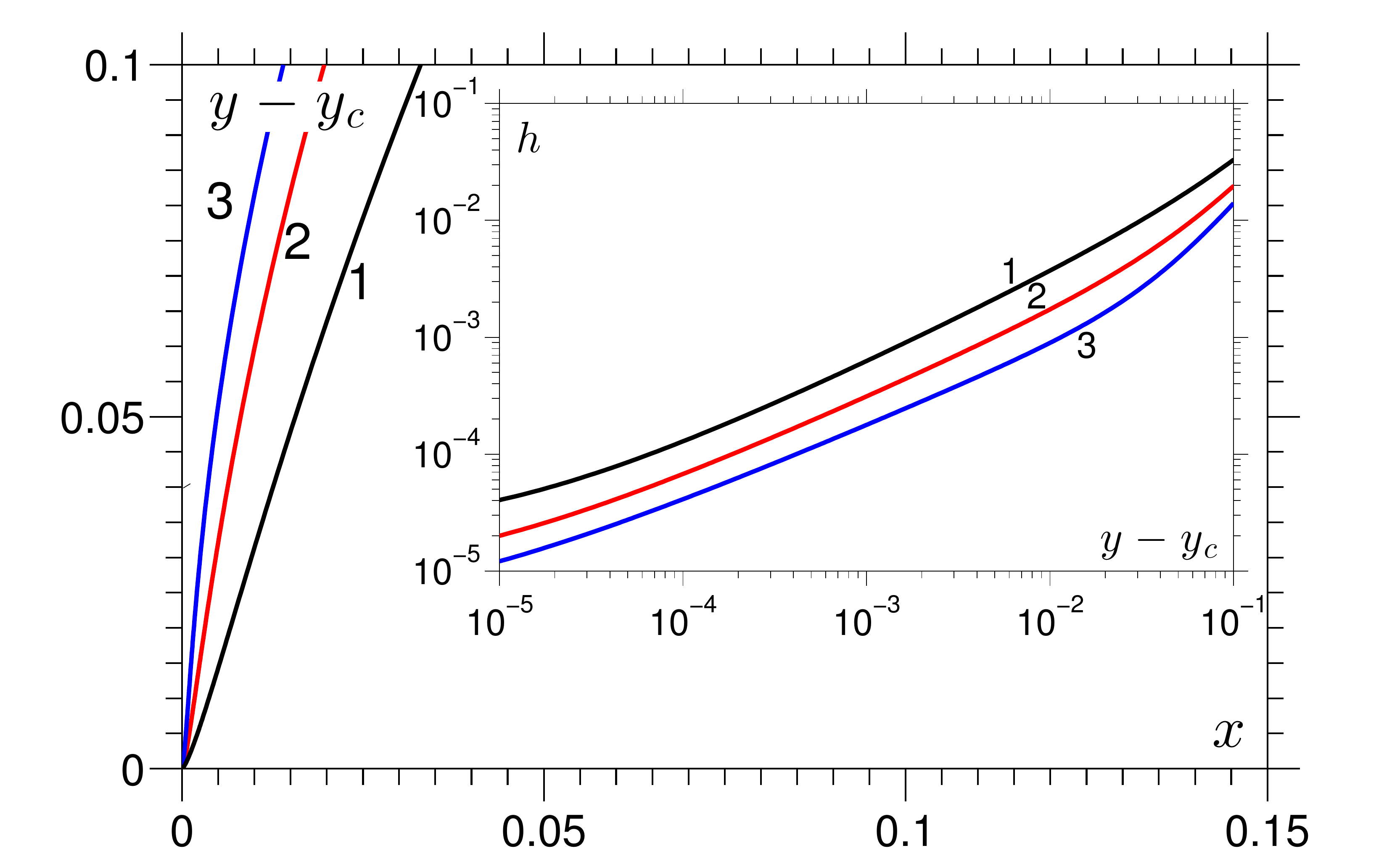}
 \caption{Free-surface shapes obtained in our base case, with $\bar{P}_g=0.014$, with curve 1: $Ca=0.4$ (in black), 2: $Ca=0.6$ (in red) and 3: $Ca=0.8$ (in blue).  The inset shows how the height of the gas film $h$ varies as a function of $y-y_c$ showing that for $10^{-4}<y-y_c<10^{-1}$ it can be considered in a lubrication setting.}
 \label{F:thinness}
\end{figure}

At high capillary numbers there is a region between the contact line and the far field in which the gas domain is a `thin film'. This region is sufficiently far from the contact line region where the free-surface bends from its equilibrium angle of $10^{\circ}$ towards an obtuse apparent angle and near enough to the contact line that gravitational forces have not started to flatten the free-surface.  For our base case,  for $Ca\ge0.4$, which includes all values of $Ca_c$, this region exists between $10^{-4}<y-y_c<10^{-1}$ as one can see from Figure~\ref{F:thinness}.  Specifically, the main figure clearly shows that the gas film is `thin' on the scale of around $y-y_c\sim0.1$ whilst the inset shows on a logarithmic plot that this behaviour continues down to $y-y_c\sim10^{-4}$.

The dynamics of the gas film are key to the air entrainment phenomenon so that one may expect that reductions in ambient pressure will only start to influence $Ca_c$ once Maxwell slip on the boundaries is large enough to affect the gas' flow characteristics.  This happens when the (dimensionless) slip length  $Kn=4.8\times 10^{-5}/\bar{P_g}$ becomes comparable to the height of the gas film $x=h(y)$.  This results in a \emph{function} $Kn_{loc}(h) = Kn/h$ which characterises the importance of non-equilibrium effects as one goes along the film.  How though, should we define a particular position along the film $h=H$ on which we can base a local Knudsen \emph{number} $Kn_{H}$?

Research into dynamic wetting/dewetting phenomena \citep{derjaguin64,vandre13,eggers04a} has long recognised that wetting failure is strongly linked to the behaviour of the inflection point on the free-surface, where the surface's curvature changes sign; this point can also be used to define an apparent contact angle \citep{tanner79}.  Given that the flow near the inflection point seems to be key to determining $Ca_c$, this point is a sensible choice for defining the characteristic film height $H$, see Figure~\ref{F:sketch}.   To do so, we must extract $H$ from our computations and determine how it depends on $Ca$ and $\bar{P}_g$.

\begin{figure}
     \centering
\includegraphics[scale=0.3]{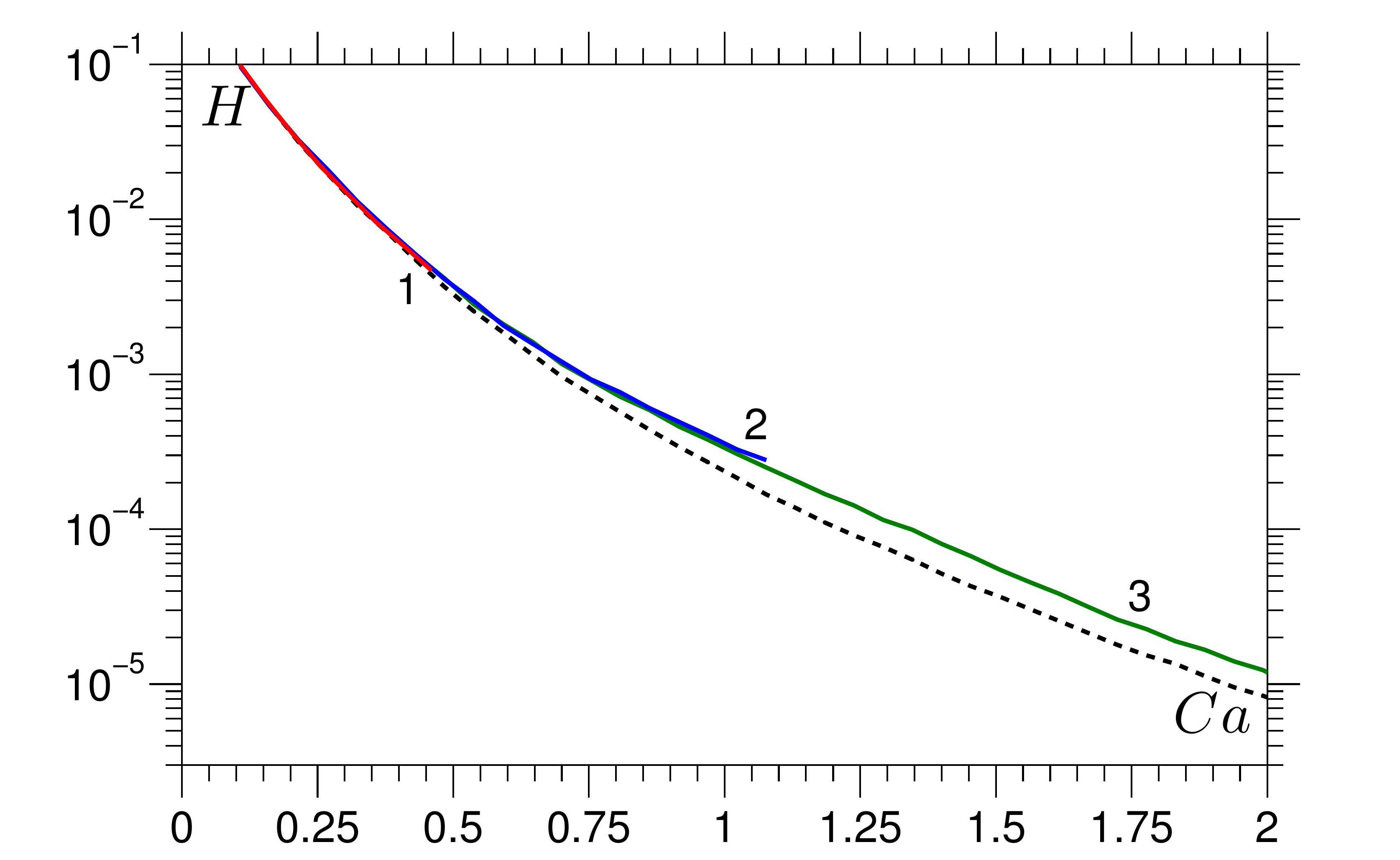}
 \caption{The dependence of the gas film height $H$, at the inflection point on the free surface, on the capillary number $Ca$.  The curve is largely independent of gas pressure $\bar{P}_g$, although the lower values of $\bar{P}_g$ allow more of the curve to be obtained. Curves corresponds to 1:~$\bar{P}_g=1$ (red), 2:~$\bar{P}_g=10^{-2}$ (blue) and 3:~$\bar{P}_g=9\times10^{-3}$ (green), with the dashed line obtained for $\bar{\mu}=0$.}.
 \label{F:H}
\end{figure}

In Figure~\ref{F:H}, we can see that although $Ca_c$ depends on $\bar{P}_g$, the value of $H$ is approximately independent of $\bar{P}_g$, so that $H\approx H(Ca)$, with smaller $\bar{P}_g$ simply revealing more of this curve.  In fact, a similar curve is obtained if the gas flow is ignored altogether ($\bar{\mu}=0$), as shown by the dashed line in Figure~\ref{F:H}, although this may not be the case at higher viscosity ratios.   Notably, $H$ decreases rapidly with increasing $Ca$ in agreement with previous experiments \citep{marchand12} and simulations \citep{vandre13}.  Reassuringly, at $Ca=Ca_c$ the film height obtained at atmospheric pressure, $H=4.2\times10^{-3}$, dimensionally corresponds to a film $H^{\star}=1$--$10~\mu$m for typical liquids, which is precisely what has been found experimentally in \cite{marchand12}.

Enhancements in $Ca_c$ at reduced pressure can be quantified by the percentage increase in $Ca_c$ from its atmospheric value $Ca_{c,atm}$ defined by $\triangle Ca_c = 100(Ca_c-Ca_{c,atm})/Ca_{c,atm}$, which is given in Table~\ref{T:dcac}.  As well as listing the pressure reduction required, values for $H$ are given from which $Kn_{H}=Kn/H$ is calculated.  The data shows that for $\bar{P}_g\sim0.1$, where significant increases in $\triangle Ca_c$ begin, corresponds to  $Kn_{H}\sim0.1$. Therefore, non-equilibrium effects in the gas, which manifest themselves through Maxwell slip, alter the flow once the slip length is around 10\% of the gas film's characteristic height and it is this mechanism which determines when variations in $\bar{P}_g$ can start to affect $Ca_c$.
\begin{table}
\begin{center}
\begin{tabular}{|c|c|c|c|}
  \hline
 $\triangle Ca_c$  & $\bar{P}_g$ & $H$  &  $Kn_{H}$ \\
 \cline{1-4}
  \hline
  $0$                & $1$ 				&	$4.2\times10^{-3}$	&	$0.011$   \\
  $10$\%          & $0.16			$ 	&	$3.1\times10^{-3}$	&	$0.1$  \\
  $20$\%          & $7\times 10^{-2}$		&	$2.3\times10^{-3}$	&	$0.3$ \\
  $50$\%          & $2.3\times10^{-2}$ 		&	$1.1\times10^{-3}$	&	$1.9$ \\
  $100$\%        & $1.2\times10^{-2}$ 		& 	$4.1\times10^{-4}$	&	$9.8$ \\
  $\ge300$\%      & $\le 9\times10^{-3}$	&	$\le 7.1\times10^{-5}$	&	$\ge 75$ \\
  \hline
\end{tabular}
\end{center} \caption{The pressure reduction $\bar{P}_g$ required for a given enhancement in capillary number $\triangle Ca_c$ with the corresponding film height $H$ and Knudsen number at the inflection point $Kn_{H}$ at this pressure.}\label{T:dcac}
\end{table}

Notably, all inflection points calculated in Table~\ref{T:dcac} fall into the thin film region apart from the entries where the critical gas pressure has been passed after which the inflection point gets close to the region of high curvature near the contact line.  This suggests that the critical pressure  $\bar{P}_{g,c}=9\times10^{-3}$ is the one which is low enough for the inflection point to no longer be located the thin film region.   From Table~\ref{T:dcac} we can see that for $\bar{P}_{g}\le \bar{P}_{g,c}$ we have $Kn_{H}\ge75$ so that Maxwell slip is so strong that the gas flow is hardly affected by the motion of its boundaries.  In this case, the gas effectively offers no resistance to the dynamic wetting process and $Ca_c$ is able to increase without bound.

\subsubsection{Limitations of the Maxwell-slip Model}

From Table~\ref{T:dcac} it is clear that the combined increases in $Kn$ and decreases in $H$  at reduced pressures both contribute to rapid changes in $Kn_{H}$.   Across a reduction in the gas pressure of $\sim10^2$, $Kn_{H}$ increases from its atmospheric value by a factor of $\sim10^4$.  Consequently, most values of $Kn_{H}$ in the Table fall well outside the `slip regime' where non-equilibrium effects can be attributed entirely to the boundary conditions.  Therefore, attributing non-equilibrium effects entirely to the boundary conditions, via Maxwell-slip, as has been considered here is not sufficient to accurately capture its behaviour.  This calls into question a number of the results which have been obtained with a gas model which goes outside its strict limits of applicability for the smaller values of $\bar{P}_g$.  In particular, is it likely that the behaviour $Ca_c\rightarrow\infty$ as $\bar{P}_{g}\rightarrow\bar{P}_{g,c}$, or even as $\bar{P}_{g}\rightarrow0$, remains robust?

To address this question we must understand more about the Knudsen layer in regimes where its effects can no longer be attributed entirely to the boundary conditions.  A useful result for this phenomenon has been obtained in \cite{lockerby05}, for pressure-driven Poiseuille flow in a channel. There, it was shown that at a Knudsen number of just $Kn=0.025$, the increased mass flow rate above that obtained for no-slip is approximately 70\% due to increased slip at the wall whilst 30\% is due to non-Newtonian effects in the Knudsen layer. This latter phenomenon is not accounted for in our work and suggests that incorporating more complex gas dynamics is likely to enhance the effects already observed with Maxwell-slip rather than suppress them.  Therefore, there is good reason to believe that with the incorporation of more complex gas models the qualitative trends observed for $Ca_c$ will remain whilst quantitatively its value could increase for a given $\bar{P}_g$.

We shall return to these points in more detail in \S\ref{S:discussion}; however, in what follows we will continue to use the problem formulation outlined in \S\ref{S:form} to give some insight into the role that non-equilibrium effects play in the gas, despite, in some cases, the model being outside its strict region of applicability.

\subsection{Lubrication Analysis}\label{S:lube}

Given the importance of the gas film's behaviour, it is of interest to see if analytic progress in a lubrication setting will shed some light on the dynamics of this film.  Figure~\ref{F:u1t}a shows that $u_{1t}$, the liquid's velocity tangential to the free-surface is within $10$\% of $u_{1t}=0.85$ throughout the thin film region. This is a useful observation, as it allows us to make analytic progress by assuming that the liquid approximately provides a constant downward velocity $V=0.85$ along the free-surface in this region.

In the lubrication setting, the steady gas flow between two impermeable surfaces, at $x=0$ and $x=h$, generated by their velocities of magnitude, respectively, $v=-1$ and $v=-V$, will, in order to conserve mass ($\int^{h}_0 v ~dx=0$), have parabolic form $v=a(x^2-h^2/3)+b(x-h/2)$ . The coefficients $a,b$ are obtained by applying the boundary conditions at $x=0,h$ which are, respectively,  the Maxwell-slip equations (\ref{sg_slip}) and (\ref{fs_slip}).  Introducing $Kn_{loc}=Kn/h$, the result is
\begin{equation}\label{lube}
a=\frac{-3\left[1+2A Kn_{loc}+V\left(2Kn_{loc}+1\right)\right]}{h^2\left[1+4 Kn_{loc}(1+A)+12 A Kn_{loc}^2\right]}\qquad\hbox{and}\qquad b=\frac{2(3-ah^2)}{3h\left(2Kn_{loc}+1\right)}.
\end{equation}

\begin{figure}
     \centering
\includegraphics[scale=0.3]{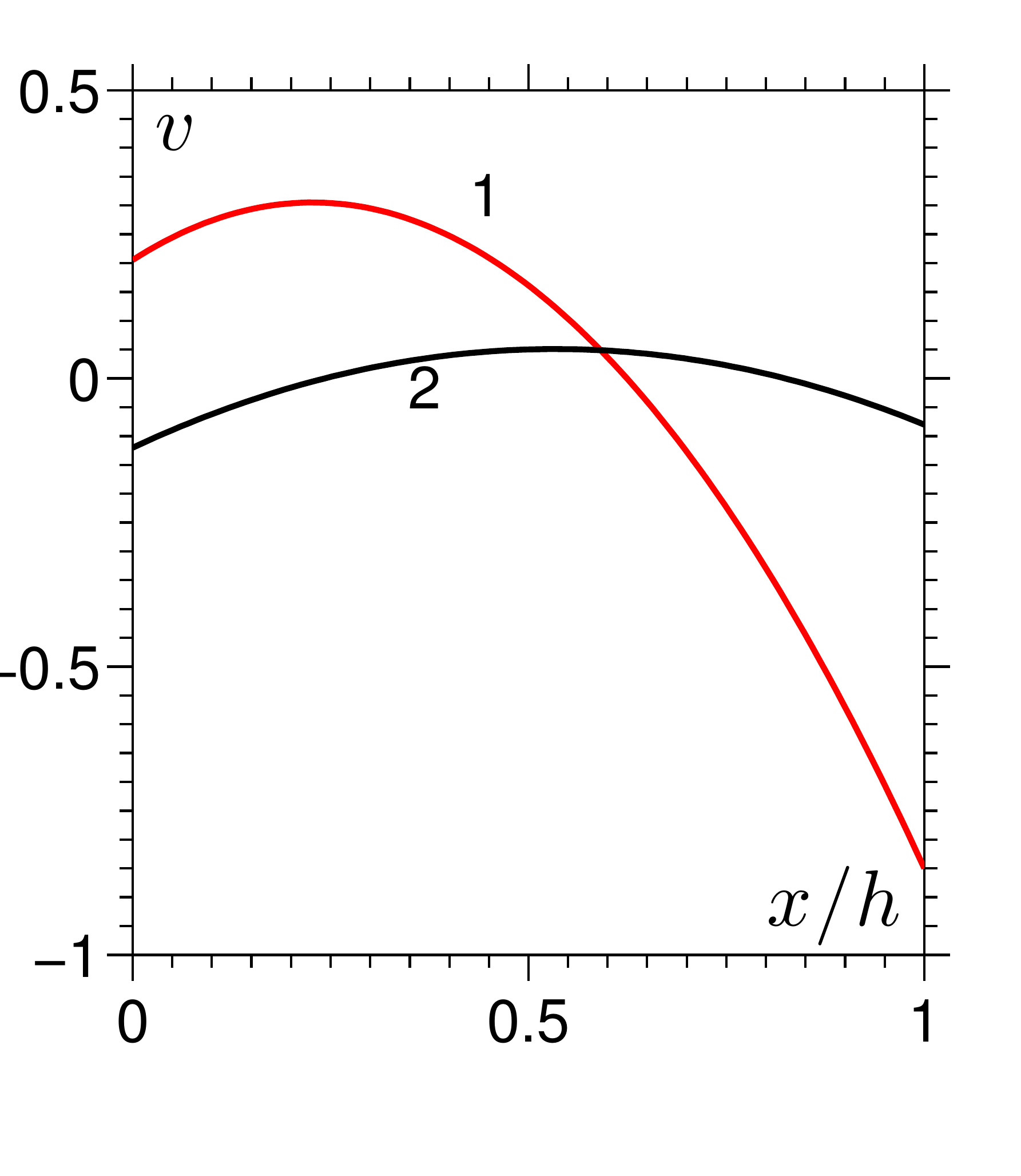}
 \caption{Vertical velocity $v$ predicted by the lubrication analysis for $h=2.5\times10^{-4}$ at $\bar{P}_g=0.14$ for the cases of $A=0$, curve 1 in red, and $A=1$, curve 2 in black.  One can see that when $A=0$ the lubrication analysis predicts the flow reversal at $x=0$ whilst this does not occur for $A=1$.  These results can be compared to those in Figure~\ref{F:stream}a,b by looking at $\tilde{y}=0.3$ where $h=2.5\times10^{-4}$.}.
 \label{F:vlube}
\end{figure}

A good test for the lubrication theory is to see whether it is able to predict flow the reversal at the solid surface observed in Figure~\ref{F:stream}a for the case of $A=0$.  To do so requires that $v(x=0)>0$, which for $A=0$ can be shown to occur when $Kn_{loc}>1/(2V)$ so that flow reversal is indeed possible if $Kn_{loc}$ is sufficiently large.  For the case in Figure~\ref{F:stream}a, using $V=0.85$ and $Kn_{loc}=3.4\times10^{-4}/h$, the lubrication analysis predicts flow reversal for $h<h_r=5.8\times 10^{-4}$ with a typical flow profile in this regime shown by curve~1 in Figure~\ref{F:vlube}. The agreement with the computed value of $h_r=6.2\times 10^{-4}$ is good and is an indication of the accuracy of our approximations in this region. 

\begin{figure}
     \centering
\includegraphics[scale=0.3]{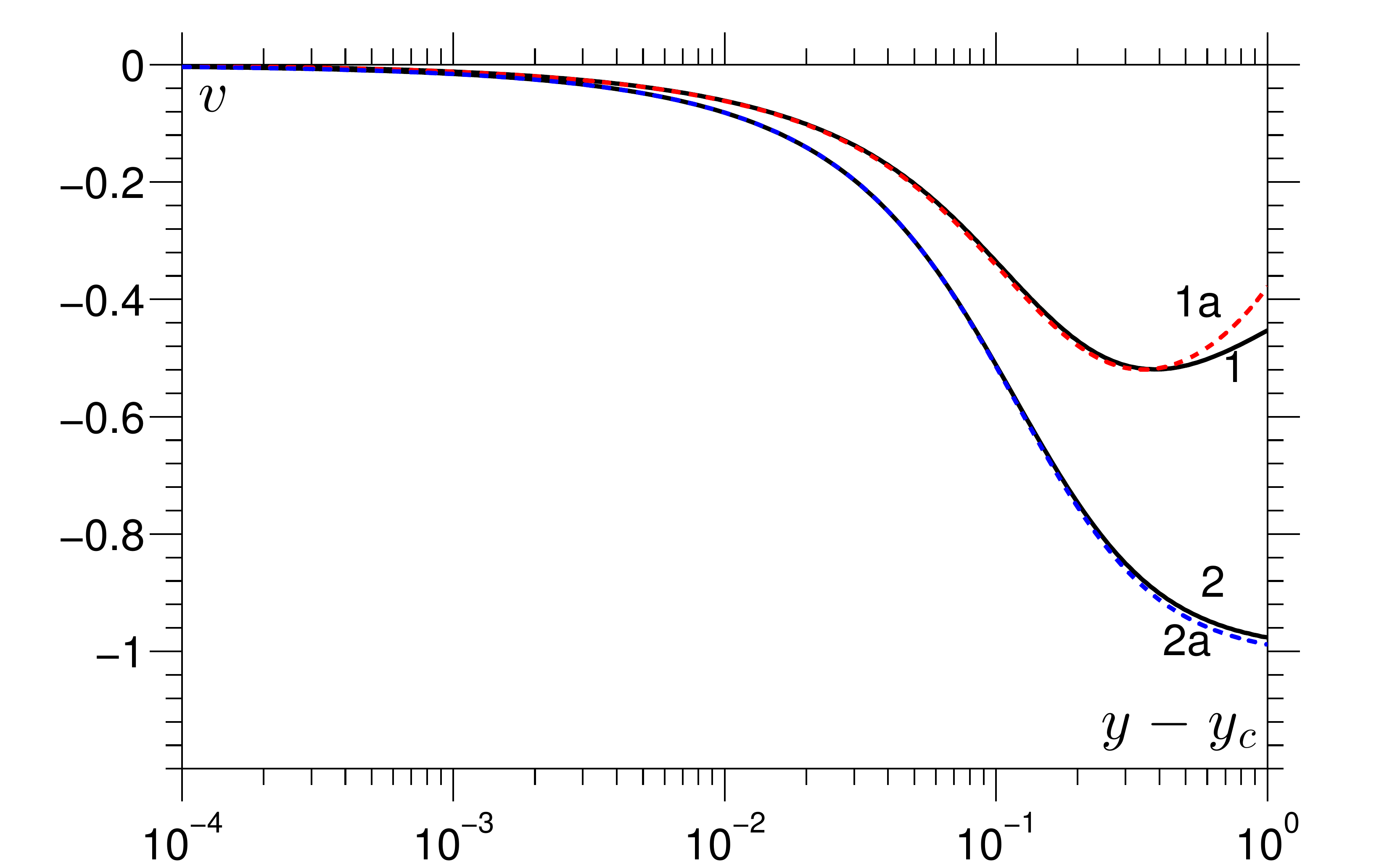}
 \caption{Comparison of the computed gas velocity \emph{tangential} to the boundaries of the gas with the \emph{vertical} velocity from the lubrication analysis for the case of $A=1$ and  $\bar{P}_g=0.014$.  Curves 1,2 are, respectively, computed values along the free-surface and solid surface whilst corresponding curves from the lubrication analysis are 1a in red and 2a in blue.  Notably, there is no flow reversal as $v\le0$ in all cases.}.
 \label{F:utlube}
\end{figure}

For $A=1$, where there is slip on each interface, we find that for $V<1$, which is always satisfied, there are no real roots for $Kn_{loc}$ so that $v(x=0)<0$ for all $h$. In other words, in this case there is no flow reversal adjacent to the solid, in agreement with the streamlines shown in Figure~\ref{F:stream}b,c.  The accuracy of the lubrication approximation for this flow is confirmed in Figure~\ref{F:utlube}, where the computed velocity \emph{tangential} to the gas' boundaries $u_t$ is plotted against the \emph{vertical} velocity predicted by (\ref{lube}). Note that here, we have used the computed value of $V$ at the free-surface, to enable a more accurate assessment of the lubrication approximation, rather than the estimate $V=0.85$ used elsewhere.

%
%

\begin{figure}
     \centering
\includegraphics[scale=0.3]{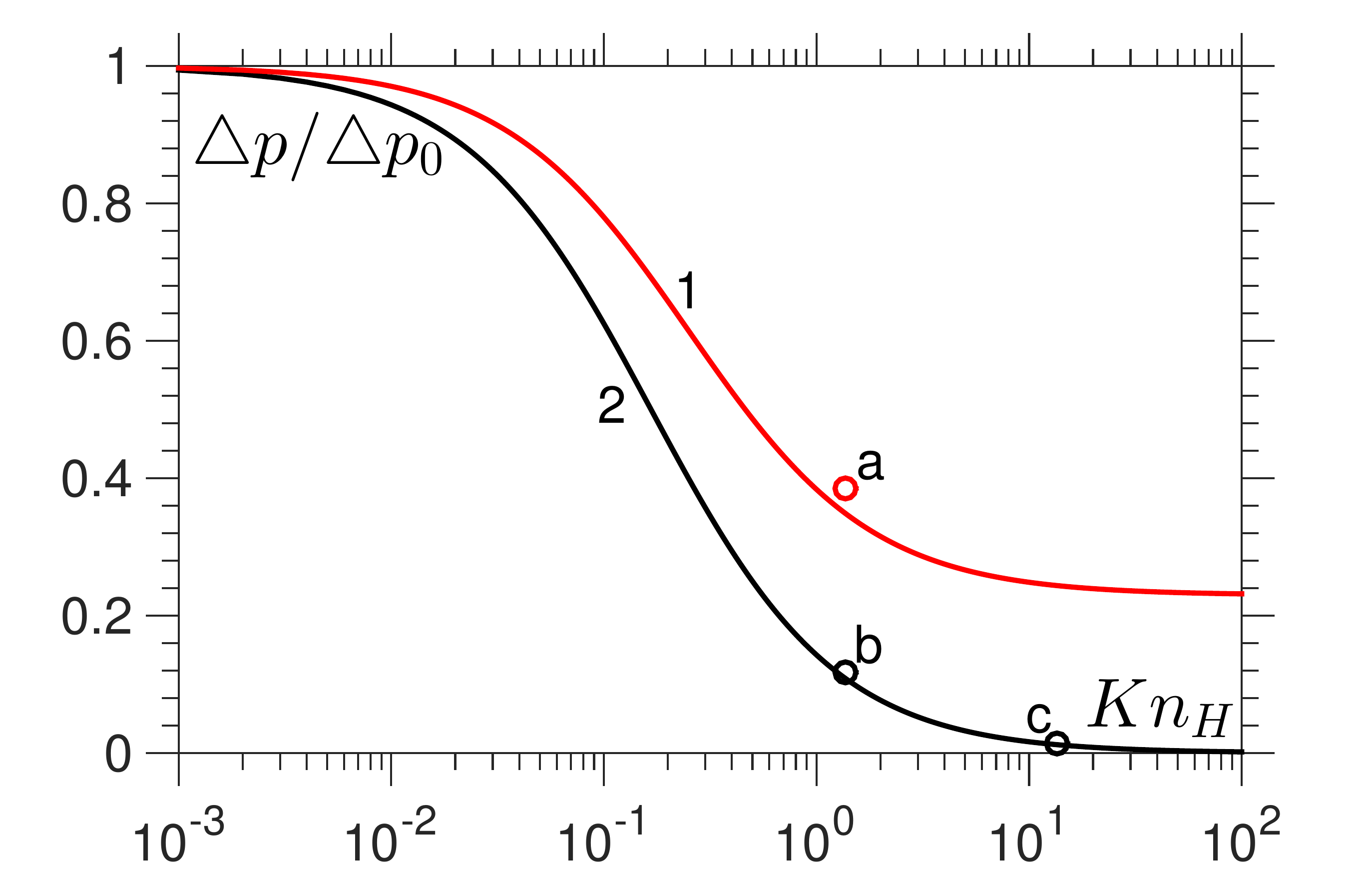}
 \caption{Normalised pressure gradient $\triangle p/\triangle p_0$ given as a function of $Kn_H$, where curve 1:~$A=0$ (in red) and curve 2:~$A=1$ (in black).  Circles labelled $a$, $b$ and $c$ are the values computed for the pressure gradient at $\tilde{y}=0.3$ for the cases shown in Figure~\ref{F:stream}a,b,c and these compare well to the predictions of the lubrication analysis.}
 \label{F:dp}
\end{figure}

In \cite{vandre13}, it is shown that air entrainment occurs when the capillary forces at the inflection point cannot sustain the pressure gradients required to remove gas from the lubricating gas film.  Here, the pressure gradient $\triangle p$ required to maintain the flow at the inflection point, where $Kn_{loc}(h=H)=Kn_H$, is given by
\begin{equation}\label{dp}
\triangle p \equiv \left.\pdiff{p}{y}\right|_{h=H}=\bar{\mu}\left.\pdiff[2]{v}{x}\right|_{h=H}=-\frac{6\bar{\mu}\left[1+2A Kn_{H}+V\left(2Kn_{H}+1\right)\right]}{H^2\left[1+4 Kn_{H}(1+A)+12 A Kn_{H}^2\right]}.\end{equation}
As $Kn_{H}\rightarrow 0$, so that one approaches no-slip on the solid surface, we have $\triangle p\rightarrow \triangle p_{0} = -6\bar{\mu}(1+V)/H^2$ and in Figure~\ref{F:dp}, the pressure gradient in (\ref{dp}), normalised by $\triangle p_0$, is given as a function of $Kn_H$ for the cases of $A=0,1$ and shown to agree well with the values from our computations.  Notably, curves 1,~2 for $A=0,1$ show that for large $Kn_H$ the normalised pressure gradient for $A=0$ asymptotes towards a non-zero value whilst for $A=1$ it tends to zero.  Determining $\triangle p/\triangle p_0$ as $Kn_H\rightarrow \infty$ from (\ref{dp}) confirms that for $A=0$ the limit is finite at $V/(2(1+V))=0.23$ whilst for $A=1$ the limit is zero.

These findings help us to understand the results shown in Figure~\ref{F:Pg_vs_cac}.  For $A=1$, as the ambient pressure is reduced ($\bar{P}_g\rightarrow 0$), and hence Maxwell slip is increased ($Kn_H\rightarrow\infty$), the pressure gradient required to pump gas from the contact line region vanishes, so that the gas offers no resistance to contact line motion and, consequently, the maximum speed of wetting appears to become unbounded $Ca_c\rightarrow\infty$.  In contrast, for $A=0$, the lack of slip at the free-surface means that gas is always being driven into the contact line region by the motion of the liquid so that however much $\bar{P}_{g}$ is reduced ($Kn_H$ increased) a pressure gradient in the gas is always required and $Ca_c$ approaches a finite value based on this.





\subsection{Evaluation of the Incompressibility Assumption}\label{S:incom}

The analytic expression derived for the pressure gradient in the thin film (\ref{dp}) can be used to estimate the validity of the assumptions we have made about the gas flow.  In particular, although the flow is clearly low Mach number, compressibility effects in thin films can still be significant, see \S4.5 of \cite{gadelhak06}.  This occurs when the large pressure gradients required to pump gas out of the thin film lead to reductions in pressure that are comparable to the ambient pressure in the gas.  As it is the gas flow at the inflection point on the free-surface that is most important for air entrainment, we will estimate whether the flow in the gas there is indeed incompressible.

The pressure gradient obtained in (\ref{dp}) takes a maximum value of $\triangle p_0= -12\bar{\mu}/H^2$, as $V\le 1$, so that the maximum pressure change along a film of length $L_f$ is $12\bar{\mu}L_f/H^2$.  Simulations show that the film length is never larger than the capillary length so that $L_f=1$ is an upper bound.  Comparing the pressure change to the ambient pressure $P_g$ which (dimensionlessly) is $P_{g}=P_g^{\star} L^{\star}/(\mu^{\star} U^{\star})$ gives us that compressibility can be neglected if $\triangle p_0 \ll P_g$ which requires
\begin{equation}\label{incom_criterion}
H \gg H_T =  3.5 \sqrt{ \mu^{\star}_g U^{\star}/(P_g^{\star}  L^{\star})},
\end{equation}
where $H_T$ is the (dimensionless) transition height below which compressible effects can no longer be neglected.  This is consistent with a similar condition derived for drop impact phenomena in \cite{mani10}. 

Taking $U^{\star}=1$~m~s$^{-1}$ gives $H_T = 1.2\times10^{-3}$ for air at atmospheric pressure whilst reducing the pressure by a factor of one hundred gives $H_T = 1.2\times10^{-2}$.  When looking at the values of $H$ obtained at various $Ca$ in Figure~\ref{F:H} it appears that compressibility may indeed be important.  However, the estimate obtained in (\ref{incom_criterion}) overpredicts $H_T$ due to (a) neglecting reductions in pressure gradients associated with slip at the interfaces and (b) approximating the film as being a channel of length $L$ and constant height $H$ whereas the film height actually increases by orders of magnitude as one moves away from the contact line, see Figure~\ref{F:thinness}.   Therefore, before abandoning incompressibility, a more accurate evaluation of this assumption is considered in which the pressure changes along the film obtained from our computations are used instead of $\triangle p_0$ from (\ref{dp}).  

\begin{figure}
     \centering
\includegraphics[scale=0.3]{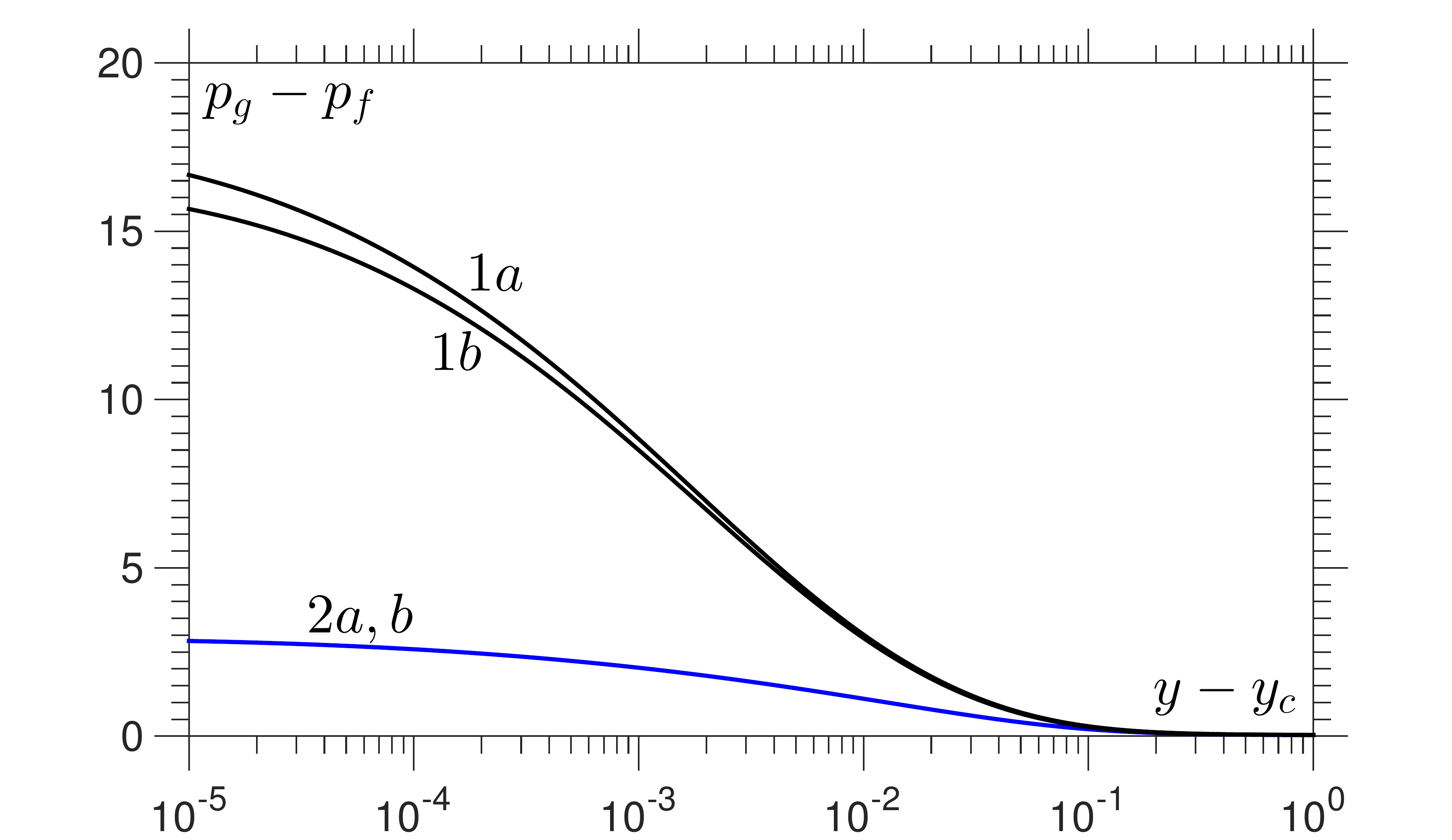}
 \caption{Deviation of the gas pressure $p_g$ from its far field value of $p_f$ as a function of distance from the contact line $y-y_c$ with curves 1 (in black) and 2 (in blue) corresponding to, respectively, cases considered in Figure~\ref{F:stream}b and Figure~\ref{F:stream}c.  The pressure along the gas-solid interface is given by curves $1a,2a$ whilst $1b,2b$ are values from the gas-liquid interface.}
 \label{F:pg_vs_y}
\end{figure}

In Figure~\ref{F:pg_vs_y} the variation in $p_g$ from its far-field value $p_f$ is plotted as a function of distance from the contact line $y-y_c$ along both the gas-solid and gas-liquid interfaces for the cases shown in Figure~\ref{F:stream}b,c, i.e.\ for $\bar{P}_g=0.14,~0.014$ with $A=1$. As one would expect in a lubrication flow, the values along the two interfaces are close and are graphically indistinguishable along curve 2.  It can be seen that the maximum change in $p_g$ in the thin film region for the cases considered is $17$ and $3$, respectively.  For the liquid associated with the base state, the substrate speed at $Ca=0.4$ is $U^{\star}=0.16$~m~s$^{-1}$ so that for $\bar{P}_g=0.014$, where the gas is most likely to be compressible, we have $P_g=260$.  Therefore, in these cases the pressure change along the film $p_{g}-p_{f}$ is  is substantially smaller than the ambient pressure $P_g$ with $(p_g-p_f)/P_g <0.012$.

Notably, although the lubrication analysis accurately predicts the pressure gradient in the film, see Figure~\ref{F:dp}, using the analytic value at the inflection point to estimate the pressure change along the \emph{entire} length of the thin film is inaccurate.  This is due to the widening of the film as one moves away from the contact line.  A more accurate estimate could have been achieved by integrating the analytically calculated pressure gradient over the computationally obtained film profile, but a simpler way is to use the computed values of $p_g$.  Doing this shows that the assumption of incompressibility is an accurate one for the base cases considered.  Further simulations confirm that even in the most extreme cases considered here, the incompressibility assumption remains a good one.

\section{Parametric Study of the System}\label{S:para}

Having fully analysed the base state, the role of the system's parameters is now established by perturbing their values about the base ones.  Henceforth, we only consider the most relevant case of $A=1$.

\subsection{Influence of Viscosity Ratio}	

The viscosity ratio $\bar{\mu}$ has long been recognised as an important parameter in coating flows as it is a measure of how much resistance the receding gas phase can produce.  Working with a dimensionless system allows us to isolate the effect of $\bar{\mu}$ on $Ca_c$, at different ambient pressures, whilst keeping all other parameters fixed at their base state. The range considered is chosen to cover all values of $\bar{\mu}$ obtained for the liquids in \cite{benkreira08} used to coat a solid in air, giving $10^{-4}\le\bar{\mu}\le10^{-2}$.
\begin{figure}
     \centering
\includegraphics[scale=0.3]{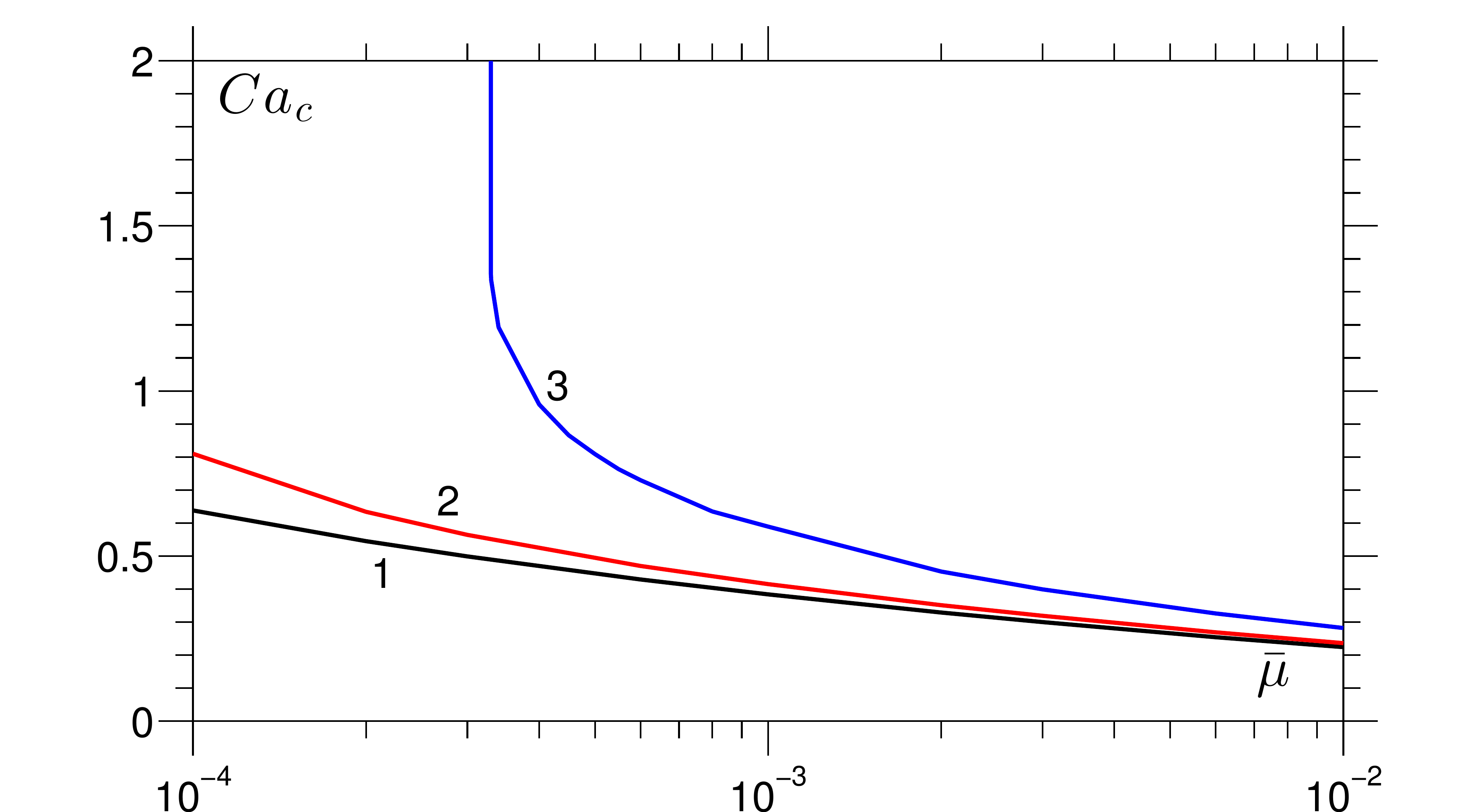}
 \caption{Dependence of the critical capillary number $Ca_c$ on the viscosity ratio $\bar{\mu}$, with all other parameters fixed at their base values, with curve 1: $\bar{P}_g=1$ (black), 2: $\bar{P}_g=0.1$ (red) and 3: $\bar{P}_g=0.01$ (blue).}
 \label{F:mubar}
\end{figure}

In Figure~\ref{F:mubar} one can see that lower values of the viscosity ratio $\bar{\mu}$ result in higher $Ca_c$ across all values of $\bar{P}_g$.  What was less expected, is that for $\bar{P}_g=0.01$ (curve 3) it appears there is a value of $\bar{\mu}$ below which $Ca_c$ increases apparently without bound.  This occurs at $\bar{\mu}=3.3\times10^{-4}$ which is slightly less than the base case's viscosity ratio ($\bar{\mu}_0=3.6\times10^{-4}$) where the critical gas pressure $P_{g,c}=9\times10^{-3}$.  Therefore, at smaller $\bar{\mu}$ the critical value $\bar{P}_{g,c}$ increases, i.e.\ less pressure reduction is required to reach the critical value. 

The effect of $\bar{\mu}$ on $\bar{P}_{g,c}$ can be understood by looking back to the lubrication analysis in the previous section and, in particular, the expression for the pressure gradient in the gas (\ref{dp}) which is proportional to $\bar{\mu}$ (in dimensional terms it would be proportional to $\mu_g^{\star}$).  Therefore, whilst the degree of Maxwell slip controls the boundary conditions to the gas flow, the bulk flow in the gas is characterised by the viscosity ratio $\bar{\mu}$, with larger values decreasing $Ca_c$ as more effort is required to remove gas from the contact line region.

\subsection{Effect of Ohnesorge Number}

\begin{figure}
     \centering
\includegraphics[scale=0.3]{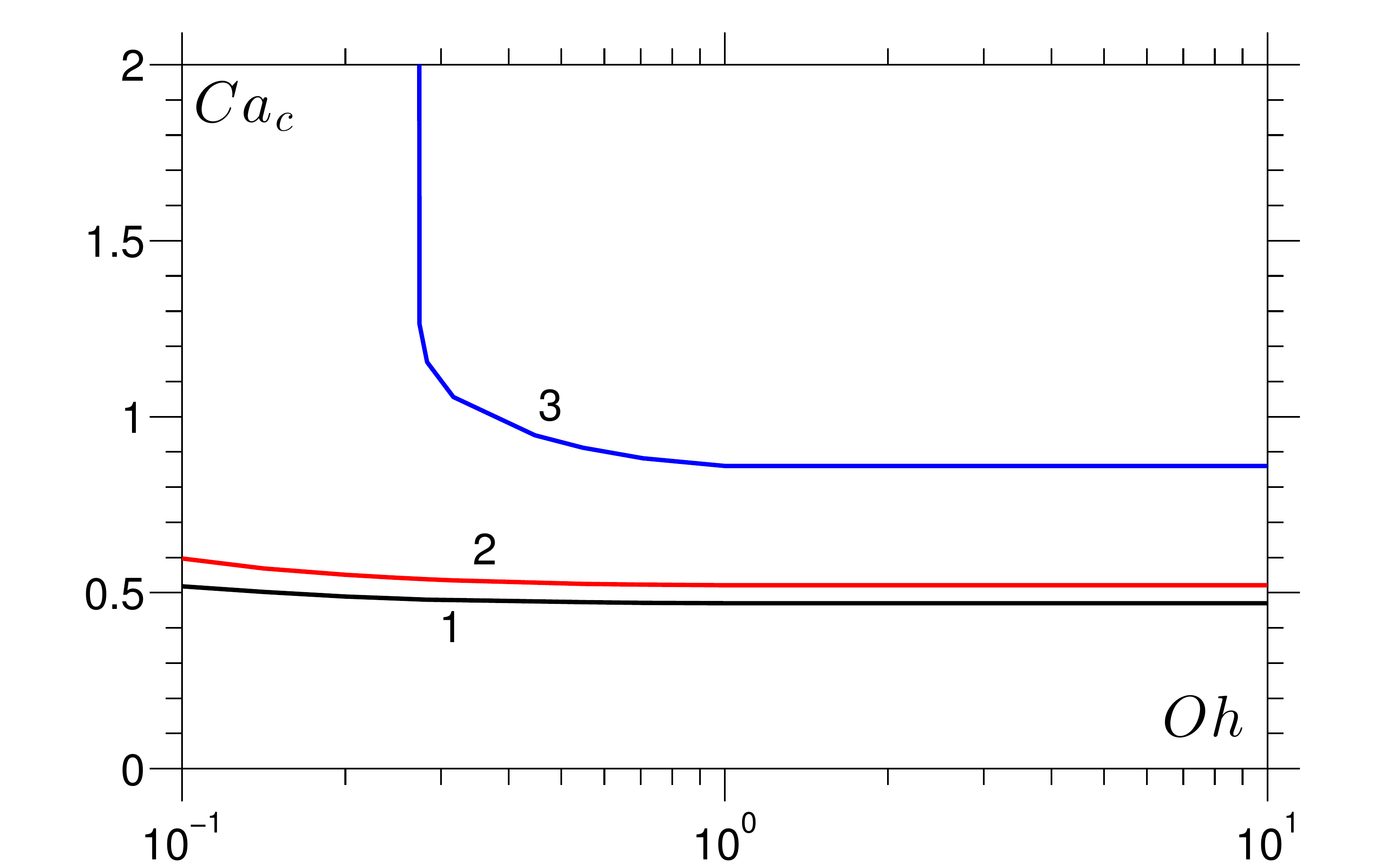}
 \caption{The dependence of the critical capillary number $Ca_c$ on the Ohnesorge number $Oh$ with all other parameters fixed at their base values, with curve 1: $\bar{P}_g=1$ (black), 2: $\bar{P}_g=0.1$ (red) and 3: $\bar{P}_g=0.01$ (blue).}
 \label{F:Oh}
\end{figure}

The effects of inertia are usually assumed to have a negligible influence on the dynamics of air entrainment, with Stokes flow, corresponding to $Oh\rightarrow \infty$, often considered.  In full computations of coating flows in \cite{vandre13} it was shown that (a) inertial effects in the gas have a negligible influence, reaffirming our assumption that $\bar{\rho}=0$ can be taken without loss of generality, and (b) inertial forces in the liquid do not alter $Ca_c$ until they are relatively large.  The results in Figure~\ref{F:Oh} at atmospheric pressure agree with these previous conclusions, with a $10^2$ reduction in $Oh$ only increasing value of $Ca_c$ by about $10$\%.  Note that $Ca_c$ at $Oh=0.1$ corresponds to $Re=52$.  

For the lowest ambient pressure $\bar{P}_g=0.01$ an entirely different behaviour is observed and there is a critical value of $Oh$ where $Ca_c$ appears to increase without bound. In contrast to the influences of $\bar{P}_g$ and $\bar{\mu}$ which could be rationalised by considering the gas flow, this effect is generated by changes in the liquid's dynamics. In particular, in \cite{vandre13} it was shown that when inertial effects become more prominent, velocity gradients in the liquid are localised to a thin boundary layer near the moving solid surface and, as a result, the free-surface becomes less deformed so that $Ca_c$ increases.  Our results show that when combined with small values of $\bar{P}_g$ this mechanism can prevent air entrainment at any $Ca$.

\subsection{Role of Substrate Wettability}\label{S:theta}

\begin{figure}
     \centering
\includegraphics[scale=0.3]{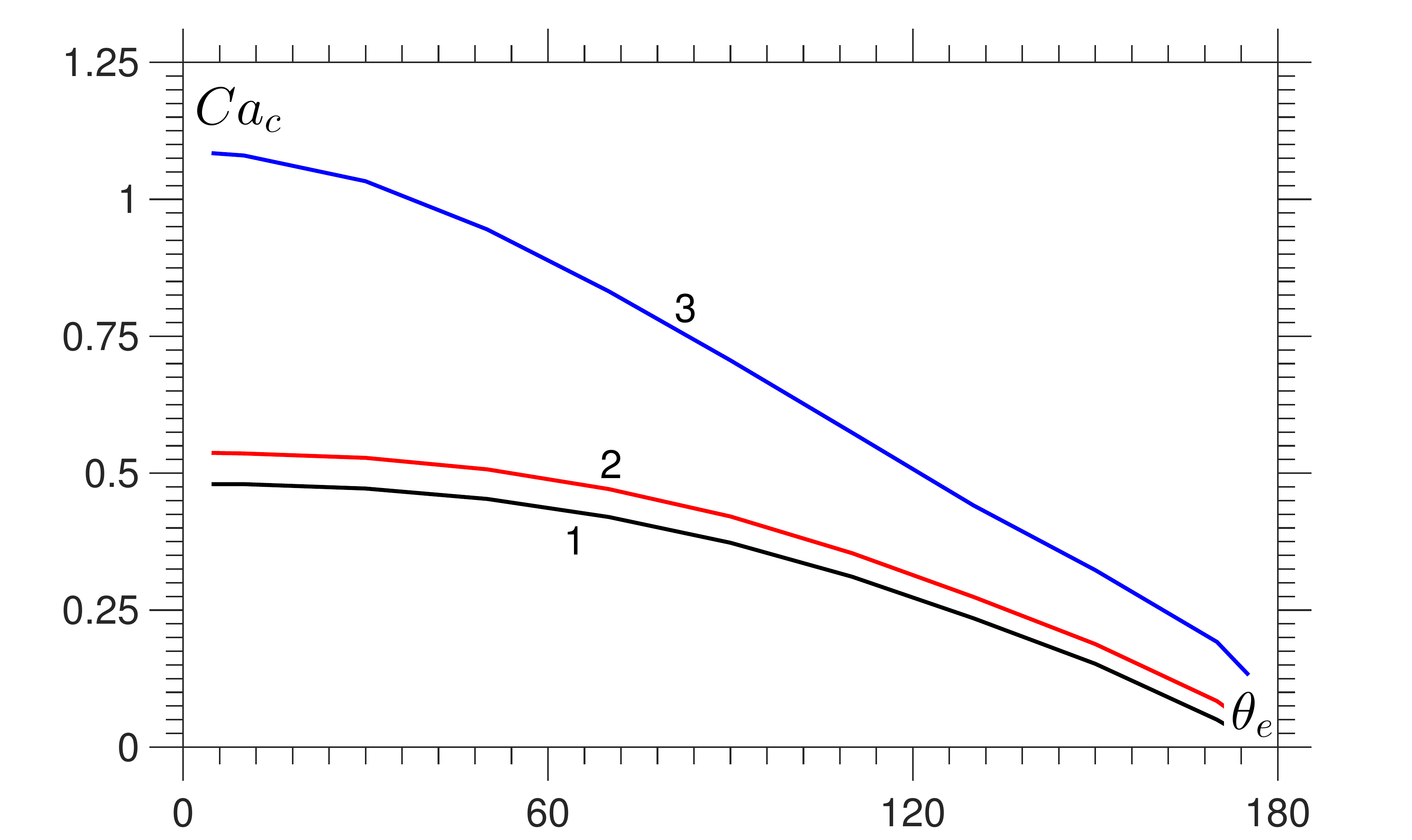}
 \caption{The dependence of the critical capillary number $Ca_c$ on the equilibrium contact angle $\theta_e$ with all other parameters fixed at their base values, with curve 1: $\bar{P}_g=1$ (black), 2: $\bar{P}_g=0.1$ (red) and 3: $\bar{P}_g=0.01$ (blue).}
 \label{F:theta}
\end{figure}

Experimentally, the wettability of the substrate is known to have an influence on the point of air entrainment both in coating flows and in impact problems, such as the impact of solid spheres on liquid baths \citep{duez07}.  In Figure~\ref{F:theta}, our results obtained at constant contact angles $\theta_d=\theta_e$, show that the more wettable the substrate, the higher $Ca_c$ is. Curves are from $5^\circ\le\theta_e\le175^\circ$ to avoid computational difficulties associated with extremely small angles in either phase, but, despite this, the trends are still clear enough.  Notably, at reduced pressures variations in $Ca_c$ with $\theta_e$ are far more dramatic: changing $\theta_e$ from $30^\circ$ to $90^\circ$ reduces $Ca_c$ by $0.32$ at $\bar{P}_g=0.01$ and compared to $0.1$ at $\bar{P}_g=1$.

The model used here predicts that $Ca_c$ monotonically decreases as $\theta_e$ increases, which is what one may intuitively expect.  However, in \cite{blake02a} the optimal value of $\theta_e$ which maximises $Ca_c$ is not at zero and can even occurs on hydrophobic substrates $\theta_e\ge90^\circ$. It is likely that such effects can only be captured by using a dynamic contact angle model and so this aspect lies beyond the scope of the present paper.  However, further experiments at both atmospheric and reduced pressure for solids of different wettabilities would be useful shed further light on these issues.

\subsection{Different Gases}

Previous studies in both dip-coating \citep{benkreira10}  and drop impact \citep{xu05} have considered the effect of using different gases on wetting failure.  The gases used tend to have a similar viscosity and density at atmospheric pressure to air but possess different molecular weights and hence mean free paths.  The fluid flow considered here is incompressible, so that changes in the speed of sound in the gas, and hence the Mach number, do not concern us.  However, changes in the mean free path at a given ambient pressure will have an effect on our results.  This effect can be accounted for in the results presented, without recomputing everything, by simply rescaling the ambient gas pressure $\bar{P}_g$.  For example, if a new gas considered has a mean free path $\ell^{\star}_{atm,n}$ which is $c$ times larger than that of air at atmospheric pressure, so that $\ell^{\star}_{atm,n} = c\ell^{\star}_{atm}$, then from (\ref{kn}) the same $Kn$ is obtained if we use a rescaled pressure $\bar{P}_{g,n}$ in the new gas which satisfies $\bar{P}_{g,n} = c\bar{P}_g$.  

In this way, all the results of the previous section can be used for any gas, rather than just air.  For example, if helium is used instead of air, approximately $\ell^{\star}_{atm,He} = 3\ell^{\star}_{atm,air}$, so that $\bar{P}_{g,He} = 3\bar{P}_g$.  Then, for the base case considered in \S\ref{S:gasp} the critical pressure was found to be $\bar{P}_g=9\times10^{-3}$ for air so that for helium this would be increased to $\bar{P}_g=2.7\times10^{-2}$.  Therefore, as seen experimentally, built into the model is the fact that gases with larger mean free paths require less pressure reduction in order to induce significant non-equilibrium gas effects that increase $Ca_c$.

\subsection{Summary of the Parametric Study}

The effects of non-equilibrium gas dynamics on air entrainment phenomena have been clarified and the role of the various dimensionless parameters on $Ca_c$ has been characterised by perturbing them about a base case.  This has allowed us to isolate the effects of parameters such as $\bar{\mu}$ which cannot easily be independently varied experimentally.  The disadvantage is, of course, that the results cannot easily be used to find, say, the effect of $\mu^{\star}$ on coating speeds as this variable comes into $Ca$, $Oh$ and $\bar{\mu}$ which have thus far been varied independently.  Therefore, in the next section we compare our results directly to experiments in \cite{benkreira08} and, in doing so, now step into a dimensional setting.

\section{Comparison to Experimental Data:  Effect of Viscosity and Gas Pressure}\label{S:exp}

Consider now whether the new model is able to account for the influence of gas pressure $P_g^{\star}$ on maximum coating speeds $U_c^{\star}$ observed in \cite{benkreira08} for different viscosity liquids.  All parameters remain unchanged from \S\ref{S:para} except those which depend on viscosity, namely $Oh$, $Ca$ and $\bar{\mu}$ which will be based on values used in the experiments of $\mu^{\star}=20,~50,~100,~200$~mPa~s.

\begin{figure}
     \centering
\subfigure[]{\includegraphics[scale=0.3]{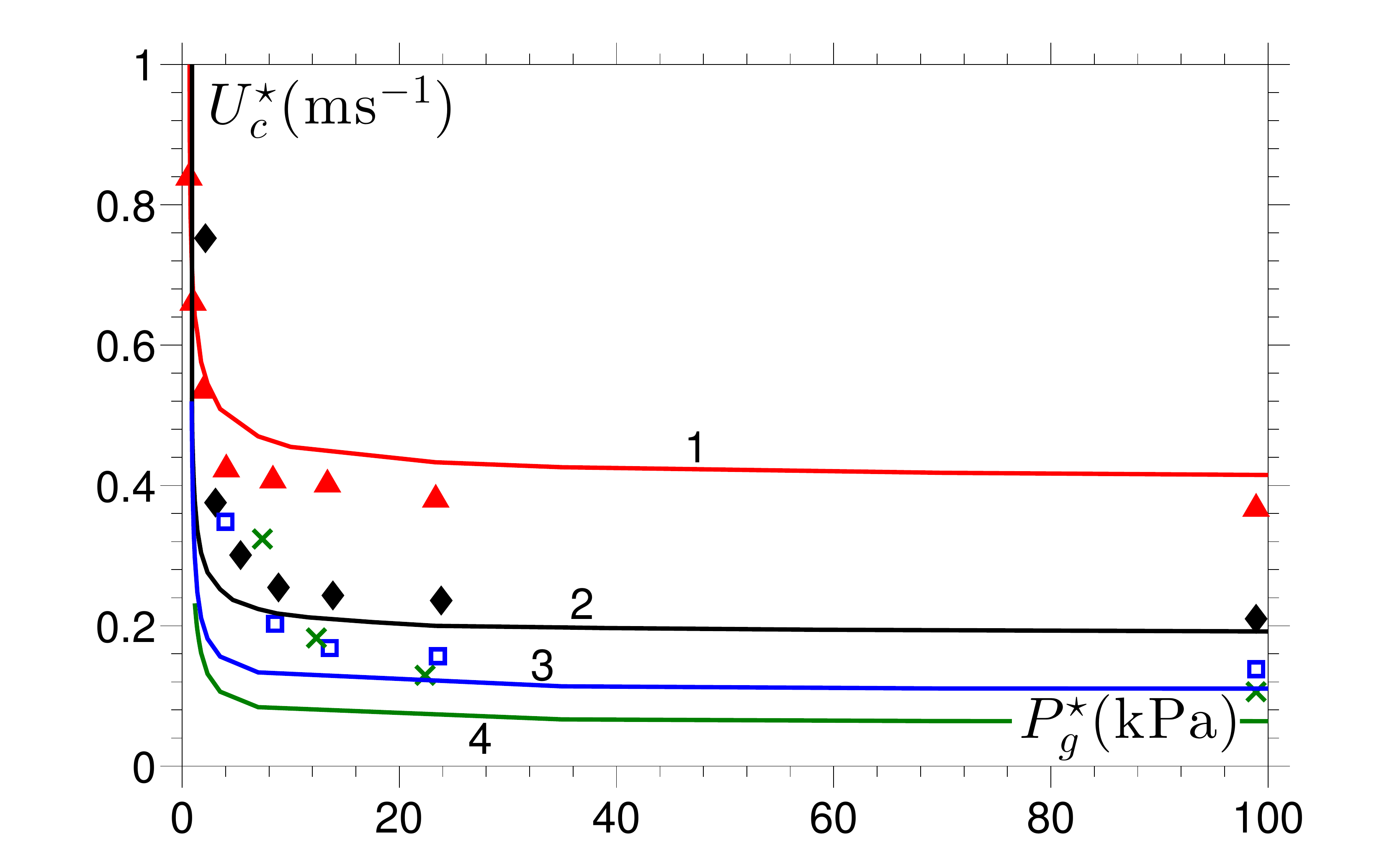}}
\subfigure[]{\includegraphics[scale=0.3]{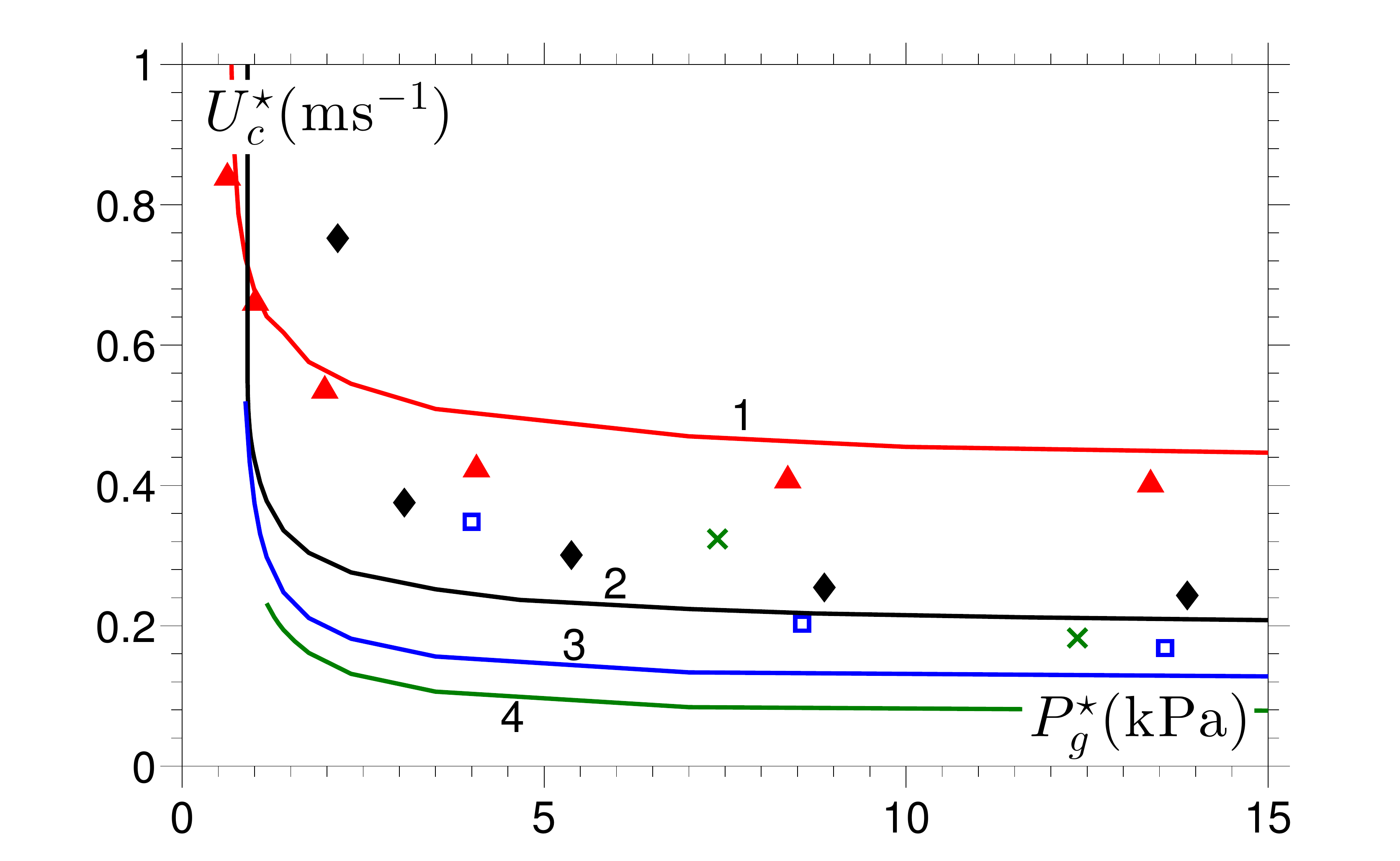}}
 \caption{A comparison of our computations for the critical wetting speed $U_c^{\star}$ as a function of gas pressure $P_g^{\star}$ to experimental results obtained in \cite{benkreira08}.  Curves are computed for liquids of different viscosities $\mu^{\star}$ with 1: 20~mPa~s (in red), 2: 50~mPa~s (in black), 3: 100~mPa~s (in blue) and 4: 200~mPa~s (in green) and the corresponding experimental data are given by, respectively, triangles, diamonds, squares and crosses.}
 \label{F:benkreira}
\end{figure}

In Figure~\ref{F:benkreira}, a direct comparison between the computations and the experimental results in \cite{benkreira08} is shown.  Notably, both predict that a significant reduction in $P_g^{\star}$ from its atmospheric value ($P_{g,atm}^{\star}=100$~kPa) is required to induce increases in $U_c^{\star}$, but that once this has been achieved, the subsequent enhancements in $U_c^{\star}$ are substantial.  At the lowest viscosity, the agreement is relatively good throughout; however, as is most clear from Figure~\ref{F:benkreira}b, one can see that at the higher viscosities, the effect of gas pressure yields enhanced coating speeds earlier in the experiments than in the computations.  For example, for the highest viscosity liquid, at $P_{g}^{\star}=12$~kPa the experimental data (crosses) shows that $U_c^{\star}$ has increased to $0.18$~m~s$^{-1}$ from its atmospheric value of $0.11$~m~s$^{-1}$ whilst the computations, shown by curve 4, only give this degree of enhancement once $P_{g}^{\star}=2$~kPa.  

Promisingly, as in the experimental results, curves 1 and 2 cross at low atmospheric pressure meaning that a higher viscosity liquid can be coated at faster speeds than a lower one, in contrast to what is observed at atmospheric pressure. Other crossovers may have been observed if higher capillary numbers could have been obtained, as suggested by the path of curve 3.  However, the position of the crossovers do not match the experimental data; whilst the computed results all appear to crossover around $P_{g}^{\star}\sim1$~kPa, the experimental results do so at higher values of $P_{g}^{\star}$ with the two highest viscosity solutions crossing over at $P_{g}^{\star}\sim10$~kPa.

In summary, whilst all of the qualitative features of the experimental results are recovered in our computations, quantitative agreement is not good over the entire range of data.  In particular, less pressure reduction is required in the experiments than in the computations for there to be a noticeable effect on the maximum wetting speed.  There are a number of possible reasons for this discrepancy, some of which we will now consider.\\
\begin{description}

\item[Roughness of the Solid:~~] In \cite{benkreira08} the ``average peak-to-valley height roughness" for the experimental results presented is $300$~nm.  Although this is `smooth' compared to the other solids used in this work, it is still quite rough compared to the solids used in some other coating flow studies.  This effect cannot be ignored as experiments in \cite{clarke02a} have demonstrated that the roughness of a solid can influence the maximum speed of wetting with, in many cases, faster coating speeds obtained with rougher surfaces.

From a modelling perspective, the level of roughness in the experiments is large when compared to the slip length at the liquid-solid interface $l_s^{\star}=10$~nm. However, simply increasing the value of $l_s^{\star}$ in an attempt to mimic the effect of the roughness does not capture the experimental data either.  It is likely that the strongest effect of the roughness will be on the actual contact angle, but a theoretical description of this mechanism at high contact-line speeds remains an open problem and is beyond the scope of this work 

\item[Dynamics of the Contact Angle:~~] Despite assuming that the contact angle is fixed at its equilibrium value, without fitting the slip length in any way we still obtain reasonable agreement for the values of $U_c^{\star}$ at atmospheric pressure.  This could be a coincidence and dynamic contact angle effects may indeed be important for quantitatively capturing this class of coating flows; however, such effects will not improve the agreement of our current model with experiments. This is because allowing the contact angle to vary will increase the angle from its equilibrium value so that, as shown by the findings of \S\ref{S:theta}, $U_c^{\star}$ will decrease for a given pressure.  Therefore, although the dynamics of the contact angle will alter our results, it will not explain why $Ca_c$ increases so quickly with reductions in $\bar{P}_g$ and is thus not our main concern here.

\item[Model for Non-Equilibrium Effects:~~] As shown in \S\ref{S:gasfilm}, the Knudsen number characterising the gas film flow indicated that its dynamics is usually outside the `slip regime' as air entrainment is approached and more complex non-equilibrium gas dynamics should be incorporated. As noted in \S\ref{S:gasfilm}, incorporation of these effects is likely to result in the same qualitative trends but enhanced quantitative ones. In particular, it is likely that less gas will be driven into the contact line region at a given pressure so that there is an enhancement in $Ca_c$ at higher ambient pressures, as observed experimentally.  Furthermore, this effect will be particularly important at smaller $\bar{\mu}$, where most of the current discrepancies between theory and experiment appear, as the smaller values of $H$ obtained there \citep{vandre13} will increase the importance of non-equilibrium effects

Models capable of predicting these effects will be discussed in \S\ref{S:discussion}.

\end{description}

The most likely cause of the observed discrepancy seems to be our inadequate description of the gas dynamics, and this aspect will be discussed further in \S\ref{S:discussion}.  Before doing so, we consider what consequences our results have for the drop impact phenomenon.

\section{Non-Equilibrium Gas Effects in Drop Impact Phenomena}\label{S:drops}

It has previously been suggested \citep{rein08},  that certain classes of splash observed in drop impact phenomena are triggered by the same mechanisms that cause wetting failure in coating flows.  This interpretation is supported by experiments such as those in \cite{driscoll11} who find ``no trapped air beneath the spreading drop outside the small central bubble; there is no significant air film beneath the drop at the time of thin-sheet ejection. This suggests that, rather than an underlying air layer, gas flow at the edge of the spreading drop is responsible for destabilizing the liquid.''  This mechanism is shown by Figure~\ref{F:drop}b. Therefore, we now consider how our results can be related to the more complex unsteady drop impact phenomenon.

\begin{figure}
     \centering
\includegraphics[scale=0.3]{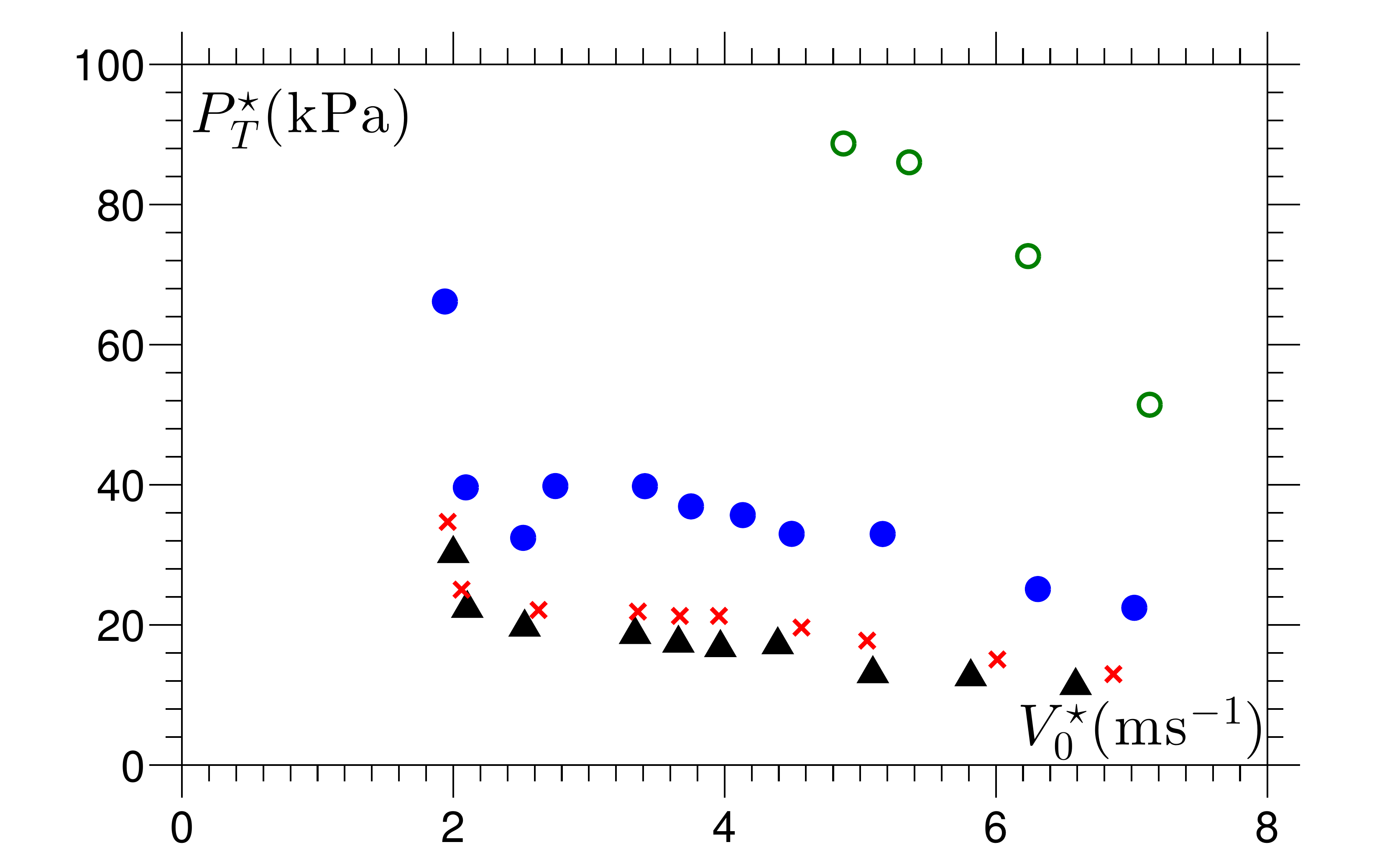}
 \caption{Critical gas pressure $P_T^{\star}$ required to suppress the splash of a drop with impact speed $V^{\star}_0$ with different gases helium (open circles in green), air (filled circles in blue), krypton (red crosses) and SF$_6$ (triangles in black).  Obtained from \cite{xu05}.}
 \label{F:xu_dim}
\end{figure}

The publication which opened up this field of research was \cite{xu05}.  This work focuses on a threshold ambient pressure $P_T^{\star}$ which must be achieved in order to suppress splashing for a particular liquid-solid-gas configuration.  What is particularly interesting is that for a given impact speed $V_0^{\star}$, the threshold pressure strongly depends on the gas, with air, helium, krypton and SF$_6$ all used.  The data is shown in Figure~\ref{F:xu_dim}, which corresponds in \cite{xu05} to the inset of their Figure~2b.  In \cite{xu05}, compressibility effects in the gas are assumed to be important when deriving a splash threshold, so that changes in the molecular mass of the gases $m_g^{\star}$, which alter the speed of sound in the gas, become important and result in a collapse of the data when $P_T^{\star}$ is scaled by $\sqrt{m_g^{\star}/m^{\star}_{air}}$, where  $m^{\star}_{air}$ is the molecular mass of air.  

An alternative explanation is that compressibility effects are negligible and that increased Maxwell slip is responsible for splash suppression at reduced pressure.  Given that the mean free path in the gas at the threshold point $\ell^{\star}_T$ is inversely proportional to both the gas pressure and the square root of the molecular mass $\ell^{\star}_T \propto (P^{\star}_T\sqrt{m_g^{\star}})^{-1}$, to achieve a given $\ell^{\star}_T$, increases in $m_g^{\star}$ require decreases in $P^{\star}_T$.  In other words, gases with higher molecular mass will need lower pressures to suppress splashing, as seen in \cite{xu05}.  

The dimensionless number characterising the level of Maxwell slip at a given threshold pressure in the gas is the Knudsen number $Kn_T=\ell^{\star}_T/R^{\star}$, where $R^{\star}=1.7$~mm is the drop radius.  Using the properties of air, the expression for $Kn_T$ required is
\begin{equation}\label{knt}
Kn_T = \frac{Kn_{air,atm}}{\sqrt{\tilde{m}_g}~ \bar{P}_T}, \qquad\hbox{where}\qquad \bar{P}_T = \frac{P^{\star}_T}{P^{\star}_{atm}}\quad \hbox{and} \quad\tilde{m}_g = \frac{m_g^{\star}}{m_{air}^{\star}},
\end{equation}
where the coefficient of proportionality $Kn_{air,atm}=4.1\times 10^{-5}$ is the Knudsen number of air at atmospheric pressure and the molecular weights in Daltons of $m^{\star}_{He}=4$, $m^{\star}_{air}=29$, $m^{\star}_{Kr}=83.8$ and $m^{\star}_{SF_6}=146$ can be taken from \cite{xu05}.  As for the coating flow analysis, the value of the Knudsen number has to be based on the global characteristics of the flow and would be larger if it could have been derived from the appropriate length scales near the moving contact line. 

Ideally, the collapse of data would use $Kn_T$ and a capillary number based on the contact line speed $U^{\star}$ at the point of air entrainment, as considered for the coating flow analysis.  However, such measurements were not made in the experimental work so we shall instead base the capillary number $Ca_{V^{\star}_0}$ on the impact speed $V^{\star}_0$, an assumption which will be discussed in \S\ref{S:discussion}.  
\begin{figure}
     \centering
\includegraphics[scale=0.3]{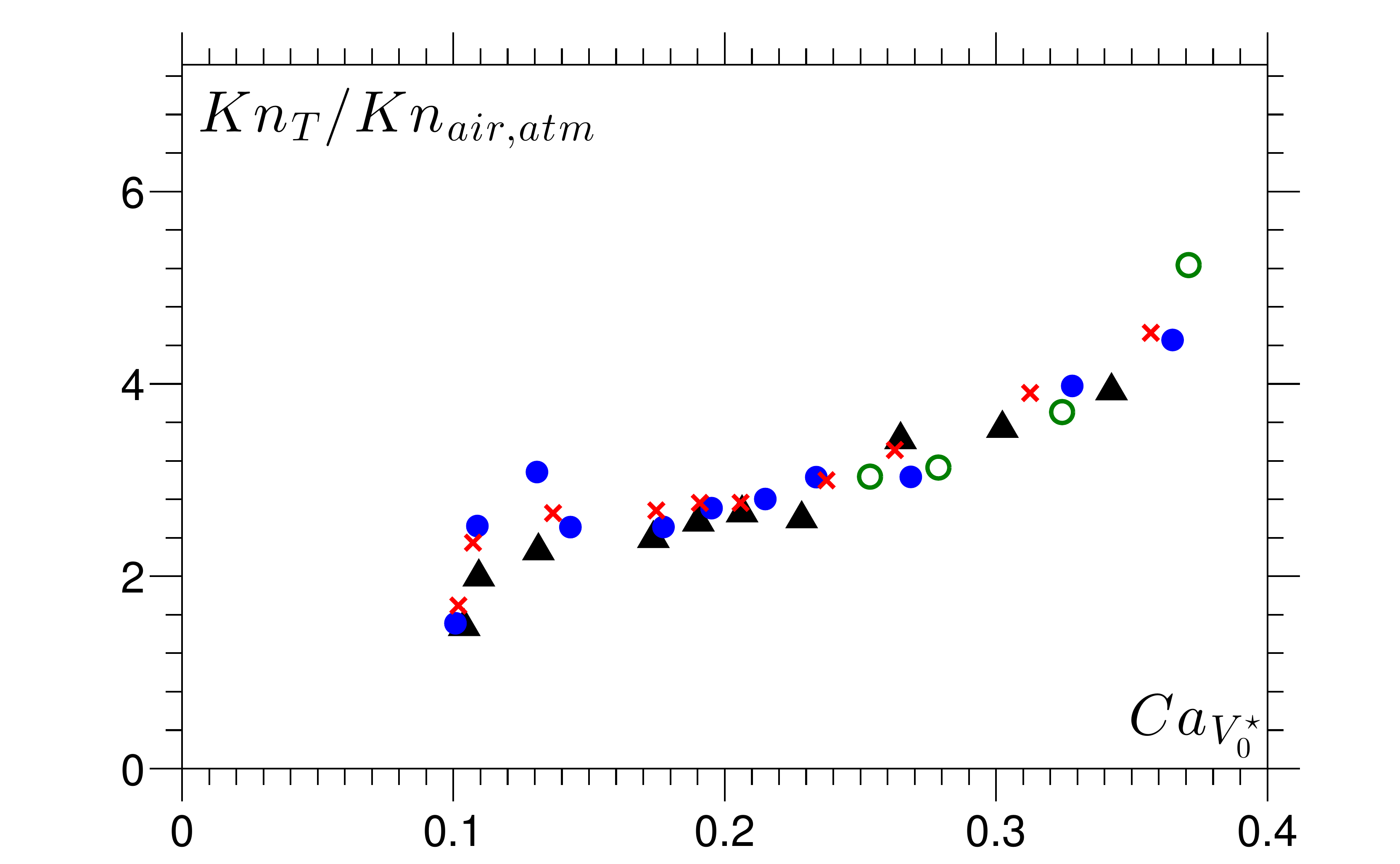}
 \caption{Critical Knudsen number $Kn_T$, scaled by its value in air at atmospheric pressure, required to suppress the splash of a drop whose capillary number based impact speed is $Ca_{V^{\star}_0}$.  The collapse of data from Figure~\ref{F:xu_dim} onto a single curve is apparent.}
 \label{F:xu_nd}
\end{figure}

In Figure~\ref{F:xu_nd}, the raw data from Figure~\ref{F:xu_dim} is shown to collapse well onto axes of $Ca_{V^{\star}_0}$ vs $Kn_T$. This is encouraging, and suggests that non-equilibrium effects in the gas may be important for drop impact phenomena.  It is particularly reassuring to see that the capillary numbers obtained are within the range one may expect for air entrainment phenomena to be occurring, although the capillary number based on the contact line speed at the entrainment point may be much larger.  Ideally, changes in the viscosity of the liquid would also be included into the data collapse, as achieved in \cite{xu05},  but this is not possible here as such variations cannot simply be absorbed into the capillary number, but will also alter many other dominant parameters such as the viscosity ratio.   

Although the data collapse looks impressive, a number of outstanding questions remain.  First, as also seen in \S\ref{S:exp}, experimentally only relatively small changes in atmospheric pressure are required to suppress splashing, with a factor of ten decrease sufficient to prevent it in all cases considered here (Figure~\ref{F:xu_dim}), and it remains to be seen whether the Maxwell-slip model is able to predict such a large effect at these reduced pressures.  Second, in this framework it is not obvious why the drop spreads for some distance before splashing, when the largest contact-line speeds are obtained earliest in the impact process.  These two issues will be addressed in \S\ref{S:discussion} and will require further experimental and theoretical work before they can be fully resolved.

%
%
%

\section{Discussion}\label{S:discussion}

It has been shown that non-equilibrium effects in a gas, influencing the flow through Maxwell slip at boundaries, can substantially increase the critical speed at which air entrainment occurs.  Notably, neglecting to use slip at the gas-liquid boundary produces qualitatively different results from those obtained when slip is allowed, predicting a finite, rather than unbounded, maximum speed of wetting as the pressure is reduced.  Importantly, the problem has been formulated locally, so that the model is valid for any shape of the gas domain and remains accurate both far away from the gas film, where non-equilibrium gas effects will have a negligible influence on the overall flow, as well as very close to the contact line where regions of high curvature invalidate the lubrication approximation.

The model proposed was shown to reproduce many of the features of the experimental results in \cite{benkreira08} but failed to accurately predict the ambient pressures at which coating speeds begin to rise for the higher viscosity liquids. These observations open up a number of potential areas for experimental and theoretical research which will now be considered.

\subsection{Extension of the Non-Equilibrium Gas Model}

General research into non-equilibrium gas dynamics has identified different `regimes' of flow which indicate the modelling approach that should be used.  Applying these to the flow in the gas film, characterised by $Kn_H$, gives that :\\
\begin{description}
  \item[Conventional regime ($Kn_H<10^{-3}$):]\quad The Navier-Stokes equations with no-slip at solid surfaces provide an accurate representation of the gas flow.
  \item[Slip regime ($10^{-3}<Kn_H<10^{-1}$):]\quad Non-equilibrium effects become significant close to the boundary so that the no slip condition must be relaxed to allow for Maxwell slip.
  \item[Transition regime ($10^{-1}<Kn_H<10$):]\quad Knudsen layers occupy significant proportions of the domain and kinetic theory is required to provide reliable predictions.
  \item[Free molecular flow regime ($Kn_H>10$):]\quad Molecular collisions become rare so that the collision operator in the Boltzmann equation can be neglected.\\
\end{description}
Clearly, identifying which regime is important for the dynamic wetting problems of interest will be key to deciding which methods should be implemented.

The Knudsen numbers recovered for the gas film flow show that in many cases the model proposed in \S\ref{S:form} has been used outside its limits of applicability.  In particular, once in the transition regime the effect of Knudsen layers can no longer be accurately captured by a first-order slip condition, as assumed here.  An entire hierarchy of approaches of varying complexity have been proposed in the published literature to overcome this inadequacy, and which of these is optimum for the class of problem considered here is, at present, unclear. A sensible first step would be to remain within a continuum framework and to incorporate more advanced non equilibrium gas effects either, in the simplest case, by using a second-order slip condition, such as that proposed in \cite{hadji03}, or, at the next level of difficulty, by modifying the bulk constitutive relations also \citep{lockerby05a}.  Research in this direction is currently underway.

The regions of applicability of the aforementioned approaches are often unclear, and the only way to test these are to compare with solutions obtained from the Boltzmann equation.  This can now be routinely achieved using a variety of methods, with the direct simulation monte carlo (DSMC) approach popular \citep{bird94}.  However, in this setting it is not clear how the liquid-gas free-surface should be treated as the Boltzmann equation cannot be applied across the boundary and into the liquid phase.  Whether some plausible assumptions about the liquid's free-surface properties could be made or whether DSMC in the gas could somehow be coupled to molecular dynamics simulations \citep{koplik95} in the bulk of the liquid remains to be seen. Progress in the latter area would shed light on not only the bulk flow of the gas, but also its interaction with boundaries formed by liquids. 

\subsection{Alternative Flow Configurations}

It is of interest to look at the effects of gas pressure on coating speeds in a range of different setups, both experimentally and theoretically.  In particular, the effects of confinement, where the contact line is isolated by a nearby boundary, have not been considered and neither have axi-symmetric geometries such as those used in fibre coating.  

The fibre coating configuration is particularly interesting as experiments in \cite{simpkins03} show that extremely high capillary numbers can be achieved before air entrainment occurs and, at present, the precise cause for this remains to be seen.  Some works claim that the same mechanism has been observed in hydrodynamic assist \citep{blake04a} whilst others have proposed that air is actually entrained into the liquid but then dissolved due to the high pressures there \citep{jacqmin02}.  Incorporating the models proposed into the computational framework developed here should allow one to distinguish between these rival interpretations.

\subsection{Development of Analytic Theory}

The theory of \cite{cox86} has widely been used to calculate an apparent contact angle from its actual value in both single and two-phase dynamic wetting phenomena.  Although this theory is based on a small capillary number asymptotics, so that it lacks accuracy for many coating flows \citep{vandre13}, it nevertheless has been a useful tool for interpreting experiments and predicts a possible mechanism by which air entrainment can occur; namely, that once the apparent angle reaches $180^\circ$, no more solutions can be obtained.  However, in its present state this theory cannot capture the phenomena investigated in this thesis.

Cox's theory contains an `inner region' where additional effects come into play, such as slip, that allows the moving contact line problem to be circumvented.  When one has the same slip length in each phase, then the theory is valid, but for the problems we have considered the slip in the gas phase is many times larger than that in the liquid.  In particular, the slip length in the gas phase is likely to become comparable with the scale of the `intermediate region', where the bending of the free-surface occurs, so that the entire asymptotic structure of the problem becomes unclear.  It would be of interest to see if Cox's approach can be generalised for this more complex class of flows.

\subsection{Parametric Experimental Study of Dip Coating at Reduced Pressures}

The results of \cite{benkreira08} have allowed us to compare our computational results to experimental data.  However, these are, at present, the only results in the published literature considering the effect of gas pressure on coating processes. Therefore, additional experimental studies would be useful in order to develop the theory of this class of problems.  Such an investigation may, for example, use a wider variety of liquids and consider solids with a range of different wettabilities and/or roughnesses.  Furthermore, considering how accelerating or decelerating the plate alters the flow may give some initial insight into more complex unsteady flows such as drop impact.  

If the aforementioned analyses could be combined with modern experimental techniques, such as interferometry measurements, then some details of the gas film's profile could potentially be recovered, as has been achieved in drop impact phenomena.  These results would provide an excellent benchmark for any potential theories to be compared to.

\subsection{Drop Impact and Spreading}

Drop impact phenomena are notoriously difficult to capture computationally due to the inherently multiscale nature of a problem which involves unsteady large deformation of the flow domain.  Although this has been achieved in the case of a single-phase flow, e.g.\ in \cite{sprittles_pof}, at present there are no codes which have captured all physical scales in (a) the dynamics of impact, where a thin gas film cushions the falling drop; (b) the topological change, when the drop first touches the solid; and (c) the spreading process, where the slip length needs to be resolved.  Building a code capable of achieving this goal is highly sought after.

A question arising from our analysis in \S\ref{S:drops} is why the drop spreads for some distance before splashing, when its highest speeds are obtained early on in the impact process.  The answer to this question is likely to involve the drop's shape and its flow field, so that one cannot simply consider an isolated moving contact line, as we have done, but must also consider the outer influence of the flow field on the point at which air entrainment occurs.  

The situation with drop impact could be similar to that of curtain coating where the point at which air entrainment occurs is dependent not only on the contact line speed, but also on the flow geometry \citep{blake94,clarke06}.  In particular, when the contact line is below the falling liquid curtain it appears to be `assisted' by the downward pressure of the curtain \citep{blake04a}, which prevents air entrainment, whilst if a heel forms in the free-surface the contact line becomes isolated from the bulk flow and there is no assist.  Similarly, in drop impact, during the initial stages of spreading the contact line is located below the drop, and thus wetting is assisted by the downward pressure of the bulk flow, until a lamella is ejected, analogous to a heel in curtain coating.  At this point, the contact line is isolated and air entrainment between the lamella and the substrate is able to lift off a sheet of liquid which creates a splash.

In order to determine if the aforementioned mechanism is the correct one for drop impact phenomena, it may still be useful to use the simpler coating flow setups.  One possibility is that the influence of an impulsively started or accelerating plate could be considered in a manner to mimic the impact process.  Another avenue of enquiry would be to alter flow domain to mimic the flux of mass arriving at the contact line in drop impact phenomena.

From an experimental perspective, it would be extremely useful to know the contact line speed at the point of air entrainment, so that the capillary numbers used in \S\ref{S:drops} can be based on this, rather than the impact speed.  In particular, if air entrainment occurs after the drop's contact line has moved a radius $r_s^{\star}$, and the contact line moves with a square root in time dependency, as shown experimentally \citep{riboux14} for the initial stages of motion, so that $r^{\star}_s/R^{\star} = A \sqrt{V_0^{\star}t^{\star}/R^{\star}}$ and $U^{\star}/V_{0}^{\star} = A/2\sqrt{R^{\star}/(V_0^{\star}t^{\star})}$, where $A$ is a constant,  we find that $U^{\star}/V_{0}^{\star}=(A^2/2) R^{\star}/r^{\star}_s$. Therefore, $Ca_{V^{\star}_0}$ is only a fixed multiple of $Ca$ if air entrainment always occurs at the same radius $r^{\star}_s$, which seems unlikely to be the case across all parameter space. Therefore, we cannot be sure that $Ca_c\propto Ca_{V_0^{\star}}$ and more experimental data is indeed required.

\section*{Appendix: Benchmark Calculations}

From a theoretical perspective, both extended lubrication theory \citep{chan13} and full computations \citep{vandre12,vandre13} have identified a critical capillary number past which no steady two-dimensional solutions exist, and it has been assumed that this corresponds to the $Ca_c$ observed in experiments.  As initially observed for the plate withdrawal problem, \i.e. the dewetting scenario \citep{snoeijer07}, the turning point at $Ca_c$ in the map of elevation $\triangle y$ (or maximum apparent contact angle $\theta_M$) versus $Ca$, separates stable and unstable branches of steady-state solutions (Figure~\ref{F:vandre_elevation}).  Consequently, when considering the entire solution space, for certain $Ca$ multiple values of elevation $\triangle y$ exist, and no function $\triangle y=\triangle y(Ca)$ exists. Notably though, along all sections of the solution path we have $Ca<Ca_c$, so that $Ca_c$ is indeed the maximum speed of wetting.

In order to determine $Ca_c$ for a given system, a number of possible solution techniques exist.  The most obvious approach is to find a solution at a relatively small $Ca$, then increase its value, using the previous computed solution as an initial guess for the next $Ca$, and repeat until no more steady state solutions exist.  This is the simplest method and reliably finds $Ca_c$, but it does not allow us to extract the solution path past the critical point where $Ca<Ca_c$, as noted in \cite{vandre12}.  A different approach, which allows the entire solution path to be captured, is to determine $Ca$, i.e.\ treat it as an unknown in the problem, for a fixed elevation of the contact line $\triangle y$. This approach works better in practice, as for every $\triangle y$, there is a unique $Ca$, i.e.\ there is a function $Ca=Ca(\triangle y)$. Consequently, curves like those in Figure~\ref{F:vandre_elevation} can be generated. 

In principle, as much of the unstable branch as required can be captured, but for $\triangle y\ll \triangle y_{crit}$, where $\triangle y_{crit}$ is the elevation when $Ca=Ca_c$, the deformation of the free-surface becomes large and this process becomes computationally intensive, particularly at small $\bar{\mu}$. Consequently, simulations were run up to the point in the solution path where the mesh failed, usually due to the finite elements becoming too distorted, and stopped there.  As one can see from Figure~\ref{F:vandre_elevation}, all simulations were extended well past $\triangle y=\triangle y_{crit}$ and it was only the smallest viscosity ratio (curve 4) where the code finished prematurely, i.e.\ before $\triangle y=-3$.

\subsection*{Calculations and Comparison with \cite{vandre13}}

In order to compare our computations to previous work in  \cite{vandre13}, consider Stokes flow $Oh^{-1}=0$; with the slip-lengths of the liquid-solid and gas-solid interfaces equal at $l_s=Kn=10^{-4}$; no-slip across the liquid-gas interface, so that $A=0$; and a constant contact angle of $\theta_d=90^\circ$. Using these parameters, the critical capillary number $Ca_c$ and maximum apparent contact angle at this point $\theta_M$, which occurs at the inflection point on the free-surface (see Figure~\ref{F:sketch}), are computed for different viscosity ratios $\bar{\mu}=10^{-1},~10^{-2},~10^{-3},~10^{-4}$ and compared to the results in \cite{vandre13}.  

\begin{figure}
     \centering
\includegraphics[scale=0.3]{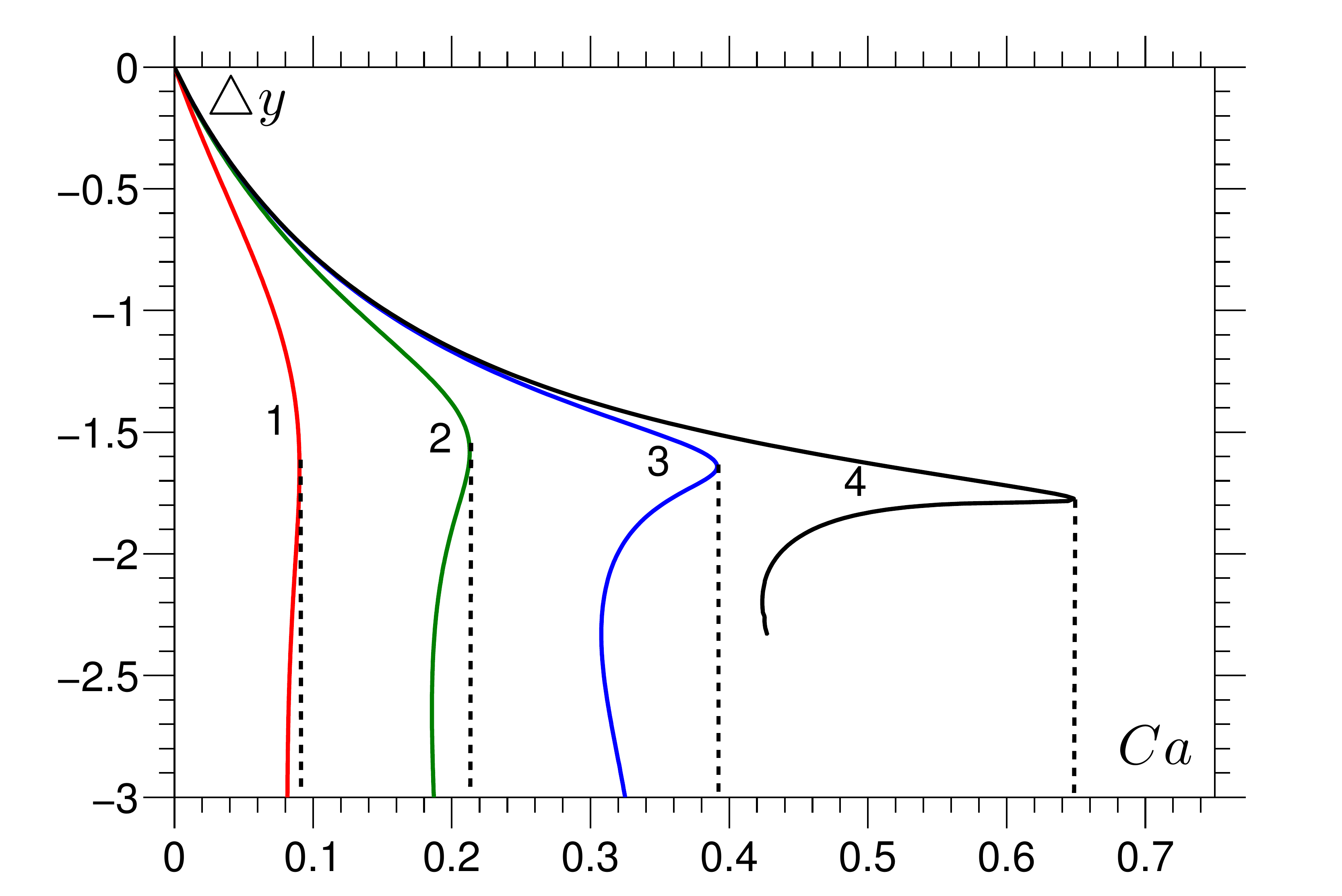}
\includegraphics[scale=0.3]{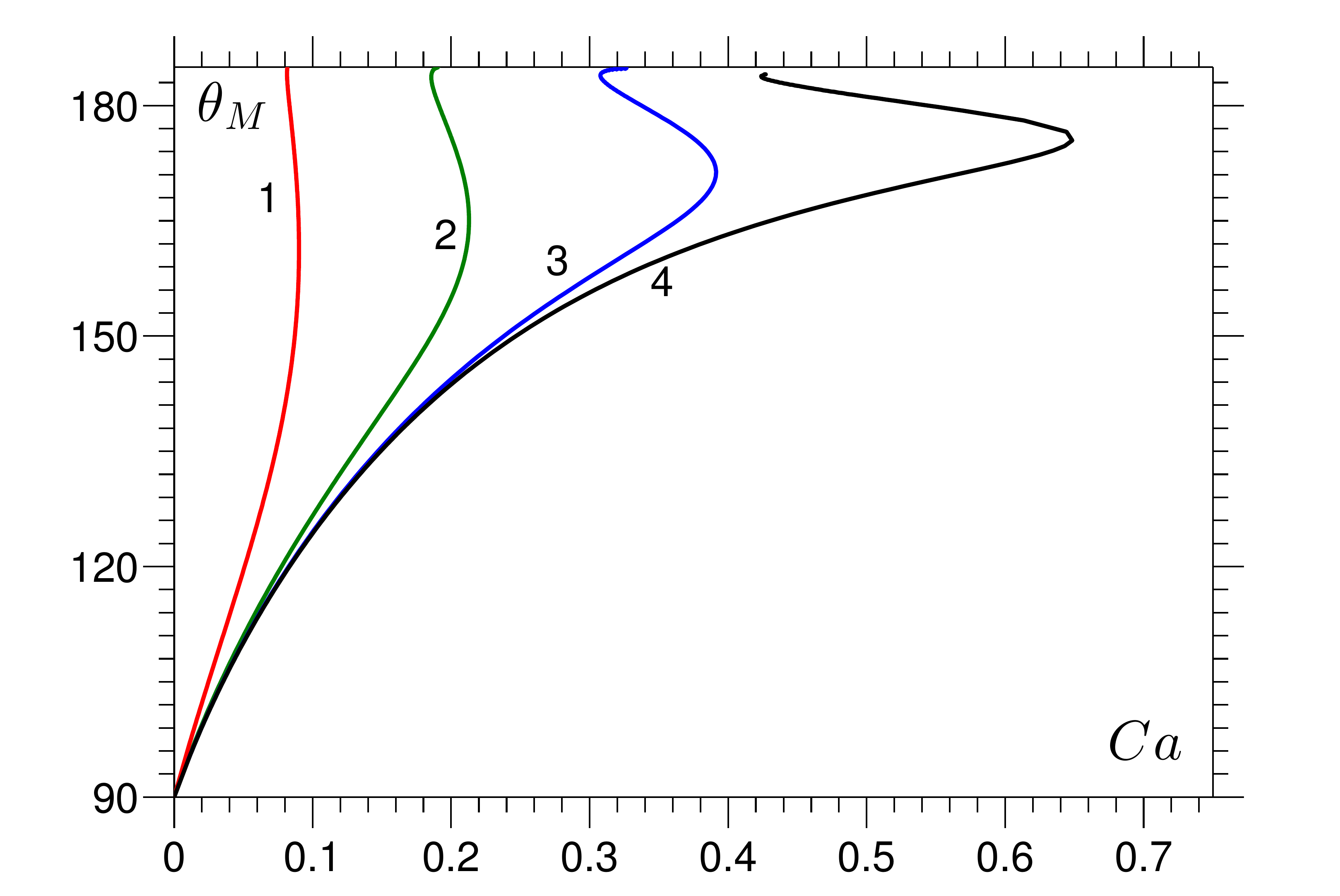}
 \caption{Dependence of the elevation $\triangle y$ and maximum apparent contact angle $\theta_M$ on the capillary number $Ca$ for $\bar{\mu}=1:10^{-1},~2:10^{-2},~3:10^{-3}~,4:10^{-4}$ and other parameters listed in the text.  Dashed lines indicate the critical capillary number $Ca=Ca_c$ past which no steady state solutions exist.}
 \label{F:vandre_elevation}
\end{figure}

In Figure~\ref{F:vandre_elevation}, the dependence of $\triangle y$ and $\theta_M$ on $Ca$ are computed.  From the curves one can clearly see the existence of a critical capillary number $Ca_c$ past which ($Ca>Ca_c$) no solutions exist.  On further decreasing $\triangle y$, so that the contact line moves further down the plate, solutions are recovered for which $Ca<Ca_c$, and the capillary number begins to oscillate around a capillary number $Ca^{*}<Ca_c$ as also observed in theoretical works on dewetting phenemena, where these solutions have been shown to correspond to a series of saddle-node bifurcations, see Figures 4 and 5 of \cite{snoeijer07}, which are related to transient states \citep{snoeijer06}. 

\begin{figure}
     \centering
\includegraphics[scale=0.3]{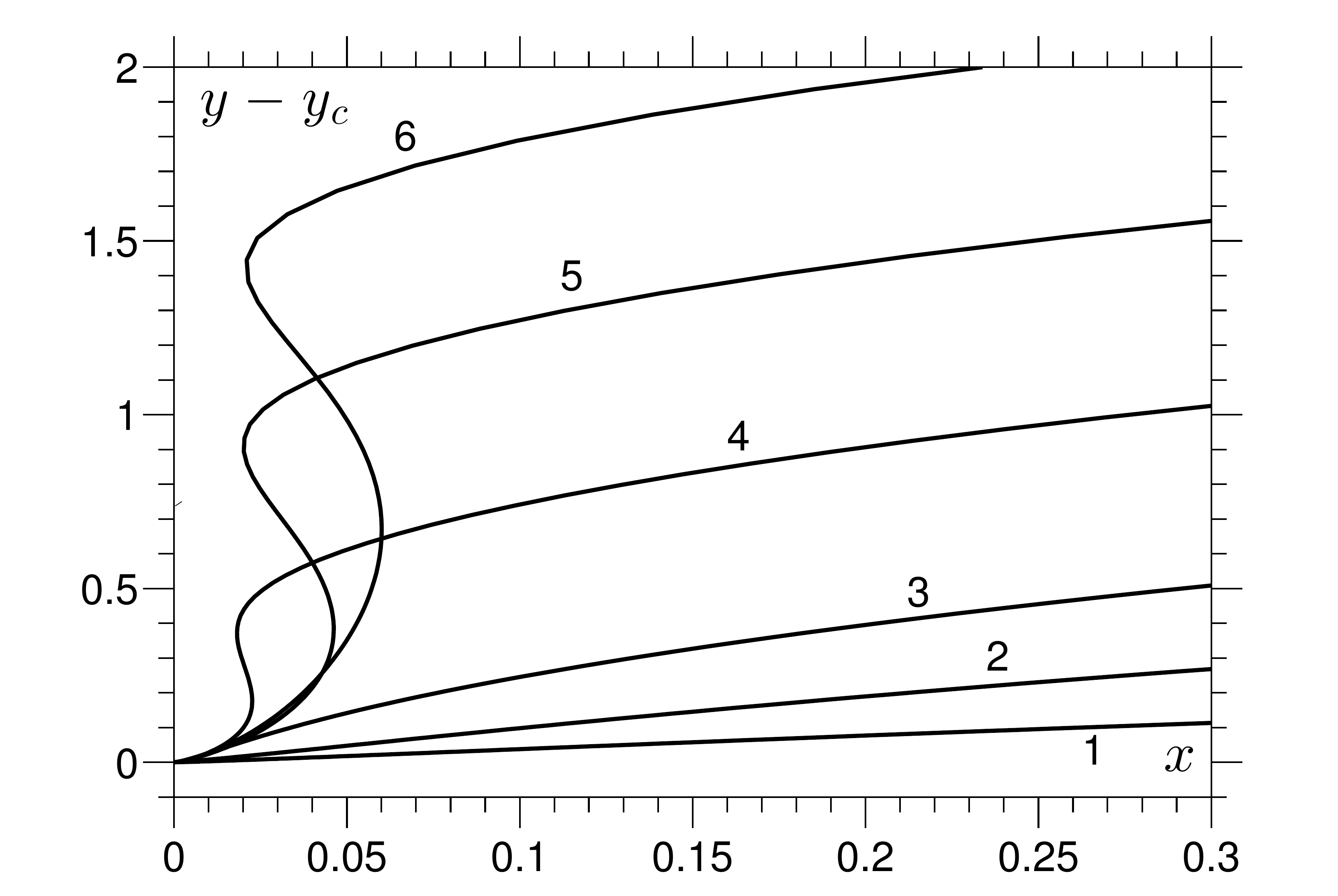}
 \caption{Free-surface profiles for the case of $\bar{\mu}=10^{-3}$ at different elevations 1:~$\triangle y=-0.5$, 2:~$\triangle y=-1$, 3:~$\triangle y=-1.5$, 4:~$\triangle y=-2$, 5:~$\triangle y=-2.5$ and 6:~$\triangle y=-3$.}
 \label{F:vandre_shapes}
\end{figure}

Similar oscillations are observed for $\theta_M$ where, notably, along the unstable branch the maximum apparent angle can increase beyond $180^\circ$.  This can be seen in Figure~\ref{F:vandre_shapes}, where between curves $3$ and $4$ a hump develops in the free-surface profile.  In this region, the gradient of the free-surface is negative so that the angle which this portion makes with the solid, i.e.\ the apparent angle, gives a maximum of $\theta_M>180^\circ$.

A comparison between the critical capillary numbers computed in \cite{vandre13} and those observed here is not likely to be exact, as different flow geometries have been used.  However, in cases where $Ca_c$ is not too large and the effect of the flow geometry is relatively weak we may expect our values to be comparable to theirs.  In Table~\ref{T:1} this is shown to be the case.  Computed values of both $Ca_c$ and  $\theta_M$ at this capillary number, which is a very sensitive measure, agree well across a large range of viscosity ratios.  It is not our intention to analyse this data in detail, as this has already been achieved in \cite{vandre13}, but this table will be useful as a benchmark for future computational work in this area.
\begin{table}
\begin{center}
\begin{tabular}{|c|c|c|c|c|}
  \hline
 \multirow{2}{*}{$\bar{\mu}$} &    \multicolumn{2}{|c|}{$Ca_c$} &   \multicolumn{2}{|c|}{$\theta_M(Ca_c)$} \\
 \cline{2-5}\\
                         & Computed &  \cite{vandre13}          &  Computed    & \cite{vandre13} \\
  \hline
  $10^{-1}$     & $0.09$ & $0.09$  & $161^\circ$   & $159^\circ$ \\
  $10^{-2}$     & $0.21$ & $0.22$  & $165^\circ$   & $166^\circ$ \\
  $10^{-3}$     & $0.39$ & $0.41$  & $171^\circ$   & $172^\circ$ \\
  $10^{-4}$     & $0.65$ & $0.65$  & $176^\circ$   & $176^\circ$ \\
  \hline

\end{tabular}
\end{center} \caption{Comparison of the critical capillary numbers $Ca_c$ and free-surface angle at the inflection point $\theta_M$ obtained from our code with those obtained in \cite{vandre13} across a range of viscosity ratios $\bar{\mu}$.}\label{T:1}
\end{table}



\section*{Acknowledgements}  JES would like to thank Terry Blake and Yulii Shikhmurzaev for numerous insightful discussions and continual encouragement over the years; Duncan Lockerby for lending his expertise with aspects of the gas dynamics; and the Referees of the paper whose valuable comments helped to improve the manuscript.

\bibliographystyle{jfm}
\bibliography{Bibliography}

\begin{thebibliography}{94}
\expandafter\ifx\csname natexlab\endcsname\relax\def\natexlab#1{#1}\fi

\bibitem[Agarwal \& Prabhu(2008)]{agrawal08}
{\sc Agarwal, A. \& Prabhu, S.~V.} 2008 Survey on measurement of tangential
  momentum accommodation coefficient. {\em Journal of Vacuum Science \&
  Technology A\/} {\bf 26}, 634--645.

\bibitem[Allen \& Raabe(1982)]{allen82}
{\sc Allen, M.~D. \& Raabe, O.~G.} 1982 Re-evaluation of {Millikan's} oil drop
  data for the motion of small particles in air. {\em Journal of Aerosol
  Science\/} {\bf 6}, 537--547.

\bibitem[Andrews \& Harris(1995)]{andrews95}
{\sc Andrews, M.~K. \& Harris, P.~D.} 1995 Damping and gas viscosity
  measurements using a microstructure. {\em Sensors and Actuators A\/} {\bf
  49}, 103--108.

\bibitem[Benkreira \& Ikin(2010)]{benkreira10}
{\sc Benkreira, H. \& Ikin, J.~B.} 2010 Dynamic wetting and gas viscosity
  effects. {\em Chemical Engineering Science\/} {\bf 65}, 1790--1796.

\bibitem[Benkreira \& Khan(2008)]{benkreira08}
{\sc Benkreira, H. \& Khan, M.~I.} 2008 Air entrainment in dip coating under
  reduced air pressures. {\em Chemical Engineering Science\/} {\bf 63},
  448--459.

\bibitem[Bird(1994)]{bird94}
{\sc Bird, G.~A.} 1994 {\em Molecular Gas Dynamics and the Direct Simulation of
  Gas Flows\/}. Clarendon Press.

\bibitem[Blake(2006)]{blake06}
{\sc Blake, T.~D.} 2006 The physics of moving wetting lines. {\em Journal of
  Colloid and Interface Science\/} {\bf 299}, 1--13.

\bibitem[Blake {\em et~al.\/}(1994)Blake, Clarke \& Ruschak]{blake94}
{\sc Blake, T.~D., Clarke, A. \& Ruschak, K.~J.} 1994 Hydrodynamic assist of
  wetting. {\em AIChE Journal\/} {\bf 40}, 229--242.

\bibitem[Blake \& {De Coninck}(2002)]{blake02a}
{\sc Blake, T.~D. \& {De Coninck}, J.} 2002 The influence of solid-liquid
  interactions on dynamic wetting. {\em Advances in Colloid and Interface
  Science\/} {\bf 96}, 21--36.

\bibitem[Blake {\em et~al.\/}(2004)Blake, Dobson \& Ruschak]{blake04a}
{\sc Blake, T.~D., Dobson, R.~A. \& Ruschak, K.~J.} 2004 Wetting at high
  capillary numbers. {\em Journal of Colloid and Interface Science\/} {\bf
  279}, 198--205.

\bibitem[Blake \& Haynes(1969)]{blake69}
{\sc Blake, T.~D. \& Haynes, J.~M.} 1969 Kinetics of liquid/liquid
  displacement. {\em Journal of Colloid and Interface Science\/} {\bf 30},
  421--423.

\bibitem[Blake \& Ruschak(1979)]{blake79}
{\sc Blake, T.~D. \& Ruschak, K.~J.} 1979 A maximum speed of wetting. {\em
  Nature\/} {\bf 282}, 489--491.

\bibitem[Blake \& Shikhmurzaev(2002)]{blake02}
{\sc Blake, T.~D. \& Shikhmurzaev, Y.~D.} 2002 Dynamic wetting by liquids of
  different viscosity. {\em Journal of Colloid and Interface Science\/} {\bf
  253}, 196--202.

\bibitem[Bouwhuis {\em et~al.\/}(2012)Bouwhuis, {van der Veen}, Tran, Keij,
  Winkels, Peters, {van der Meer}, Sun, Snoeijer \& Lohse]{bouwhuis12}
{\sc Bouwhuis, W., {van der Veen}, R. C.~A., Tran, T., Keij, D.~L., Winkels,
  K.~G., Peters, I.~R., {van der Meer}, D., Sun, C., Snoeijer, J.~H. \& Lohse,
  D.} 2012 Maximal air bubble entrainment at liquid-drop impact. {\em Physical
  Review Letters\/} {\bf 109}, 264501.

\bibitem[Burley \& Kennedy(1976)]{burley76}
{\sc Burley, R. \& Kennedy, B.~S.} 1976 An experimental study of air
  entrainment at a solid/liquid/gas interface. {\em Chemical Engineering
  Science\/} {\bf 31}, 901--911.

\bibitem[Bussmann {\em et~al.\/}(2000)Bussmann, Chandra \&
  Mostaghimi]{bussmann00}
{\sc Bussmann, M., Chandra, S. \& Mostaghimi, J.} 2000 Modeling the splash of a
  droplet impacting a solid surface. {\em Physics of Fluids\/} {\bf 12},
  3121--3132.

\bibitem[Cercignani(2000)]{cercignani00}
{\sc Cercignani, C.} 2000 {\em Rarefied Gas Dynamics: From Basic Concepts to
  Actual Calculations\/}. Cambridge University Press.

\bibitem[Chan {\em et~al.\/}(2013)Chan, Srivastava, Marchand, Andreotti,
  Biferale, Toschi \& Snoeijer]{chan13}
{\sc Chan, T.~S., Srivastava, S., Marchand, A., Andreotti, B., Biferale, L.,
  Toschi, F. \& Snoeijer, J.~H.} 2013 Hydrodynamics of air entrainment by
  moving contact lines. {\em Physics of Fluids\/} {\bf 25}, 074105.

\bibitem[Chapman \& Cowling(1970)]{chapman70}
{\sc Chapman, S. \& Cowling, T.~G.} 1970 {\em The Mathematical Theory of
  Non-Uniform Gases: An Account of the Kinetic Theory of Viscosity, Thermal
  conduction and Diffusion in Gases\/}. Cambridge University Press.

\bibitem[Clarke(2002)]{clarke02a}
{\sc Clarke, A.} 2002 Coating on a rough surface. {\em AIChE Journal\/} {\bf
  48}, 2149--2156.

\bibitem[Clarke \& Stattersfield(2006)]{clarke06}
{\sc Clarke, A. \& Stattersfield, E.} 2006 Direct evidence supporting nonlocal
  hydrodynamic influence on the dynamic contact angle. {\em Physics of
  Fluids\/} {\bf 18}, 048106.

\bibitem[Cox(1986)]{cox86}
{\sc Cox, R.~G.} 1986 The dynamics of the spreading of liquids on a solid
  surface. {Part 1. Viscous flow}. {\em Journal of Colloid and Interface
  Science\/} {\bf 168}, 169--194.

\bibitem[{De Coninck} \& Blake(2008)]{deconinck08}
{\sc {De Coninck}, J. \& Blake, T.~D.} 2008 Wetting and molecular dynamics
  simulations of simple fluids. {\em Annual Review of Materials Research\/}
  {\bf 38}, 1--22.

\bibitem[{de Ruiter} {\em et~al.\/}(2015){de Ruiter}, Lagraauw, Ende \&
  Mugele]{ruiter15}
{\sc {de Ruiter}, J., Lagraauw, R., Ende, D. \& Mugele, F.} 2015
  Wettability-independent bouncing on flat surfaces mediated by thin air films.
  {\em Nature Physics\/} {\bf 11}, 48--53.

\bibitem[{de Ruiter} {\em et~al.\/}(2012){de Ruiter}, Oh, Ende \&
  Mugele]{ruiter12}
{\sc {de Ruiter}, J., Oh, J.~M., Ende, D. \& Mugele, F.} 2012 Dynamics of
  collapse of air films in drop impact. {\em Physical Review Letters\/} {\bf
  108}, 074505.

\bibitem[Dejaguin \& Levi(1964)]{derjaguin64}
{\sc Dejaguin, B.~V. \& Levi, S.~M.} 1964 {\em Film Coating Theory\/}. Focal
  Press, London.

\bibitem[Derby(2010)]{derby10}
{\sc Derby, B.} 2010 Inkjet printing of functional and structural materials:
  fluid property requirements, feature stability and resolution. {\em Annual
  Review of Materials Research\/} {\bf 40}, 395--414.

\bibitem[Driscoll \& Nagel(2011)]{driscoll11}
{\sc Driscoll, M.~M. \& Nagel, S.~R.} 2011 Ultrafast interference imaging of
  air in splashing dynamics. {\em Physical Review Letters\/} {\bf 107}, 154502.

\bibitem[Duchemin \& Josserand(2012)]{duchemin12}
{\sc Duchemin, L. \& Josserand, C.} 2012 Rarefied gas correction for the bubble
  entrapment singularity in drop impacts. {\em Comptes Rendus M\'{e}canique\/}
  {\bf 340}, 797--803.

\bibitem[Duez {\em et~al.\/}(2007)Duez, Ybert, Clanet \& Bocquet]{duez07}
{\sc Duez, C., Ybert, C., Clanet, C. \& Bocquet, L.} 2007 Making a splash with
  water repellency. {\em Nature Physics\/} {\bf 3}, 180--183.

\bibitem[{Dussan V}(1976)]{dussan76}
{\sc {Dussan V}, E.~B.} 1976 The moving contact line: the slip boundary
  condition. {\em Journal of Fluid of Mechanics\/} {\bf 77}, 665--684.

\bibitem[{Dussan V}(1977)]{dussan77}
{\sc {Dussan V}, E.~B.} 1977 Immiscible liquid displacement in a capillary
  tube: the moving contact line. {\em AIChE Journal\/} {\bf 23}, 131--133.

\bibitem[{Dussan V} \& Davis(1974)]{dussan74}
{\sc {Dussan V}, E.~B. \& Davis, S.~H.} 1974 On the motion of a fluid-fluid
  interface along a solid surface. {\em Journal of Fluid Mechanics\/} {\bf 65},
  71--95.

\bibitem[{Dussan V} {\em et~al.\/}(1991){Dussan V}, Ram\'{e} \&
  Garoff]{dussan91}
{\sc {Dussan V}, E.~B., Ram\'{e}, E. \& Garoff, S.} 1991 On identifying the
  appropriate boundary conditions at a moving contact line: an experimental
  investigation. {\em Journal of Fluid Mechanics\/} {\bf 230}, 97--116.

\bibitem[Eggers(2004)]{eggers04a}
{\sc Eggers, J.} 2004 Hydrodynamic theory of forced dewetting. {\em Physical
  Review Letters\/} {\bf 93}, 094502.

\bibitem[Eggers {\em et~al.\/}(2010)Eggers, Fontelos, Josserand \&
  Zaleski]{eggers10}
{\sc Eggers, J., Fontelos, M.~A., Josserand, C. \& Zaleski, S.} 2010 Drop
  dynamics after impact on a solid wall: Theory and simulations. {\em Physics
  of Fluids\/} {\bf 22}, 062101.

\bibitem[{Gad-el-Hak}(2006)]{gadelhak06}
{\sc {Gad-el-Hak}, M.} 2006 Flow physics. In {\em MEMS: Introduction and
  Fundamentals\/} (ed. M.~{Gad-el-Hak}), pp. 4--1 -- 4--36. CRC Press.

\bibitem[Hadjiconstantinou(2003)]{hadji03}
{\sc Hadjiconstantinou, N.~G.} 2003 Comment on {Cercignani's} second-order slip
  coefficient. {\em Physics of Fluids\/} {\bf 15}, 2352--2354.

\bibitem[Hocking(1976)]{hocking76}
{\sc Hocking, L.~M.} 1976 A moving fluid interface on a rough surface. {\em
  Journal of Fluid Mechanics\/} {\bf 76}, 801--807.

\bibitem[Hoffman(1975)]{hoffman75}
{\sc Hoffman, R.~L.} 1975 A study of the advancing interface. {I}. {Interface}
  shape in liquid-gas systems. {\em Journal of Colloid and Interface Science\/}
  {\bf 50}, 228--241.

\bibitem[Huh \& Mason(1977)]{huh77}
{\sc Huh, C. \& Mason, S.~G.} 1977 The steady movement of a liquid meniscus in
  a capillary tube. {\em Journal of Fluid Mechanics\/} {\bf 81}, 401--409.

\bibitem[Huh \& Scriven(1971)]{huh71}
{\sc Huh, C. \& Scriven, L.~E.} 1971 Hydrodynamic model of steady movement of a
  solid/liquid/fluid contact line. {\em Journal of Colloid and Interface
  Science\/} {\bf 35}, 85--101.

\bibitem[Jacqmin(2002)]{jacqmin02}
{\sc Jacqmin, D.} 2002 Very, very fast wetting. {\em Journal of Fluid
  Mechanics\/} {\bf 455}, 347--358.

\bibitem[Kistler(1993)]{kistler93}
{\sc Kistler, S.~F.} 1993 Hydrodynamics of wetting. In {\em Wettability\/} (ed.
  J.~C. Berg), pp. 311--429. Marcel Dekker.

\bibitem[Kistler \& Scriven(1983)]{kistler83}
{\sc Kistler, S.~F. \& Scriven, L.~E.} 1983 {Coating flows}. In {\em
  Computational Analysis of Polymer Processing\/} (ed. J.~R.~A. Pearson \&
  S.~M. Richardson), pp. 243--299. Applied Science Publishers London and New
  York.

\bibitem[Kolinski {\em et~al.\/}(2014)Kolinski, Mahadevan \&
  Rubinstein]{kolinski14}
{\sc Kolinski, J.~M., Mahadevan, L. \& Rubinstein, S.~M.} 2014 Drops can bounce
  from perfectly hydrophilic surfaces. {\em Europhysics Letters\/} {\bf 108},
  24001.

\bibitem[Kolinski {\em et~al.\/}(2012)Kolinski, Rubinstein, Mandre, Brenner,
  Weitz \& Mahadevan]{kolinski12}
{\sc Kolinski, J.~M., Rubinstein, S.~M., Mandre, S., Brenner, M.~P., Weitz,
  D.~A. \& Mahadevan, L.} 2012 Skating on a film of air: drops impacting on a
  surface. {\em Physical Review Letters\/} {\bf 108}, 074503.

\bibitem[Koplik \& Banavar(1995)]{koplik95}
{\sc Koplik, J. \& Banavar, J.~R.} 1995 Continuum deductions from molecular
  hydrodynamics. {\em Annual Review of Fluid Mechanics\/} {\bf 27}, 257--292.

\bibitem[Lauga {\em et~al.\/}(2007)Lauga, Brenner \& Stone]{lauga07}
{\sc Lauga, E., Brenner, M. \& Stone, H.~A.} 2007 Microfluidics: the no-slip
  boundary condition. {\em Springer Handbook of Experimental Fluid Mechanics\/}
  pp. 1219--1240.

\bibitem[Lockerby {\em et~al.\/}(2004)Lockerby, Reese, Emerson \&
  Barber]{lockerby04}
{\sc Lockerby, D.~A., Reese, J.~M., Emerson, D.~R. \& Barber, R.~W.} 2004
  Velocity boundary condition at solid walls in rarefied gas calculations. {\em
  Physical Review E\/} {\bf 70}, 017303.

\bibitem[Lockerby {\em et~al.\/}(2005{\natexlab{{\em a\/}}})Lockerby, Reese \&
  Gallis]{lockerby05a}
{\sc Lockerby, D.~A., Reese, J.~M. \& Gallis, M.~A.} 2005{\natexlab{{\em a\/}}}
  Capturing the knudsen layer in continuum-fluid models of nonequilibrium gas
  flows. {\em AIAA Journal\/} {\bf 43}, 1391--1393.

\bibitem[Lockerby {\em et~al.\/}(2005{\natexlab{{\em b\/}}})Lockerby, Reese \&
  Gallis]{lockerby05}
{\sc Lockerby, D.~A., Reese, J.~M. \& Gallis, M.~A.} 2005{\natexlab{{\em b\/}}}
  The usefulness of higher-order constitutive relations for describing the
  {Knudsen} layer. {\em Physical of Fluids\/} {\bf 17}, 100609.

\bibitem[Mandre \& Brenner(2012)]{mandre12}
{\sc Mandre, S. \& Brenner, M.~P.} 2012 The mechanism of a splash on a dry
  solid surface. {\em Journal of Fluid Mechanics\/} {\bf 690}, 148--172.

\bibitem[Mani {\em et~al.\/}(2010)Mani, Mandre \& Brenner]{mani10}
{\sc Mani, M., Mandre, S. \& Brenner, M.~P.} 2010 Events before droplet
  splashing on a solid surface. {\em Journal of Fluid Mechanics\/} {\bf 647},
  163--185.

\bibitem[Marchand {\em et~al.\/}(2012)Marchand, Chan, Snoeijer \&
  Andreotti]{marchand12}
{\sc Marchand, A., Chan, T.~S., Snoeijer, J.~H. \& Andreotti, B.} 2012 Air
  entrainment by contact lines of a solid plate plunged into a viscous fluid.
  {\em Physical Review Letters\/} {\bf 108}, 204501.

\bibitem[Maxwell(1867)]{maxwell67}
{\sc Maxwell, J.~C.} 1867 On the dynamical theory of gases. {\em Philosophical
  Transactions of the Royal Society of London\/} {\bf 157}, 49--88.

\bibitem[Maxwell(1879)]{maxwell79}
{\sc Maxwell, J.~C.} 1879 On stresses in rarified gases arising from
  inequalities of temperature. {\em Philosophical Transactions of the Royal
  Society of London\/} {\bf 170}, 231--256.

\bibitem[Millikan(1923)]{millikan23}
{\sc Millikan, R.~A.} 1923 The general law of fall of a small spherical body
  through a gas, and its bearing upon the nature of molecular reflection from
  surfaces. {\em Physical Review\/} {\bf 22}, 1--23.

\bibitem[Mues {\em et~al.\/}(1989)Mues, Hens \& Boiy]{mues89}
{\sc Mues, W., Hens, J. \& Boiy, L.} 1989 Observation of a dynamic wetting
  process using laser-{Doppler} velocimetry. {\em AIChE Journal\/} {\bf 35},
  1521--1526.

\bibitem[Navier(1823)]{navier23}
{\sc Navier, C. L. M.~H.} 1823 M\'emoire sur les lois mouvement des fluides.
  {\em M\'em. de l'Acad. de Sciences l'Inst. de France\/} {\bf 6}, 389--440.

\bibitem[Ngan \& {Dussan V}(1982)]{ngan82}
{\sc Ngan, C.~G. \& {Dussan V}, E.~B.} 1982 On the nature of the dynamic
  contact angle: an experimental study. {\em Journal of Fluid Mechanics\/} {\bf
  118}, 27--40.

\bibitem[Ram\'{e} \& Garoff(1996)]{rame96}
{\sc Ram\'{e}, E. \& Garoff, S.} 1996 Microscopic and macroscopic dynamic
  interface shapes and the interpretation of dynamic contact angles. {\em
  Journal of Colloid and Interface Science\/} {\bf 177}, 234--244.

\bibitem[Rein(1993)]{rein93}
{\sc Rein, M.} 1993 Phenomena of liquid drop impact on solid and liquid
  surfaces. {\em Fluid Dynamics Research\/} {\bf 12}, 61--93.

\bibitem[Rein \& Delplanque(2008)]{rein08}
{\sc Rein, M. \& Delplanque, J.~{-P}.} 2008 The role of air entrainment on the
  outcome of drop impact on a solid surface. {\em Acta Mechanica\/} {\bf 201},
  105--118.

\bibitem[Riboux \& Gordillo(2014)]{riboux14}
{\sc Riboux, G. \& Gordillo, J.~M.} 2014 Experiments of drops impacting a
  smooth solid surface: a model of the critical impact speed for drop
  splashing. {\em Physical Review Letters\/} {\bf 113}, 024507.

\bibitem[Ruschak(1980)]{ruschak80}
{\sc Ruschak, K.~J.} 1980 A method for incorporating free boundaries with
  surface tension in finite element fluid-flow simulators. {\em International
  Journal for Numerical Methods in Engineering\/} {\bf 15}, 639--648.

\bibitem[Schroll {\em et~al.\/}(2010)Schroll, Josserand, Zaleski \&
  Zhang]{schroll10}
{\sc Schroll, R.~D., Josserand, C., Zaleski, S. \& Zhang, W.~W.} 2010 Impact of
  a viscous liquid drop. {\em Physical Review Letters\/} {\bf 104}, 034504.

\bibitem[Shikhmurzaev(1997)]{shik97}
{\sc Shikhmurzaev, Y.~D.} 1997 Moving contact lines in liquid/liquid/solid
  systems. {\em Journal of Fluid Mechanics\/} {\bf 334}, 211--249.

\bibitem[Shikhmurzaev(2006)]{shik06}
{\sc Shikhmurzaev, Y.~D.} 2006 {Singularities at the moving contact line.
  Mathematical, physical and computational aspects}. {\em Physica D\/} {\bf
  217}, 121--133.

\bibitem[Shikhmurzaev(2007)]{shik07}
{\sc Shikhmurzaev, Y.~D.} 2007 {\em Capillary Flows with Forming Interfaces\/}.
  Chapman \& Hall/CRC, Boca Raton.

\bibitem[Simpkins \& Kuck(2003)]{simpkins03}
{\sc Simpkins, P.~G. \& Kuck, V.~J.} 2003 On air entrainment in coatings. {\em
  Journal of Colloid and Interface Science\/} {\bf 263}, 562--571.

\bibitem[Smith {\em et~al.\/}(2003)Smith, Li \& Wu]{smith03}
{\sc Smith, F.~T., Li, L. \& Wu, G.~X.} 2003 Air cushioning with a
  lubrication/inviscid balance. {\em Journal of Fluid Mechanics\/} {\bf 482},
  291--318.

\bibitem[Snoeijer \& Andreotti(2013)]{snoeijer13}
{\sc Snoeijer, J.~H. \& Andreotti, B.} 2013 Moving contact lines: Scales,
  regimes, and dynamical transitions. {\em Annual Review of Fluid Mechanics\/}
  {\bf 45}, 269--292.

\bibitem[Snoeijer {\em et~al.\/}(2007)Snoeijer, Andreotti, Delon \&
  Fermigier]{snoeijer07}
{\sc Snoeijer, J.~H., Andreotti, B., Delon, G. \& Fermigier, M.} 2007
  Relaxation of a dewetting contact line. {Part 1.} {A} full-scale hydrodynamic
  calculation. {\em Journal of Fluid Mechanics\/} {\bf 579}, 63--83.

\bibitem[Snoeijer {\em et~al.\/}(2006)Snoeijer, Delon, Fermigier \&
  Andreotti]{snoeijer06}
{\sc Snoeijer, J.~H., Delon, G., Fermigier, M. \& Andreotti, B.} 2006 Avoided
  critical behavior in dynamically forced wetting. {\em Physical Review
  Letters\/} {\bf 96}, 174504.

\bibitem[Sprittles \& Shikhmurzaev(2012{\natexlab{{\em a\/}}})]{sprittles_pof2}
{\sc Sprittles, J.~E. \& Shikhmurzaev, Y.~D.} 2012{\natexlab{{\em a\/}}}
  Coalescence of liquid drops: different models versus experiment. {\em Physics
  of Fluids\/} {\bf 24}, 122105.

\bibitem[Sprittles \& Shikhmurzaev(2012{\natexlab{{\em b\/}}})]{sprittles_pof}
{\sc Sprittles, J.~E. \& Shikhmurzaev, Y.~D.} 2012{\natexlab{{\em b\/}}} The
  dynamics of liquid drops and their interaction with solids of varying
  wettabilities. {\em Physics of Fluids\/} {\bf 24}, 082001.

\bibitem[Sprittles \& Shikhmurzaev(2012{\natexlab{{\em
  c\/}}})]{sprittles_ijnmf}
{\sc Sprittles, J.~E. \& Shikhmurzaev, Y.~D.} 2012{\natexlab{{\em c\/}}} A
  finite element framework for describing dynamic wetting phenomena. {\em
  International Journal for Numerical Methods in Fluids\/} {\bf 68},
  1257--1298.

\bibitem[Sprittles \& Shikhmurzaev(2013)]{sprittles_jcp}
{\sc Sprittles, J.~E. \& Shikhmurzaev, Y.~D.} 2013 Finite element simulation of
  dynamic wetting flows as an interface formation process. {\em Journal of
  Computational Physics\/} {\bf 233}, 34--65.

\bibitem[Sprittles \& Shikhmurzaev(2014{\natexlab{{\em
  a\/}}})]{sprittles14_pre}
{\sc Sprittles, J.~E. \& Shikhmurzaev, Y.~D.} 2014{\natexlab{{\em a\/}}}
  Dynamics of liquid drops coalescing in the inertial regime. {\em Physical
  Review E\/} {\bf 89}, 063006.

\bibitem[Sprittles \& Shikhmurzaev(2014{\natexlab{{\em
  b\/}}})]{sprittles14_jfm2}
{\sc Sprittles, J.~E. \& Shikhmurzaev, Y.~D.} 2014{\natexlab{{\em b\/}}} A
  parametric study of the coalescence of liquid drops in a viscous gas. {\em
  Journal of Fluid Mechanics\/} {\bf 753}, 279--306.

\bibitem[Sundararajakumar \& Koch(1996)]{sundararajakumar96}
{\sc Sundararajakumar, R.~R. \& Koch, D.~L.} 1996 Non-continuum lubrication
  flows between particles colliding in a gas. {\em Journal of Fluid
  Mechanics\/} {\bf 313}, 283--308.

\bibitem[Tanner(1979)]{tanner79}
{\sc Tanner, L.~H.} 1979 The spreading of silicone drops on horizontal
  surfaces. {\em Journal of Physics D: Applied Physics\/} {\bf 12}, 1473--1484.

\bibitem[Thoroddsen {\em et~al.\/}(2005)Thoroddsen, Etoh \&
  Takehara]{thoroddsen05a}
{\sc Thoroddsen, S.~T., Etoh, T.~G. \& Takehara, K.} 2005 The air bubble
  entrapped under a drop impacting on a solid surface. {\em Journal of Fluid
  Mechanics\/} {\bf 545}, 203--212.

\bibitem[Vandre {\em et~al.\/}(2012)Vandre, Carvalho \& Kumar]{vandre12}
{\sc Vandre, E., Carvalho, M.~S. \& Kumar, S.} 2012 Delaying the onset of
  dynamic wetting failure through meniscus confinement. {\em Journal of Fluid
  Mechanics\/} {\bf 707}, 496--520.

\bibitem[Vandre {\em et~al.\/}(2013)Vandre, Carvalho \& Kumar]{vandre13}
{\sc Vandre, E., Carvalho, M.~S. \& Kumar, S.} 2013 On the mechanism of wetting
  failure during fluid displacement along a moving substrate. {\em Physics of
  Fluids\/} {\bf 25}, 102103.

\bibitem[Vandre {\em et~al.\/}(2014)Vandre, Carvalho \& Kumar]{vandre14}
{\sc Vandre, E., Carvalho, M.~S. \& Kumar, S.} 2014 Characteristics of air
  entrainment during dynamic wetting failure along a planar substrate. {\em
  Journal of Fluid Mechanics\/} {\bf 747}, 119--140.

\bibitem[Velarde(2011)]{velarde11}
{\sc Velarde, M.~G.} 2011 {\em Discussion and Debate: Wetting and Spreading
  Science - quo vadis?\/}. The European Physical Journal Special Topics.

\bibitem[Voinov(1976)]{voinov76}
{\sc Voinov, O.~V.} 1976 Hydrodynamics of wetting. {\em Fluid Dynamics\/} {\bf
  11}, 714--721.

\bibitem[Weinstein \& Ruschak(2004)]{weinstein04}
{\sc Weinstein, S.~J. \& Ruschak, K.~J.} 2004 Coating flows. {\em Annual Review
  of Fluid Mechanics\/} {\bf 36}, 29--53.

\bibitem[Wilson {\em et~al.\/}(2006)Wilson, Summers, Shikhmurzaev, Clarke \&
  Blake]{wilson06}
{\sc Wilson, M. C.~T., Summers, J.~L., Shikhmurzaev, Y.~D., Clarke, A. \&
  Blake, T.~D.} 2006 Nonlocal hydrodynamic influence on the dynamic contact
  angle: slip models versus experiment. {\em Physical Review E\/} {\bf 83},
  041606.

\bibitem[Xu(2007)]{xu07}
{\sc Xu, L.} 2007 Liquid drop splashing on smooth, rough, and textured
  surfaces. {\em Physical Review E\/} {\bf 75}, 056316.

\bibitem[Xu {\em et~al.\/}(2005)Xu, Zhang \& Nagel]{xu05}
{\sc Xu, L., Zhang, W. \& Nagel, S.~R.} 2005 Drop splashing on a dry smooth
  surface. {\em Physical Review Letters\/} {\bf 94}, 184505.

\bibitem[Yarin(2006)]{yarin06}
{\sc Yarin, A.~L.} 2006 Drop impact dynamics: Splashing, spreading, receding,
  bouncing. . . {\em Annual Review of Fluid Mechanics\/} {\bf 38}, 159--192.

\end{thebibliography}

\end{document}